\documentclass[11pt,twoside,letterpaper]{article} 
\usepackage{times,fancyhdr}
\usepackage[dvips]{graphicx}
\usepackage{epsfig}
\usepackage{amssymb}
\usepackage{amsmath}
\usepackage{amsfonts}
\usepackage{amsthm,amscd}
\usepackage{amsbsy}
\usepackage{latexsym}
\usepackage{bm}
\usepackage{url} 			
\usepackage{layout}
\usepackage{pslatex}
\usepackage{cite}
\usepackage{fleqn} 		
\usepackage{makeidx}
\makeindex 						
\usepackage{layout} 	
\usepackage{epstopdf}	
\usepackage{color}
\usepackage{hyperref}
\usepackage[latin1]{inputenc}
\usepackage[T1]{fontenc}
\usepackage{poligraf} 
\usepackage[letter,cam,center]{crop} 
\usepackage{type1cm} 	
\usepackage{courier}	
\usepackage{lscape} 	

\usepackage{mathrsfs}

\newcommand{\beq}{\begin{equation}}
\newcommand{\eeq}{\end{equation}}
\newcommand{\bea}{\begin{eqnarray}}
\newcommand{\eea}{\end{eqnarray}}

\def\<{\langle}
\def\>{\rangle}


\makeatletter 
\def\cleardoublepage{\clearpage\if@twoside \ifodd\c@page\else%
    \hbox{}%
    \thispagestyle{empty}%
    \newpage%
    \if@twocolumn\hbox{}\newpage\fi\fi\fi} 
\makeatother

\def\figurename{Figure}
\makeatletter
\renewcommand{\fnum@figure}[1]{\figurename~\thefigure.}
\makeatother

\def\tablename{Table}
\makeatletter
\renewcommand{\fnum@table}[1]{\tablename~\thetable.}
\makeatother

\begin{document}
\title{
\bfseries\scshape Threshold Model for Triggered Avalanches on Networks}
\author{\bfseries\itshape M. Ausloos \thanks{Marcel.Ausloos@ulg.ac.be}\\
GRAPES,  SUPRATECS, Universit\'e de Li\`ege,\\
 B5 Sart-Tilman, B-4000 Li\`ege, Euroland\\
\bfseries\itshape F. Petroni\thanks{fpetroni@unica.it}\\
Dipartimento di Scienze Economiche ed Aziendali,\\
Universit\`a degli studi di Cagliari, 09123 Cagliari, Italy}
\date{}
\maketitle
\thispagestyle{empty}
\setcounter{page}{1}

\begin{abstract}
Based on a theoretical model for opinion spreading on a network, through avalanches, the effect of external field  is now considered, by using methods from non-equilibrium statistical mechanics. The original part contains the implementation that the avalanche is only triggered when a local variable (a so called awareness) reaches and goes above a threshold. The dynamical rules are constrained to be as simple as possible, in order to sort out the basic features, though more elaborated variants are proposed. Several results are obtained for a  Erd\"os-R\'enyi network and interpreted through simple analytical laws, scale free or logistic map-like, i.e., (i) the sizes, durations, and number of  avalanches, including the respective distributions, (ii)  the number of times the external field is applied to one possible node before all nodes are found to be above the threshold, (iii)
  the number of nodes still below the threshold and
  the number of hot nodes (close to threshold) at each time step.
\end{abstract}

\section{Introduction}

From a statistical physics point of view, opinion formation is similar to epidemic or forest fire spreading, landslide, crystal growth, fracture, percolation and pertains to the abundant literature on phase transition studies. The modeling of such  phenomena, i.e. for reproducing stylized features,  is highly important in order to connect statistical physics with the socio-economic world. Some conceptual difficulty has been resolved due to the abandon of (Bravais or random) lattice based concepts in favor of studies of cases on more elaborate networks in which the statistical characteristics, like the number of  sites, neighbors, distances, weights and directivity of links are not trivial.

Euler invented network theory \cite{Euler}, for the K\"onigsberg (or kaliningrad) bridge crossing problem,    in the 1780s, but this subject remained a form of abstract mathematics. Graph Theory was born later to study similar problems, including molecular bonding \cite{Biggs,3}. An increased interest in the topics can be registered during the last decade, particularly due to their potential for an apparently quite  unbounded area of applications. Indeed, the inter-disciplinary (or rather trans-disciplinary) concept of ``network'' is frequently met in all scientific research areas  \cite{linked,pastor1,pastor2}, its covering field spanning from computer science to medicine and social psychology, e.g. see some work in   classical random graph study \cite{4}, non-equilibrium growing networks \cite{5},    WWW and Internet structure properties \cite{pastor1,8,newman1,newman3,newman4,newman2}, social networks of scientific collaborations \cite{newman1,newman3,newman4,newman2,10,ebeling}, paper citations \cite{ebeling,13},  collective listening habits and music genres \cite{lambi2,lambi3}, language \cite{rodgers}, and even finance \cite{glma,marl,dimatteo}. Relevant questions pertain to the critical dynamics of properties, not only on the network, but also  about the network structure itself.   Recent results on the dynamics of social networks   \cite{froncszak1,froncszak2} suggest  the occurrence of   phase transitions in a large class of models   \cite{lambiNbod}.   This is similar to percolation and  nucleation-growth problems.

As a paradigm for large-scale networks, colleagues often consider co-authorship or citation  networks  \cite{newman1,newman3,newman4,newman2,10,ebeling}, namely networks where nodes represent scientists, and where a link is drawn between them if they co-authored a common paper or cite some paper. Other examples, where a system (nodes) has an evolution between two
 phases on a network are  the
 unanimity rule on random networks \cite{galam1,galam2,galam3,sznajd,staufferJASSS}. Lattice based cases for social evolutions are on the contrary numerous, like prey-predator problems \cite{pekalskiPPreview}.

 Let us stress that the identification of the mechanisms responsible for diffusion and, possibly leading to
scientific avalanches   \cite{marsili} is primordial in a very general sense in order to understand the scientific response to external political decisions, and to develop efficient policy recommendations. 
 It is important to point that science spreading is usually modeled by master equations with auto-catalytic processes \cite{andrea}, or by epidemic models on static networks \cite{epidem,boguna,x,holyst}. In this article, however, we present a novel approach where the opinion formation and spreading is controlled by a triggering field and occurs only if a threshold is reached. Without going into comparative details the model can be thought to be connected to epidemic \cite{epidem} and forest fire spreading \cite{staufferJPhysF,Duarte,Perry,turcotte,miller}, and landslide \cite{malamud1,malamud2,nicodemi1,nicodemi2}. However the process as discussed here below occurring on a network seems to be novel and of crucial interest and differs from already known models like the Galam \cite{galam3} and/or Sznajd \cite{sznajd} model and their extensions. We focus on the development of neighbouring (scientific or not) ``opinions'' in the course of time, thereby eyeing the spreading of  ideas in the scientific community. We will indicate the possibility of variants. 
 
For our present framework, let us recall   that two 
decades ago  people looked at forest fires  \cite{Duarte,Perry}  
in particular in the case when a tree is ignited only if a neighbouring tree burns long enough.
This fact that wet wood does not yet burn corresponds to our awareness
below the threshold $\phi$.  We also agree that avalanche problems are similar to epidemics with threshold, see a recent paper \cite{boguna},
 when there is some self-organized information propagation, restricted to realistic ``social'' constraints.
 Finally no need to repeat that most of the ideas here below developed can encompass many networks, not only  those formed by scientists.  
 
\section{Avalanche on networks}

\subsection{Modelization}

Let a network be given and statistically characterized by a number $N$ of nodes (also called sites) and a number $L$ of links; as usual one call $k_i$ the degree of the node $i$, i.e. the number of links attached to the node $i$. The network considered below is chosen to be a so called Erd\"os-R\'enyi network,  i.e. each node $i$ $(i=1,\dots,N)$ has a probability equal to $p$ to be connected to another node $j$ $(j\neq i, j=1,\dots,N)$. Studies on other networks are opened investigations.

Each site $i$ can be considered to be an agent, e.g. a scientist or not, which has an $awareness$  $a_i$  about an opinion; $a_i$  is hereby taken from a uniform distribution between 0 and 1 at time $t_0$; this can be modified in further studies, so is the number of opinions
 
 Suppose that there is a predefined threshold of opinion $\phi$ at time $t_0$, ($\phi = 1$, here to reduce the number of parameters in the model; this can be modified in further papers as well). 
 When some $a_i$  is above  the threshold we consider that the site/agent spreads its {\it awareness} over its neighbors, as should be intuitively the case practically. Notice that in view of the hypothesis on the initial distribution of $a_i$ 's here above, no spontaneous spreading occurs at $t_0$, in this paper. Some triggering is needed. Consider that due to some cause, at time $t_1$,  a randomly chosen site has an increase of awareness such that $a_i(t_1) = a_i(t_0) + \nu $. The initial {\it increase of awareness}  value $\nu$ could be chosen in many various ways, i.e. $i$ and $t$ dependent; let it be a forever constant in the present discussion.  We consider that the spreading dynamics of the awareness could be restricted to a (to be decided in later variants) number $n_i$ of nodes which have necessarily an awareness below the threshold.  In the following we let $n_i$ be all nearest neighbor nodes which are not yet aware.
Some of these $n_i$ neighbors, due to the incoming information $a_i/n_i$  from $i$,  may reach the awareness threshold $\phi$; next, these, newly fully aware sites (agents)   also spread $information$, during the same time step, again to their own neighbors which have an awareness below threshold. The process can go one and lasts until none of the newly aware neighbor sites goes above the threshold; thus this avalanche stops.
 
We define the $size$ of the (first) $avalanche$ as the number of nodes  (here $N_1$) which have been reached by the $information$ spreading; the $duration$ of the avalanche is the number of ``levels'', away from the original site, i.e. $n_{T_1}$ which has been encountered in order to reach this stop during the time step. Sometimes this avalanche size is the number of nodes  $N$ in the network; in such a case the time duration of the avalanche is $T_1$.  

However it is possible  that the process stops even though not all nodes have been ``infected'' by the spreading opinion, but there is a new awareness value distribution in the network; these are new ``initial conditions''. We let the dynamical process repeat itself, under the same rules:
one arbitrary but a {\it not fully aware} node $i'$  is picked up;  its awareness is increased $a_{i'}(t_{h,1})= a_{i'}(t_{h,0}) +\nu$, for simplicity through the same $\nu$ value which led to the first triggering; we insist that we consider that the same $\nu$ value is used in order to reduce the number of parameters, but this can also be modified in further studies.

Since there is a new time origin for the subsequent process, it is useful to introduce a notation $h$ labeling the number of times a triggering has been activated;   the value $h = 2$ indicates the second sequence in time  ($t_2$) for which there is an external (or not) $field$  (implying a $\nu$ ) which modifies the opinion awareness of some site, here $i'$.

Thereafter, the spreading goes on again with the same rules as above up to the network is so called {\it fully aware}, i.e. when all nodes have $a_i \ge \phi$.
In so doing one can measure for a given $N$, $\nu$ and $p$  (and for this $\phi$)
\begin{enumerate}
\item the size of each avalanche
\item the time duration of each avalanche
\item the number of times the external field is applied to one possible node before all nodes are found to be above the threshold
\item which is equivalent to ... the number of avalanches $N_A$ before all nodes are above the threshold
\item the number of times the external field is added before an avalanche starts
\item the number of nodes still below the threshold at each time step
\item the number of hot nodes (close to threshold) at each time step.
\end{enumerate}

\begin{figure}
\centering
\includegraphics[height=5cm,width=4cm]{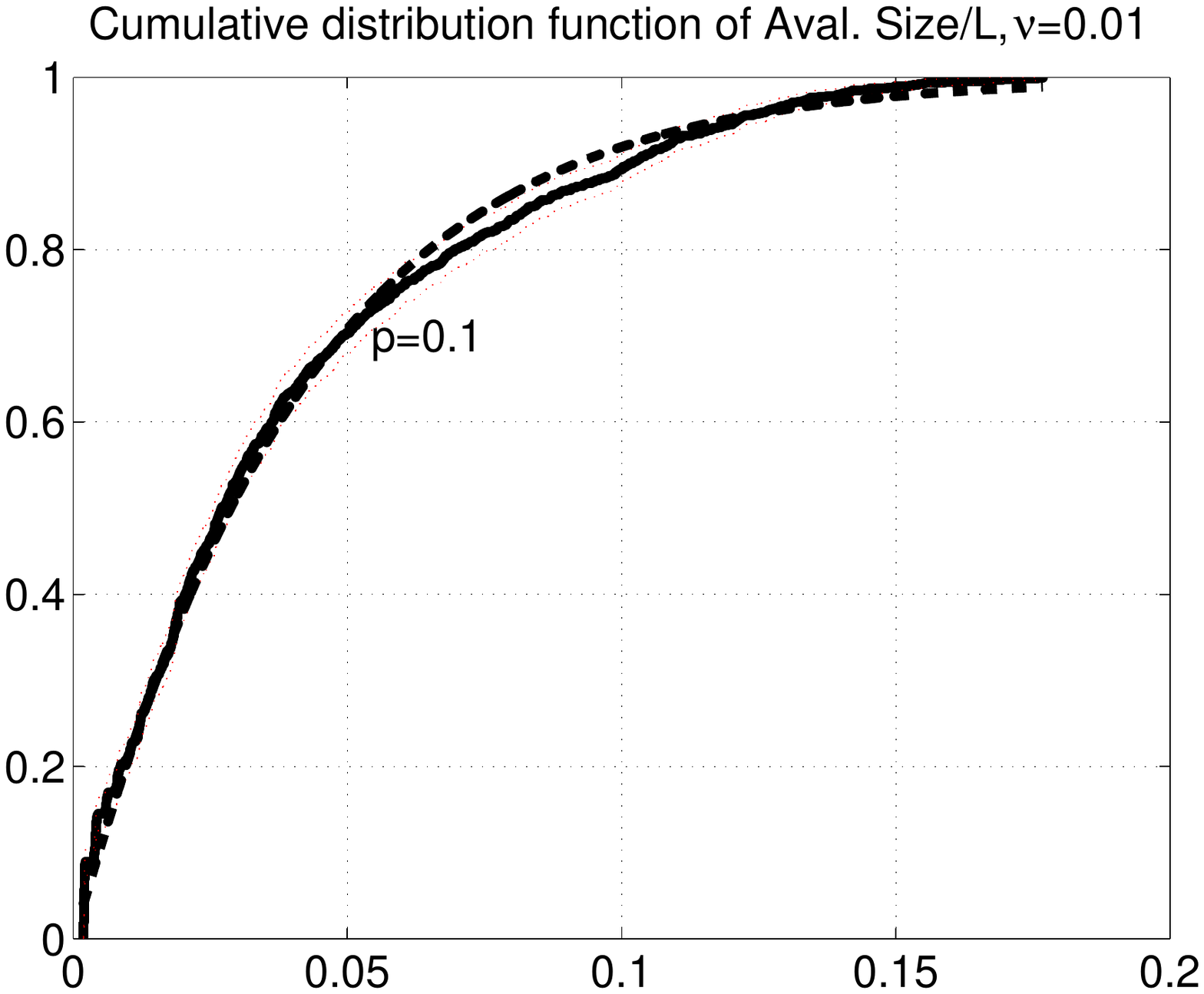}
\includegraphics[height=5cm,width=4cm]{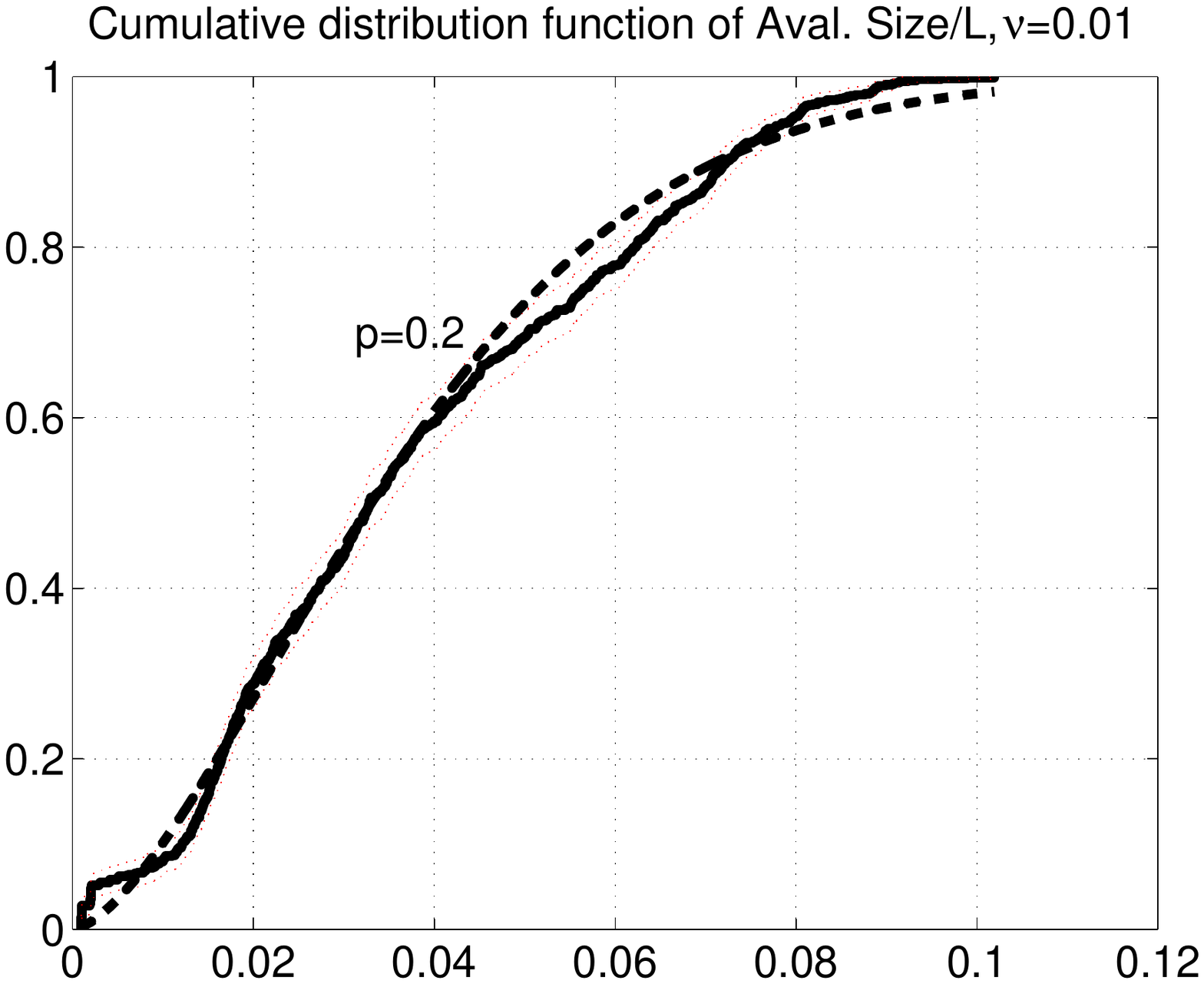}
\includegraphics[height=5cm,width=4cm]{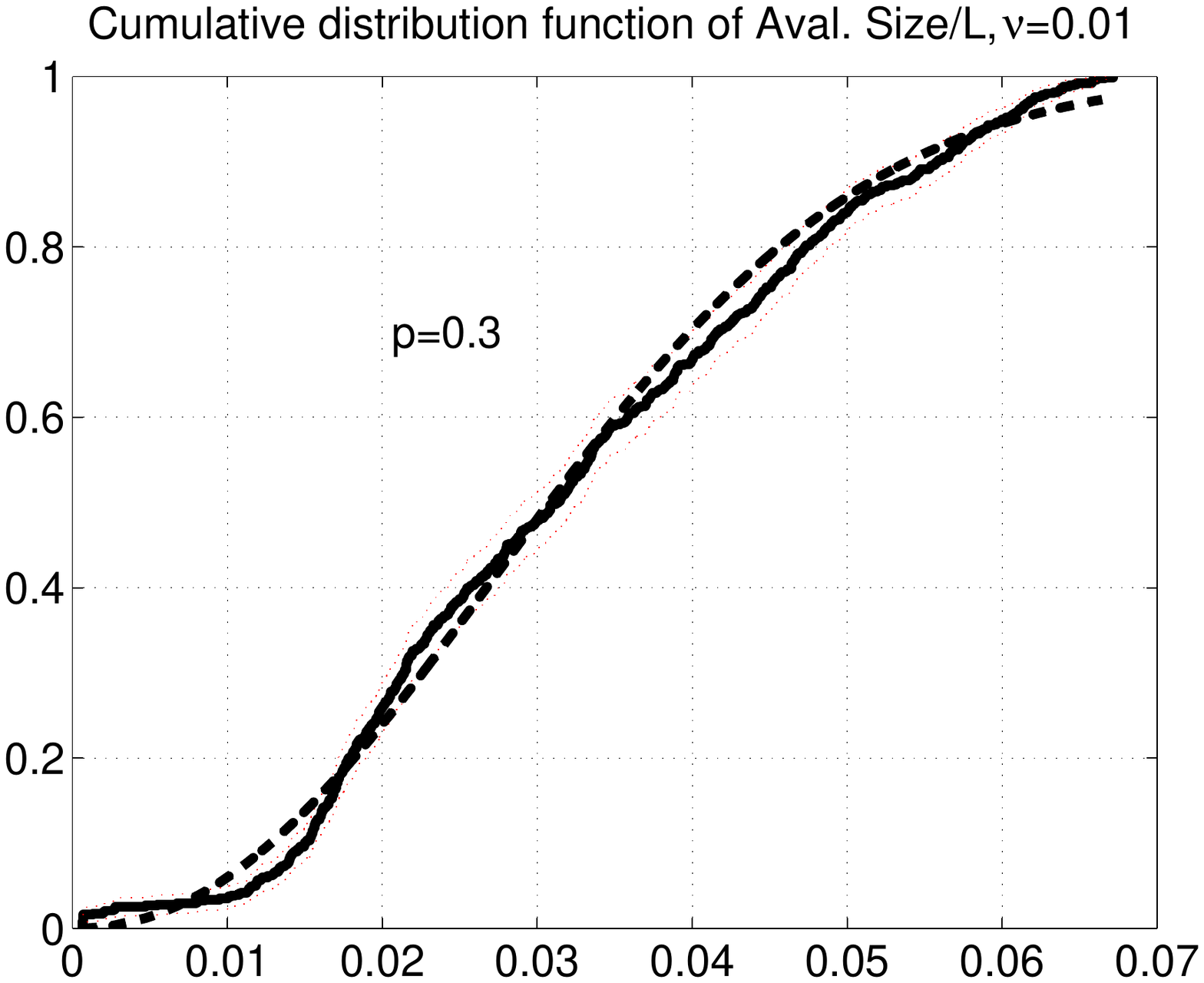}
\includegraphics[height=5cm,width=4cm]{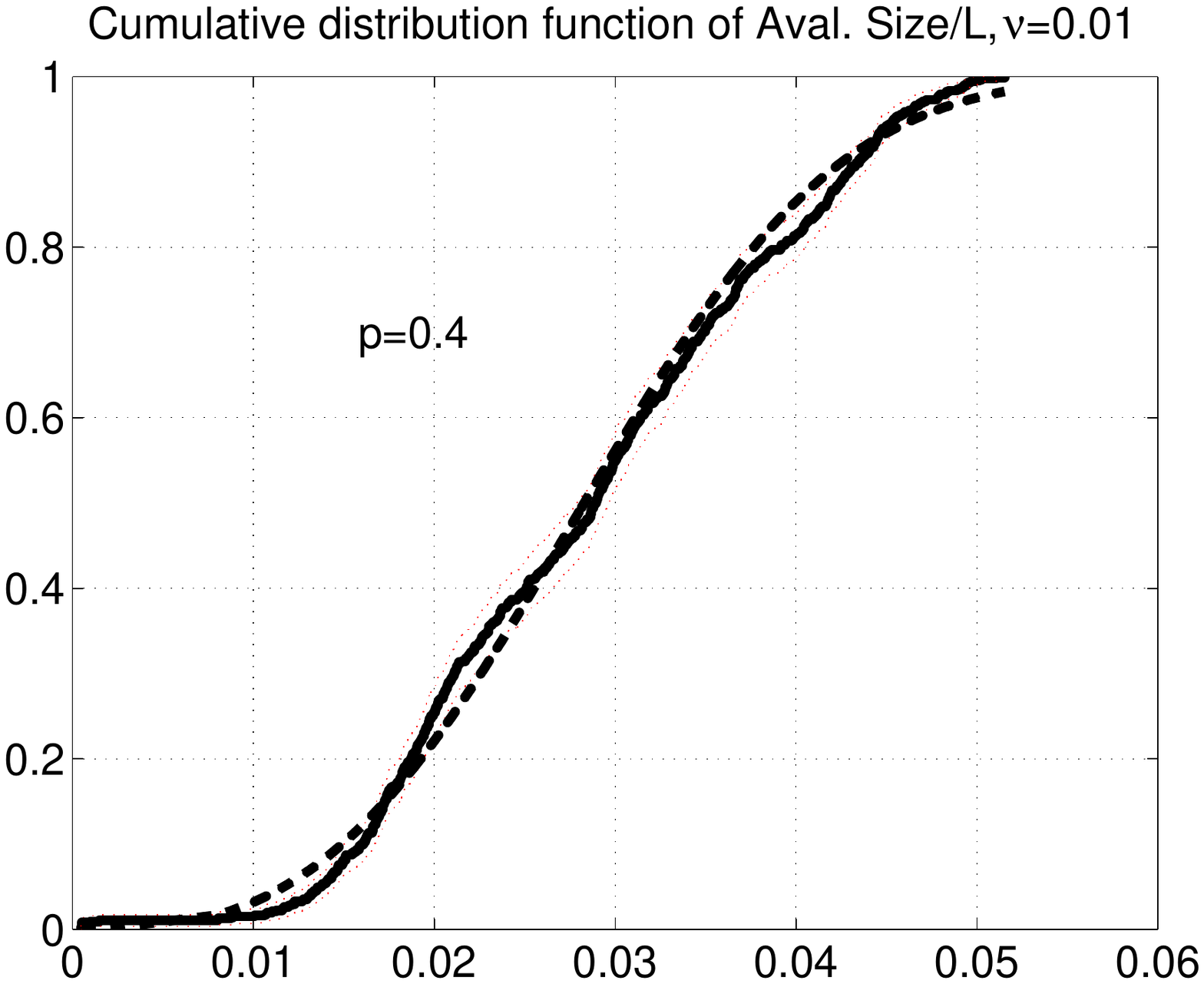}
\includegraphics[height=5cm,width=4cm]{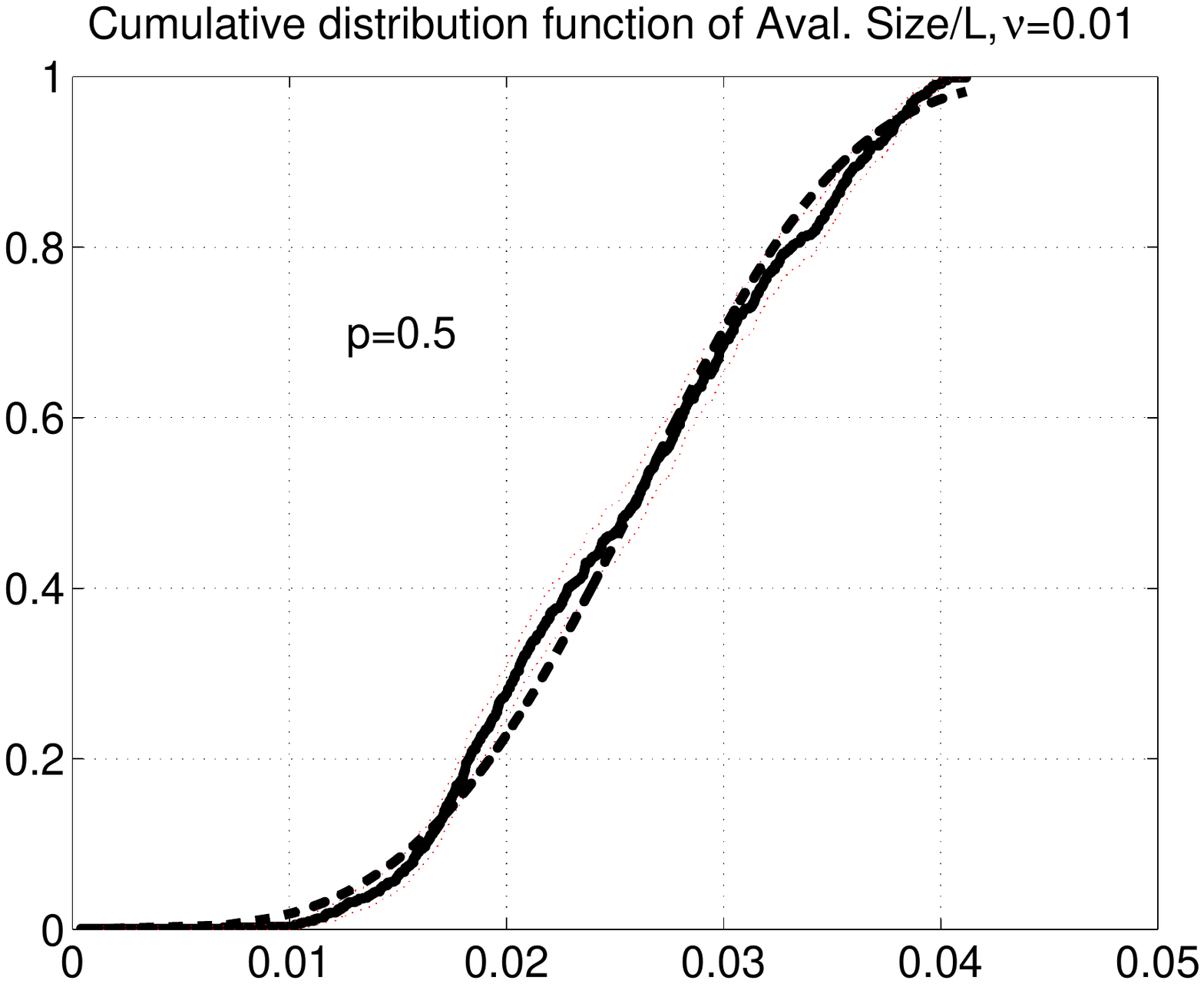}
\includegraphics[height=5cm,width=4cm]{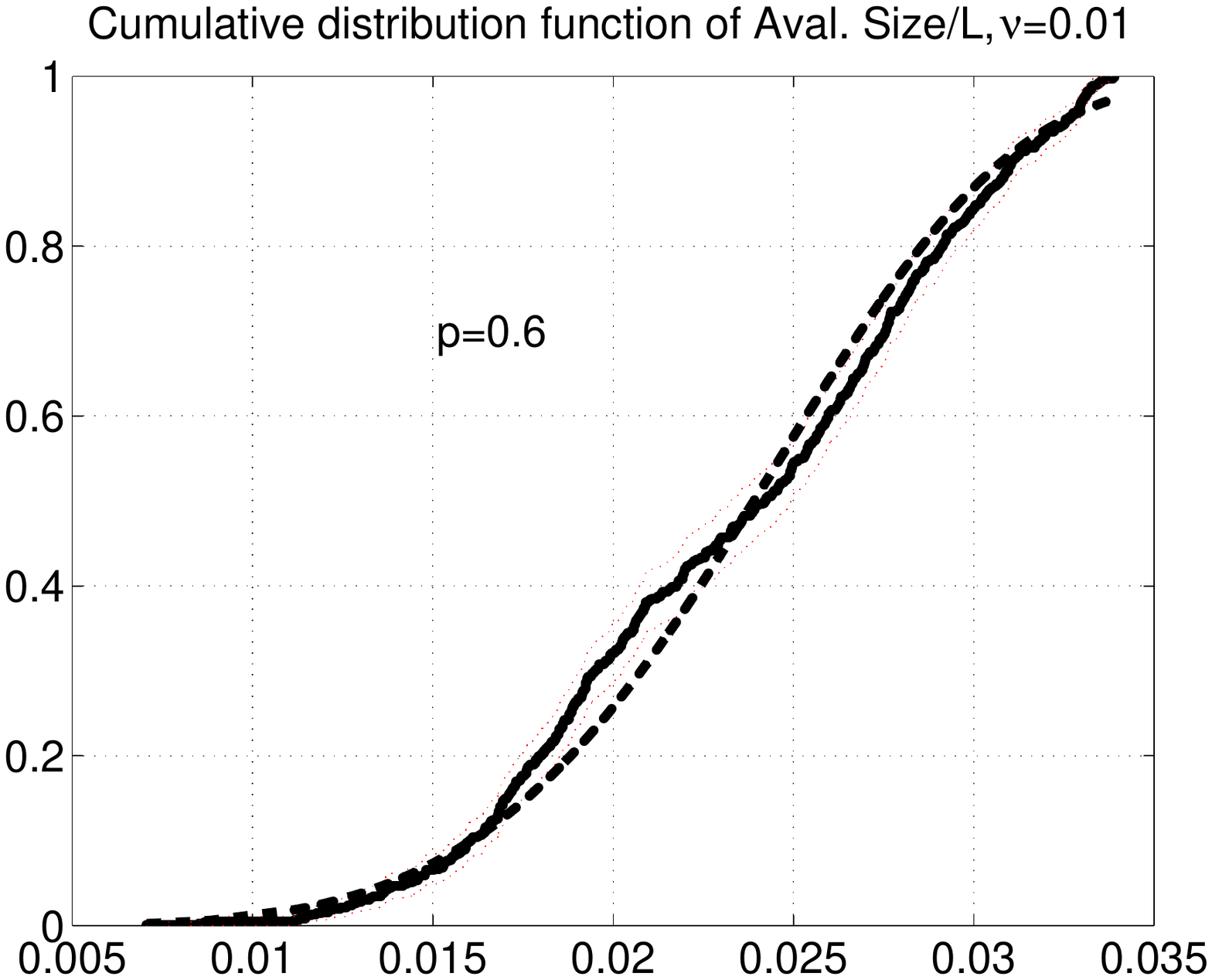}
\includegraphics[height=5cm,width=4cm]{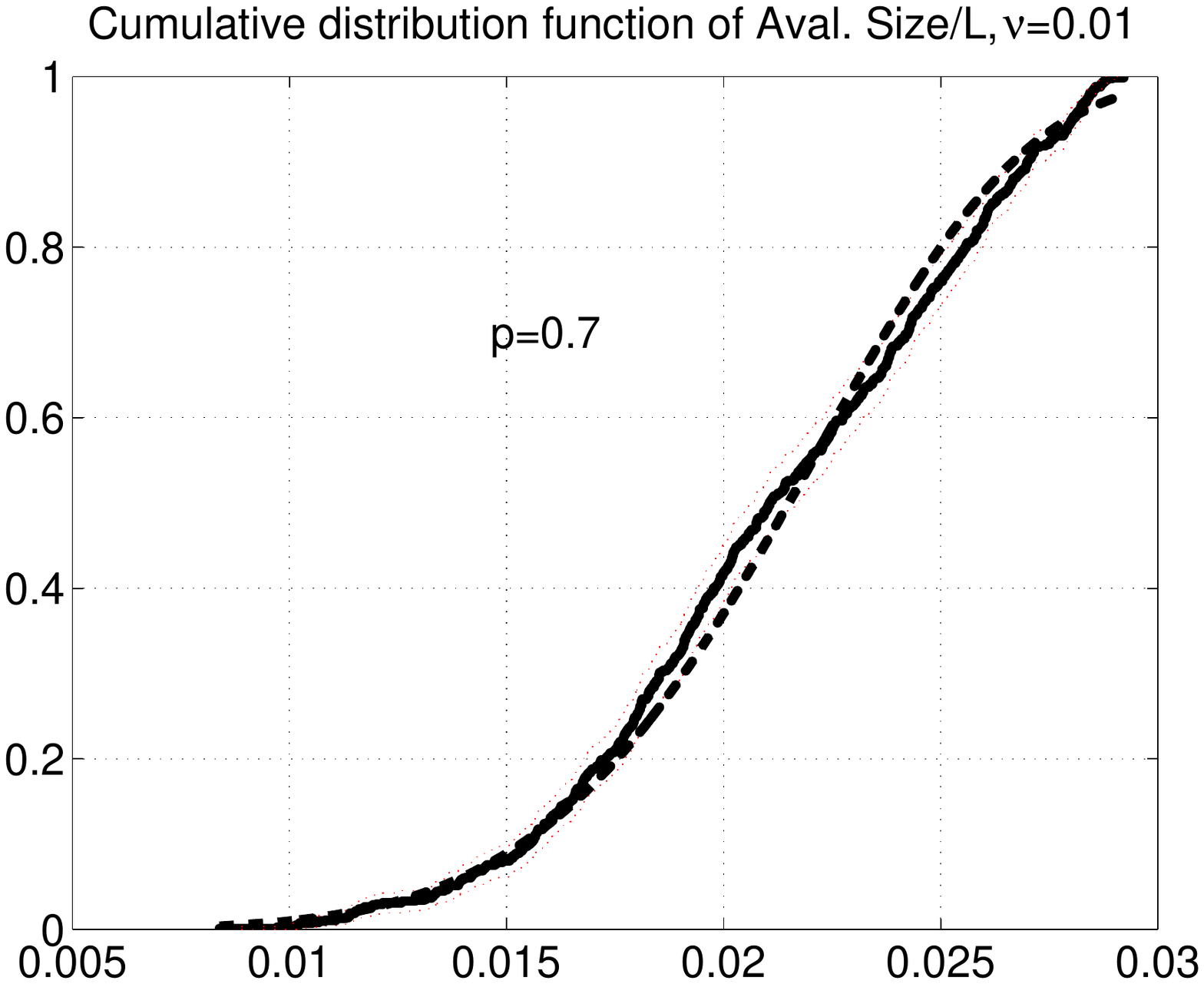}
\includegraphics[height=5cm,width=4cm]{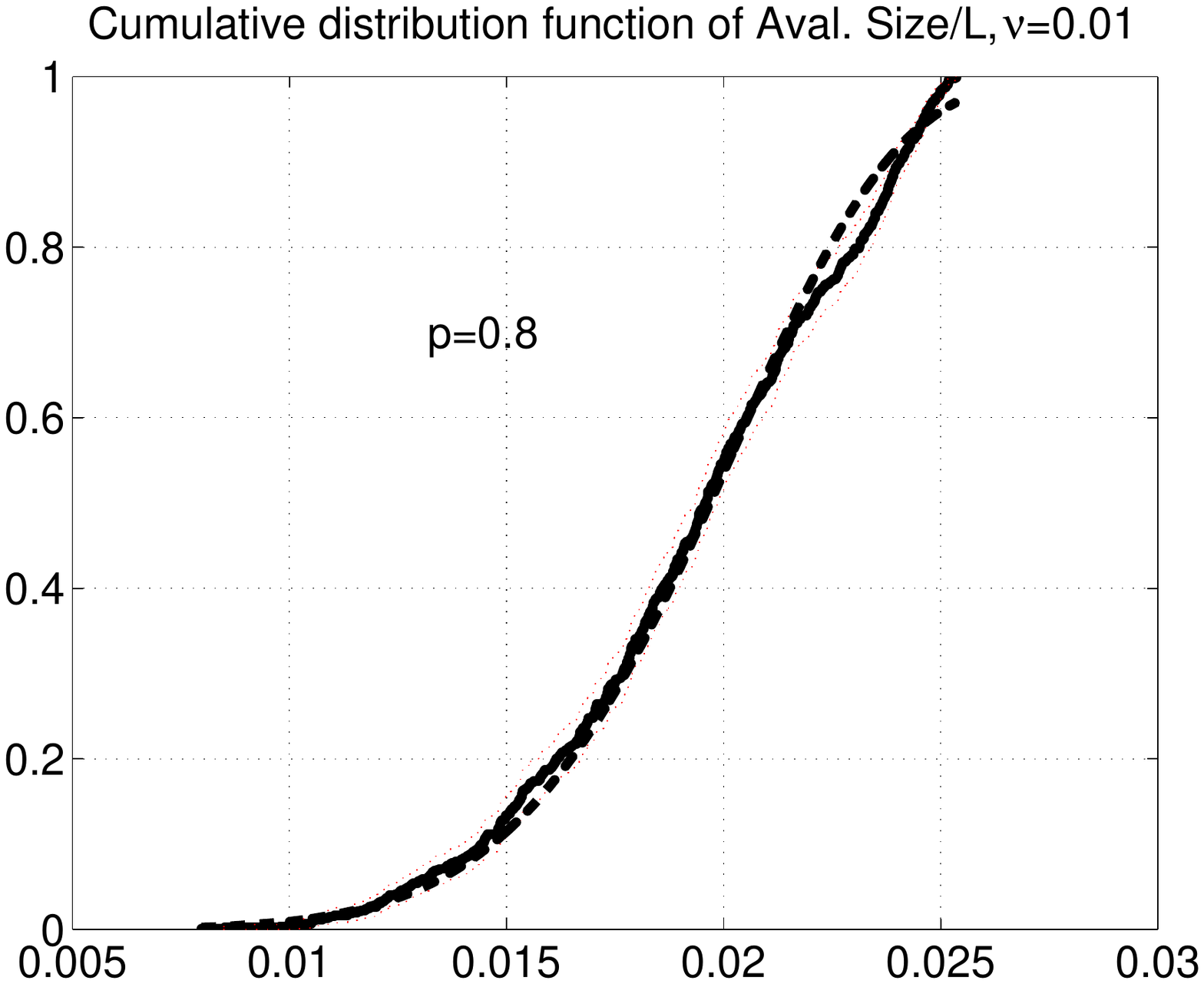}
\includegraphics[height=5cm,width=4cm]{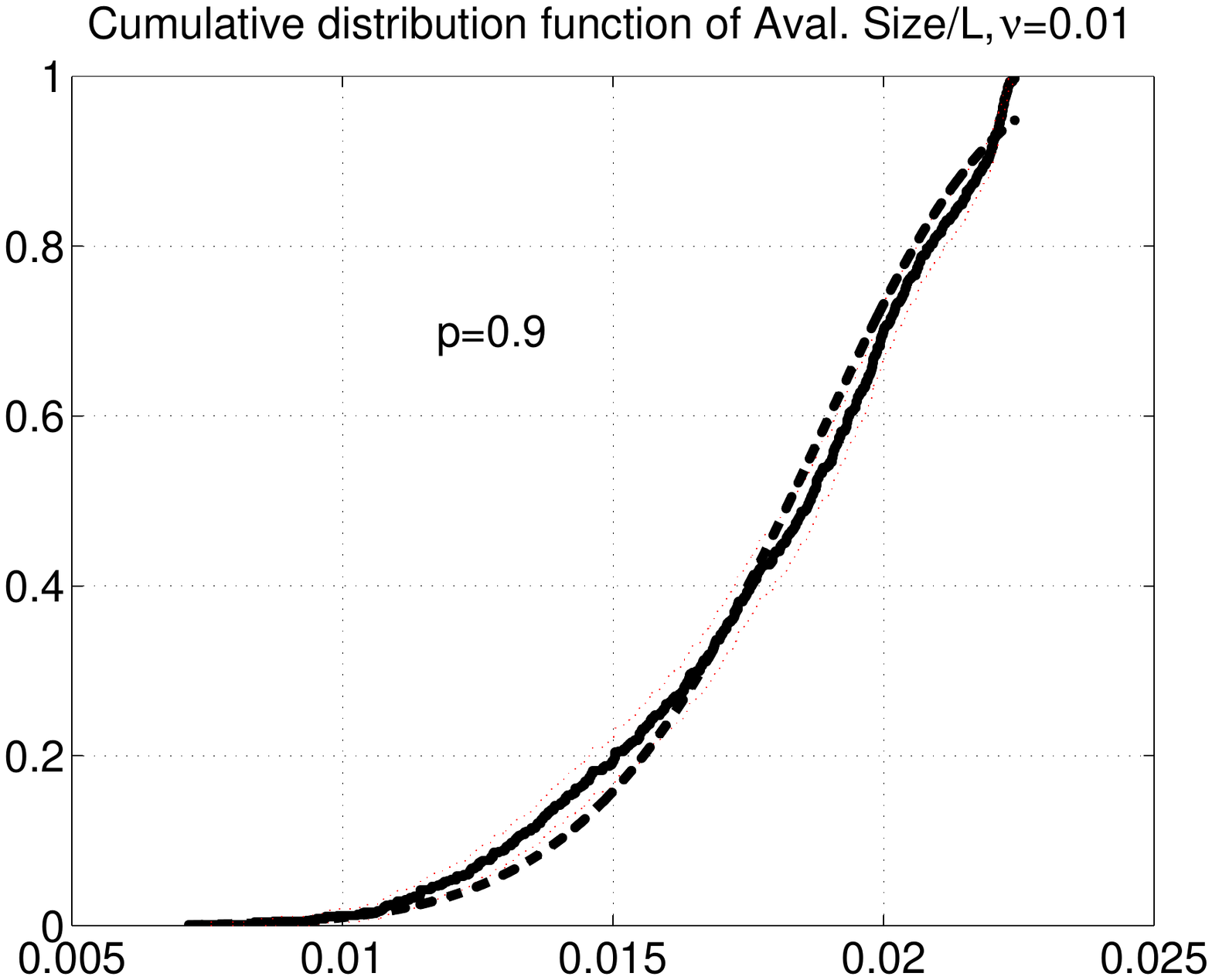}
\includegraphics[height=5cm,width=4cm]{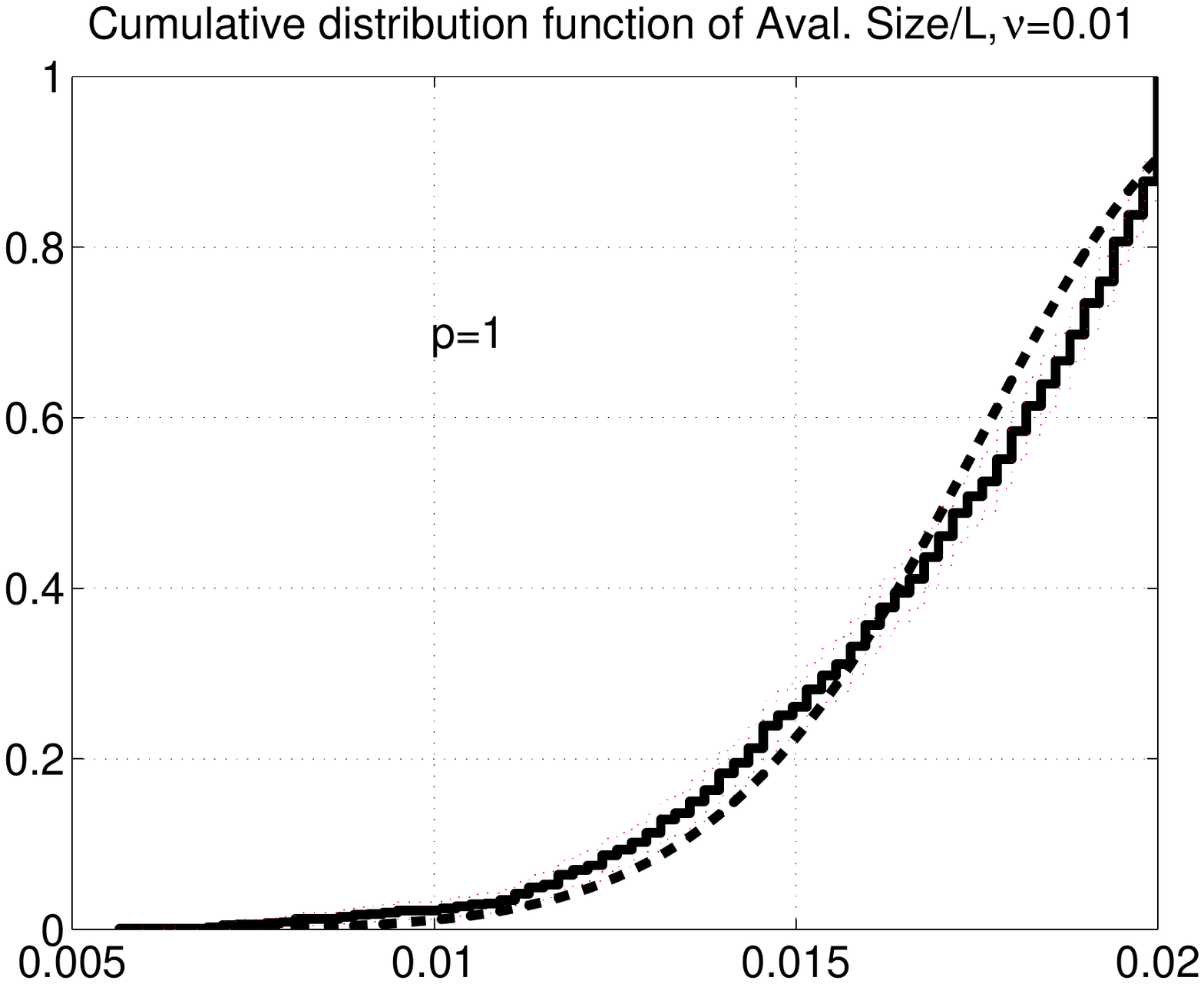}
\caption{Cumulative distribution function of the avalanche size fitted with a Weibull distribution. In all plots $\nu=0.01$ and for each plot the value of the connection probability $p$ is indicated.}\label{fig1a}
\end{figure} 
  
\begin{figure}
\centering
\includegraphics[height=5cm,width=4cm]{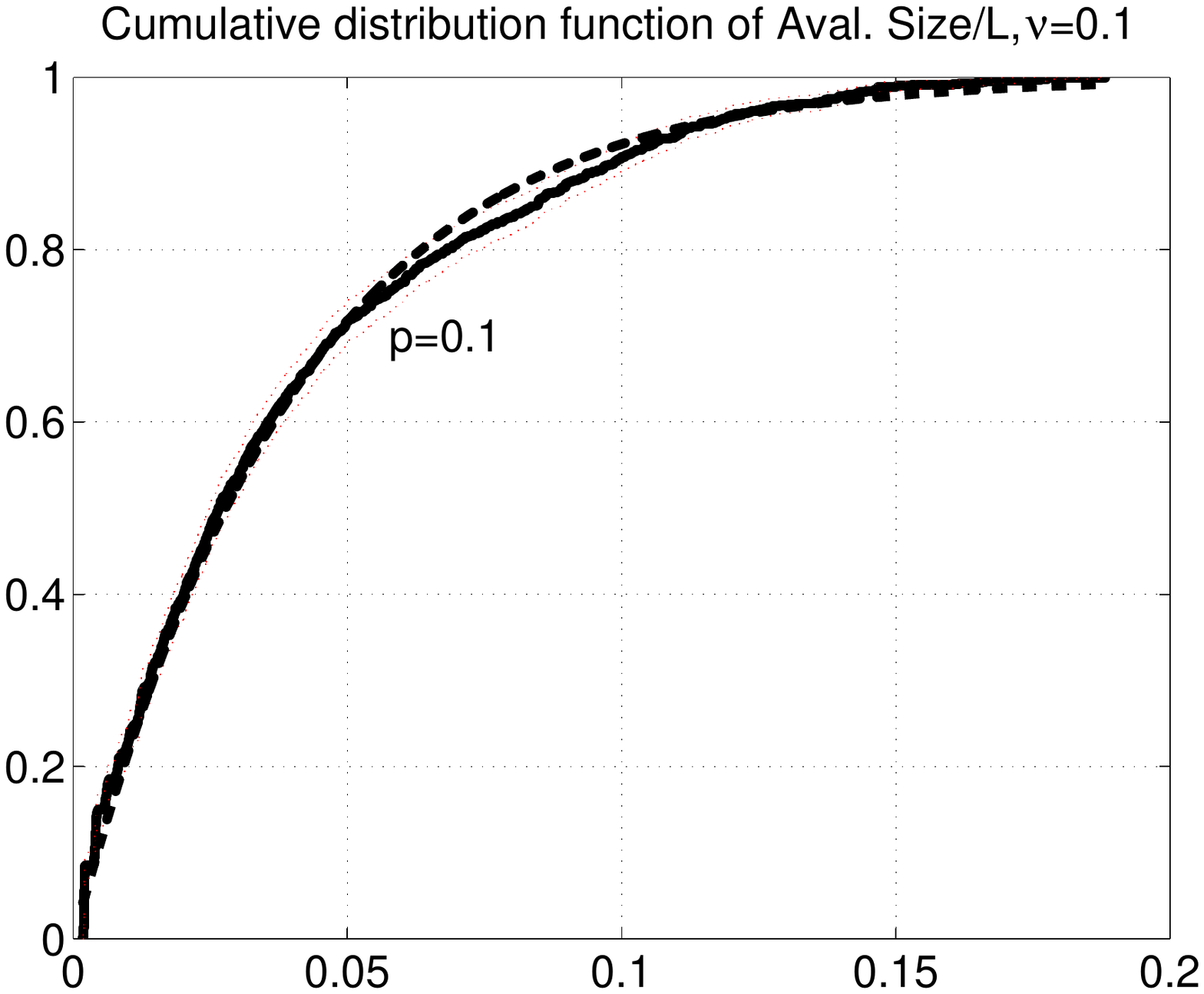}
\includegraphics[height=5cm,width=4cm]{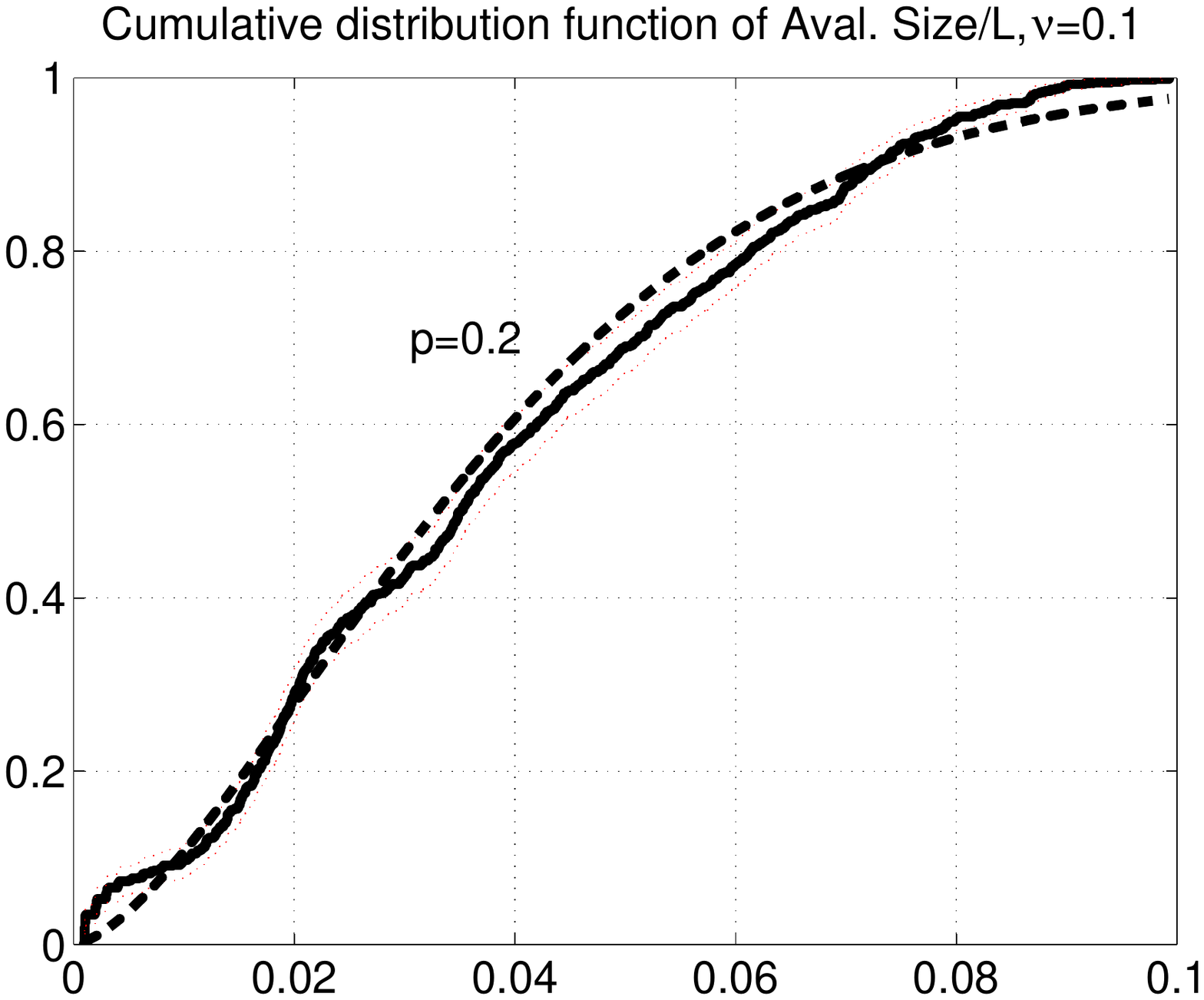}
\includegraphics[height=5cm,width=4cm]{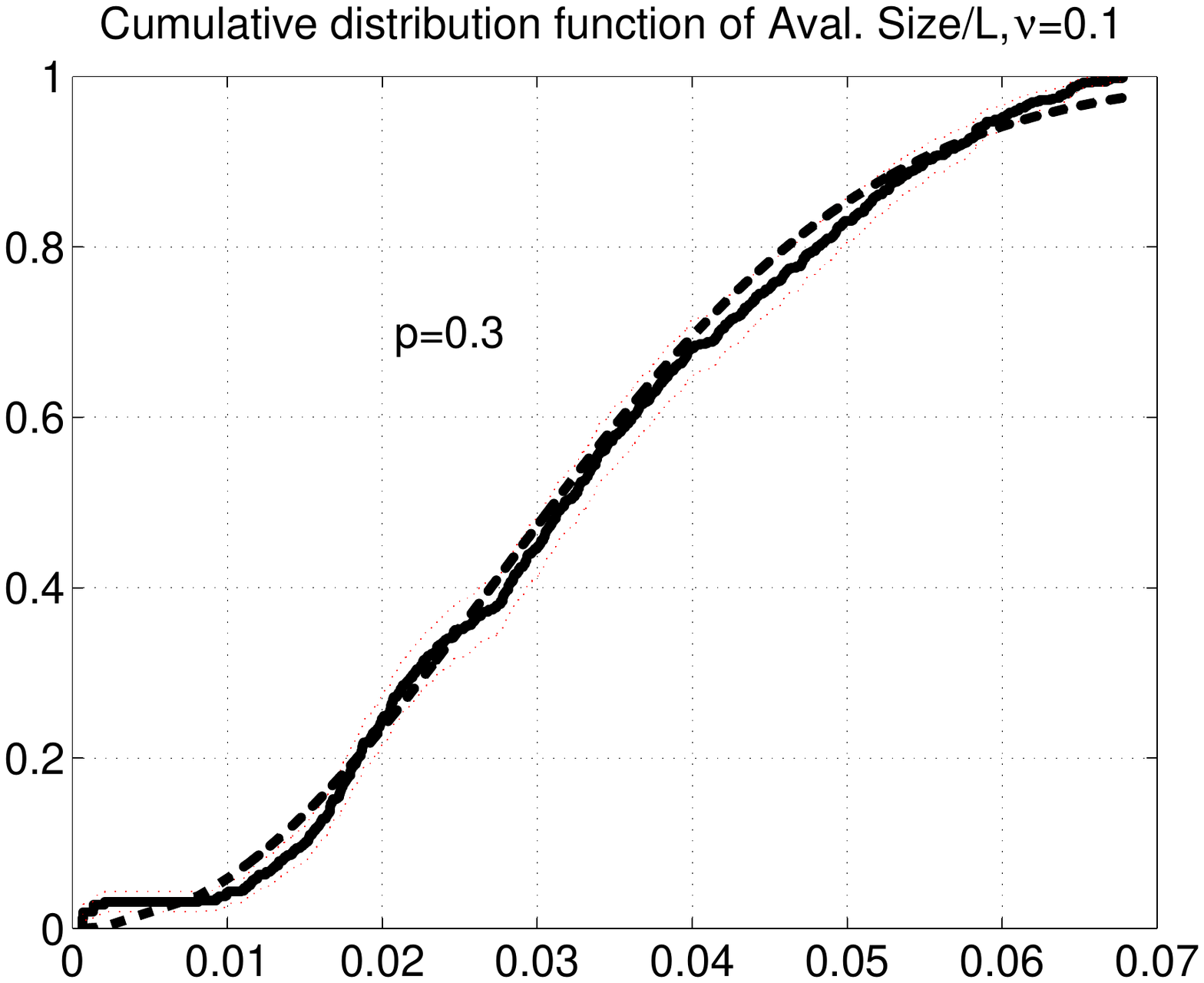}
\includegraphics[height=5cm,width=4cm]{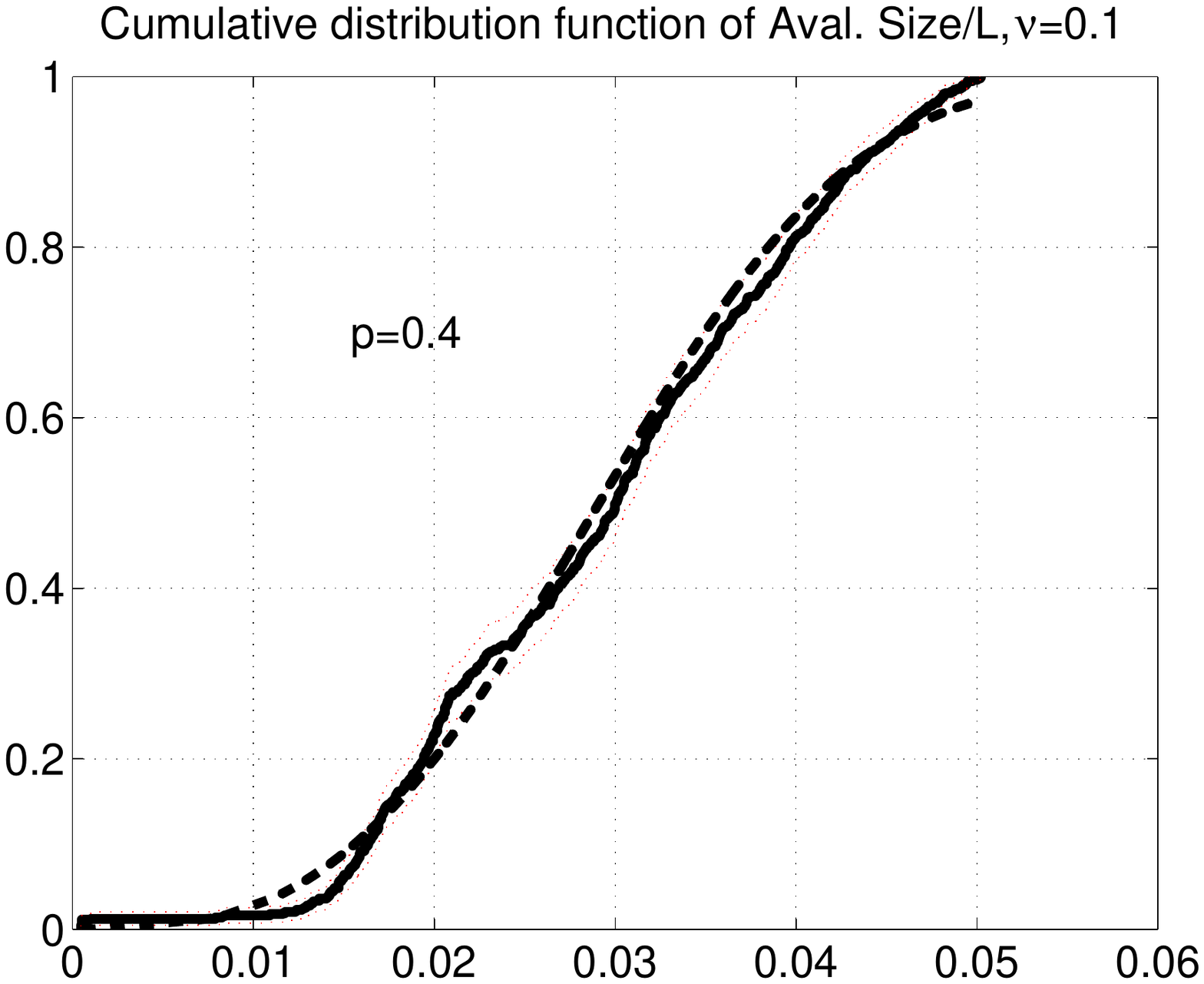}
\includegraphics[height=5cm,width=4cm]{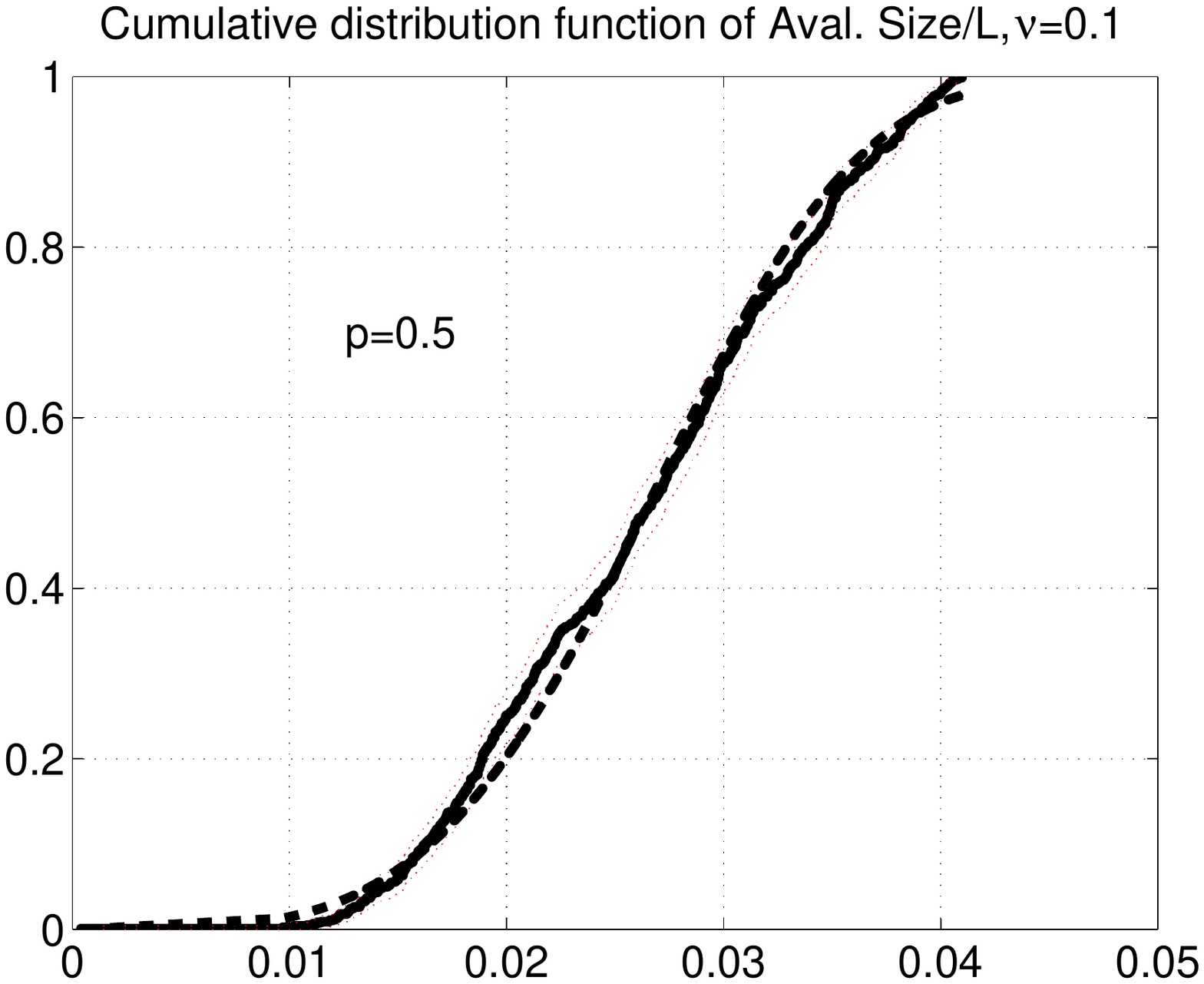}
\includegraphics[height=5cm,width=4cm]{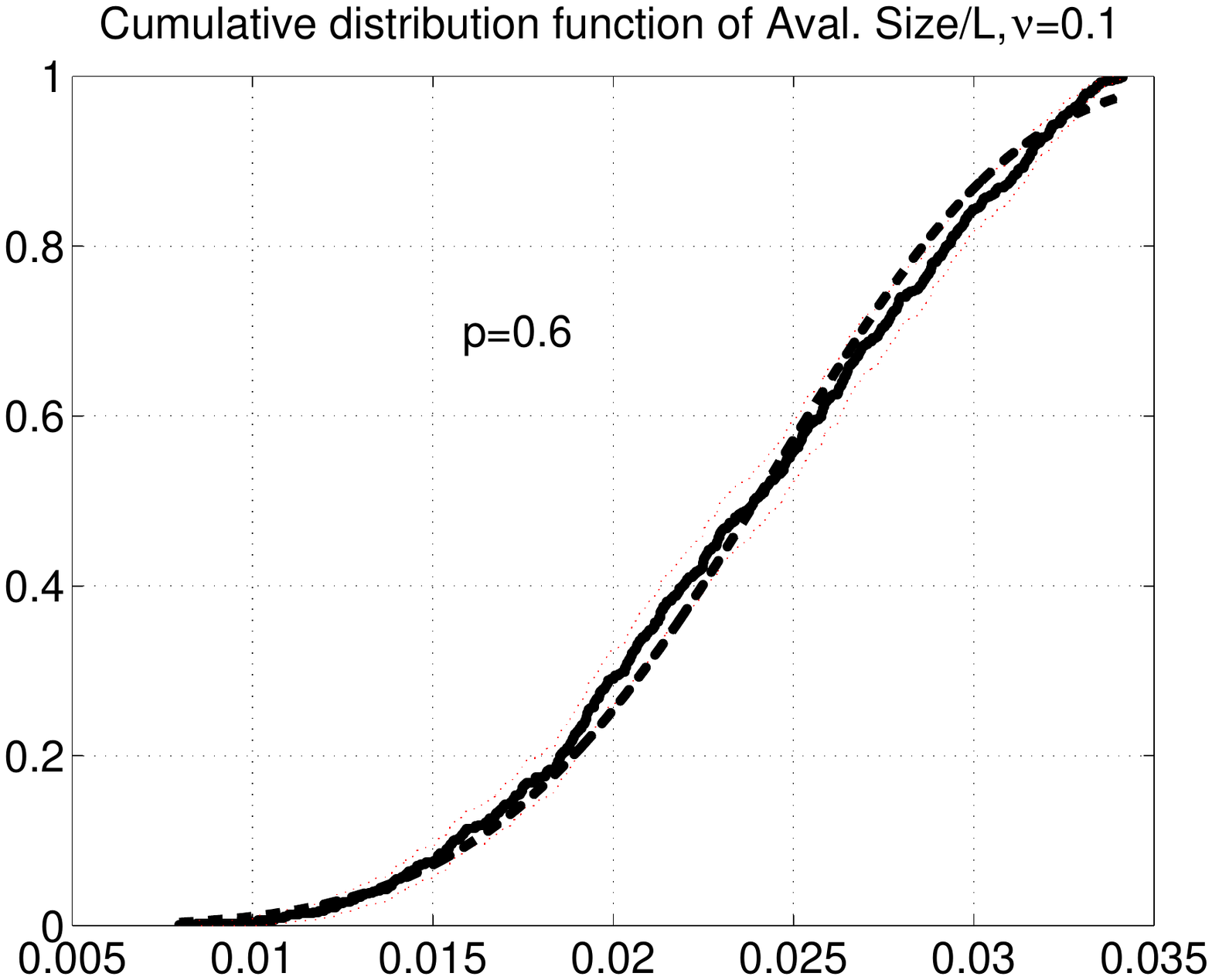}
\includegraphics[height=5cm,width=4cm]{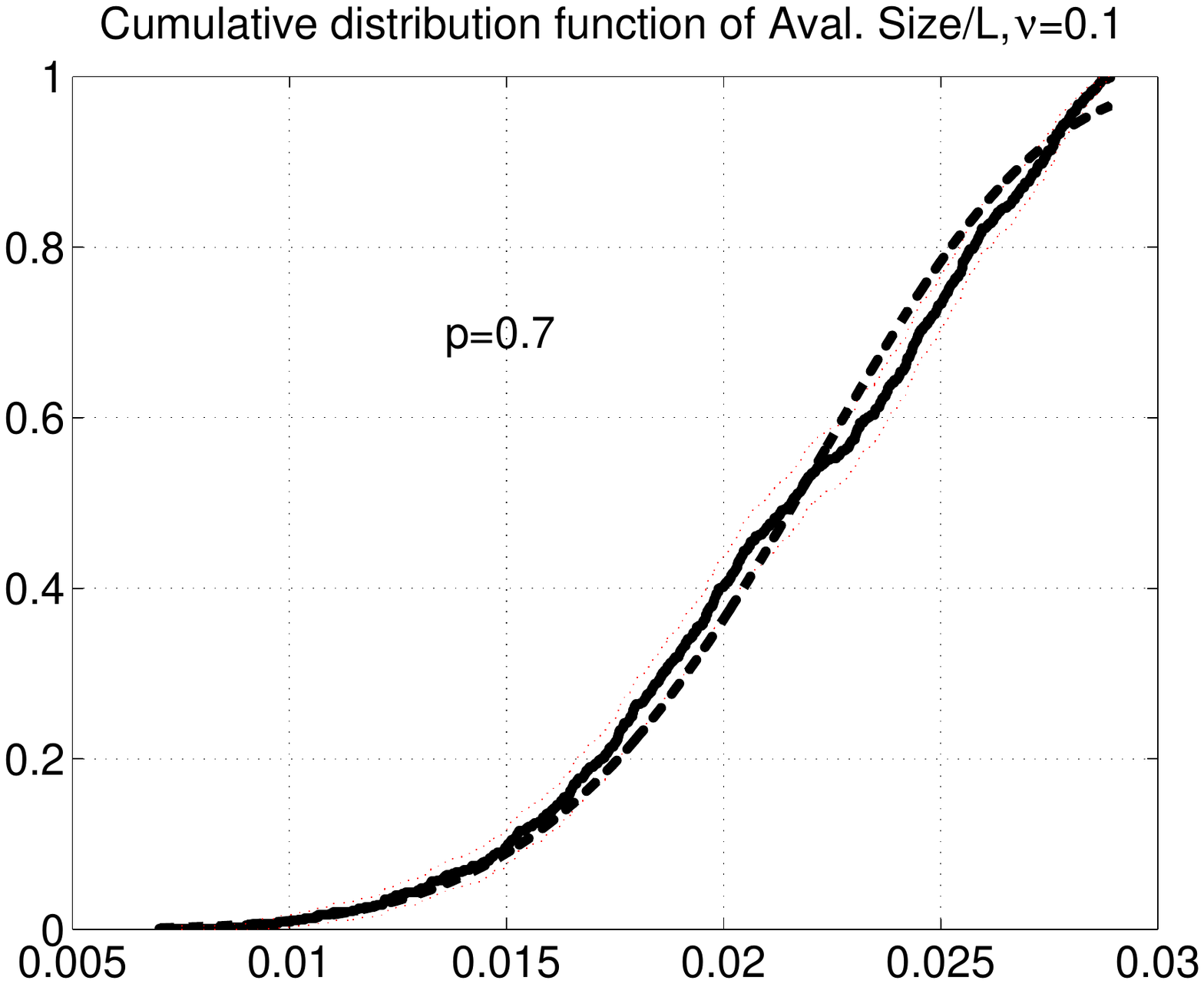}
\includegraphics[height=5cm,width=4cm]{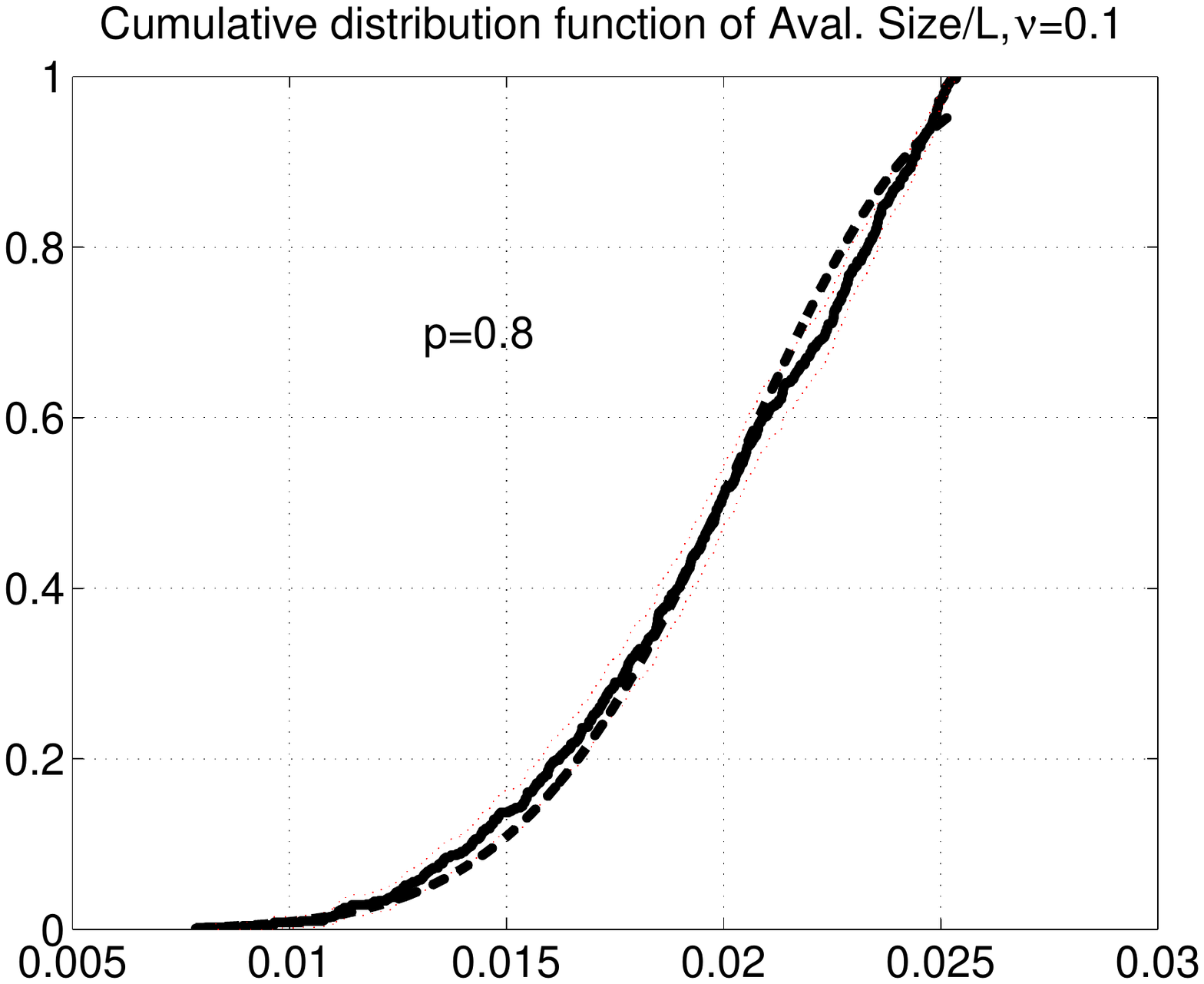}
\includegraphics[height=5cm,width=4cm]{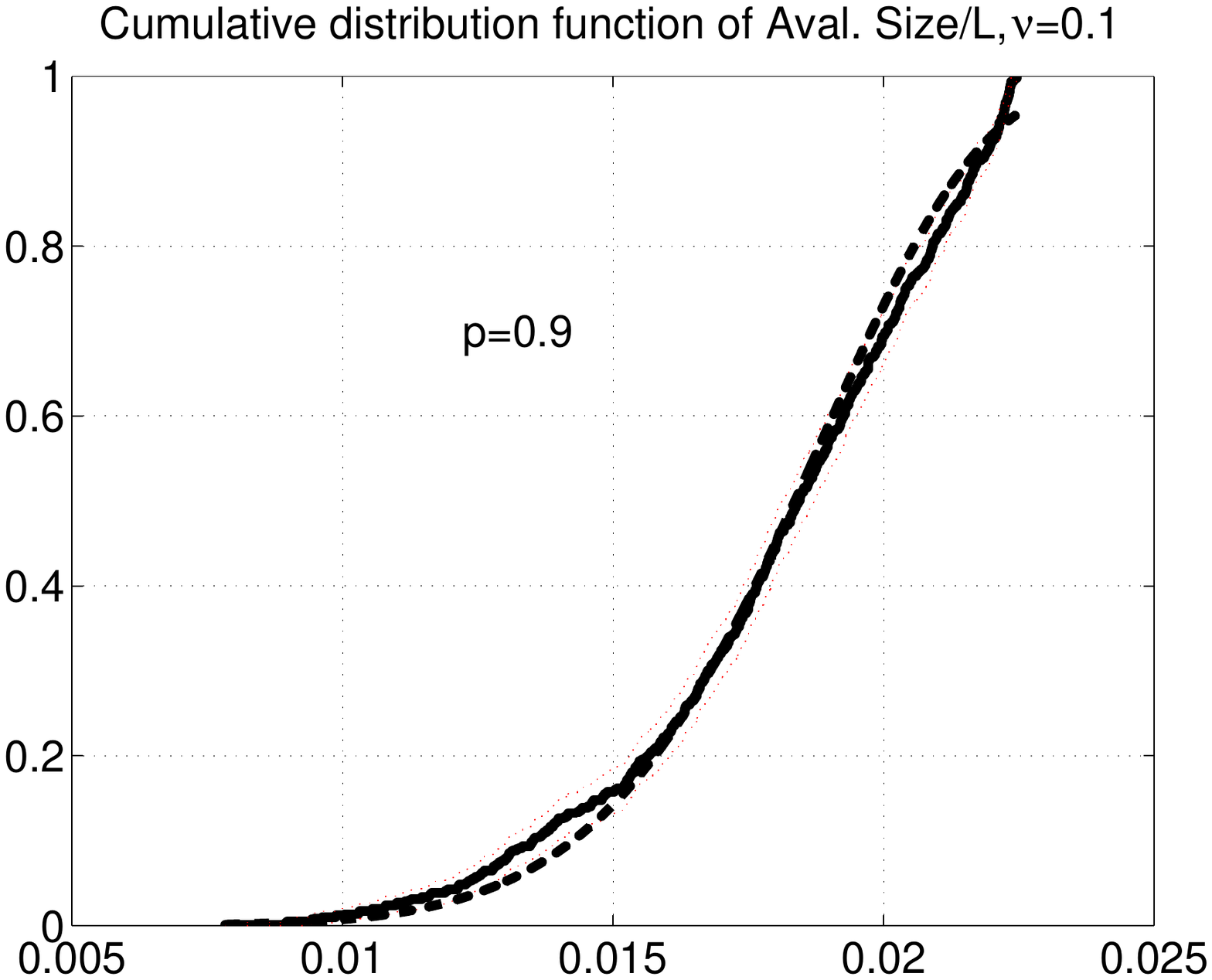}
\includegraphics[height=5cm,width=4cm]{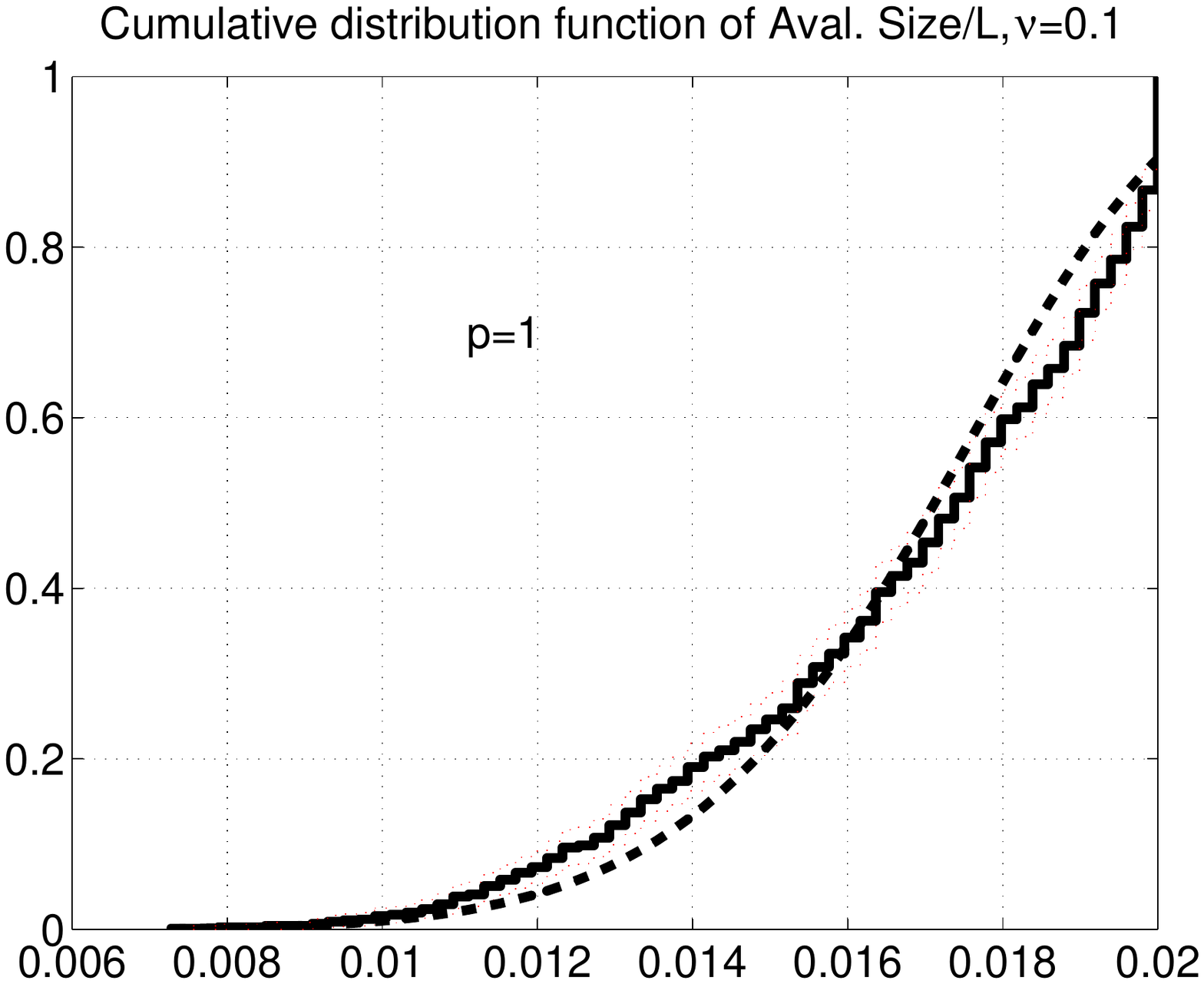}
\caption{Cumulative distribution function of the avalanche size fitted with a Weibull distribution. In all plots $\nu=0.10$ and for each plot the value of the connection probability $p$ is indicated.}\label{fig1b}
\end{figure} 

\begin{figure}
\centering
\includegraphics[height=4cm,width=4cm]{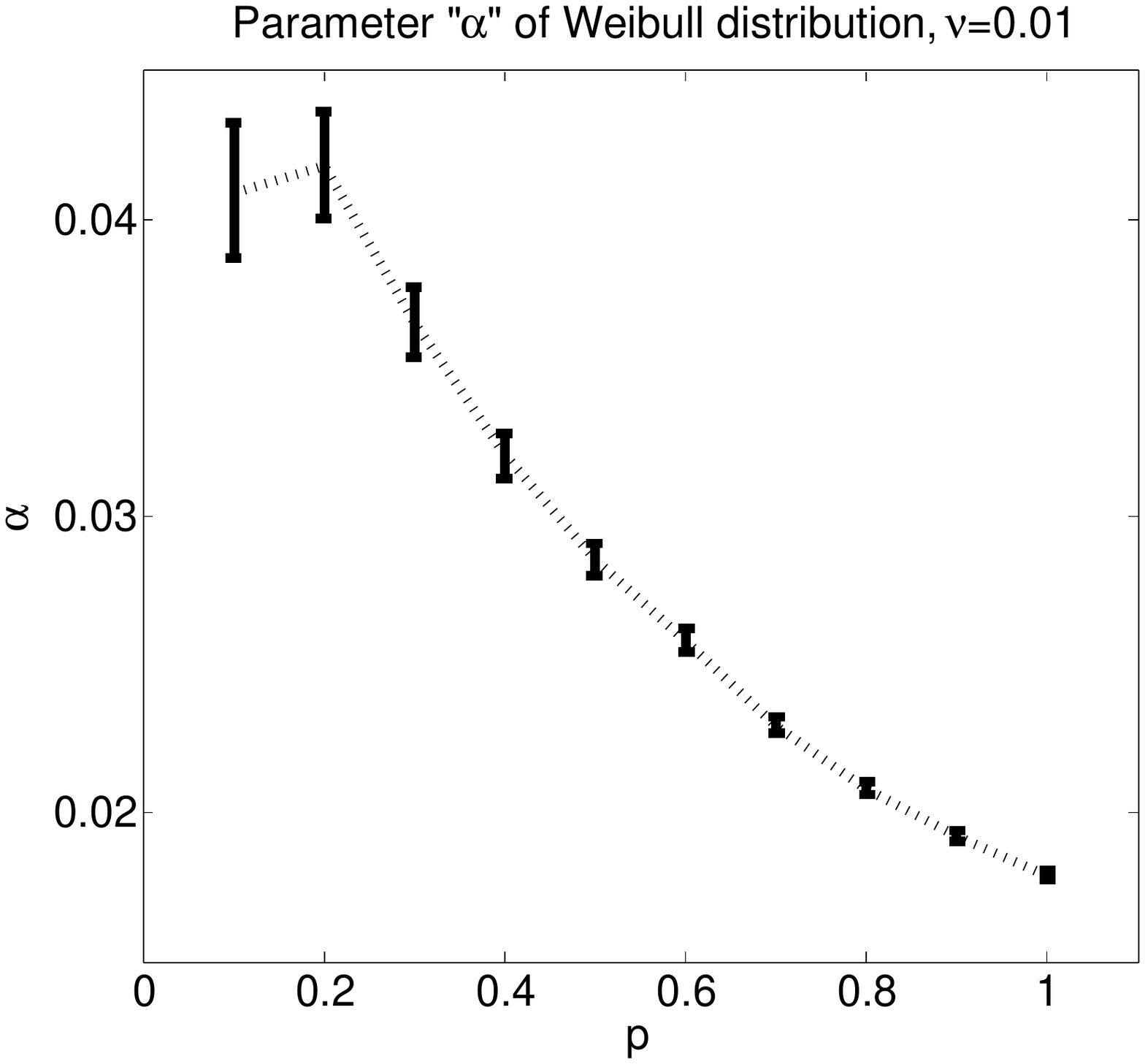}
\includegraphics[height=4cm,width=4cm]{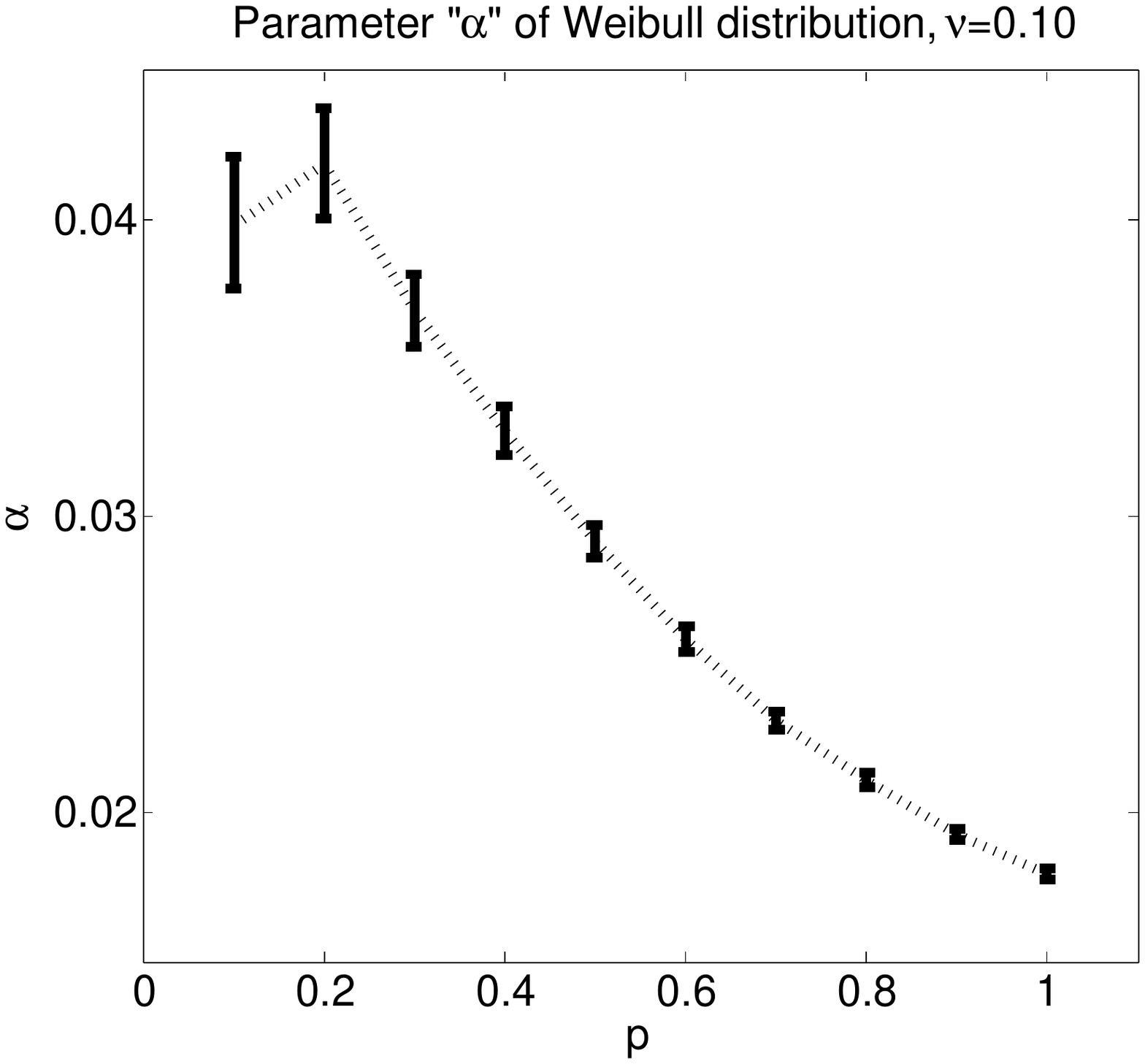}
\includegraphics[height=4cm,width=4cm]{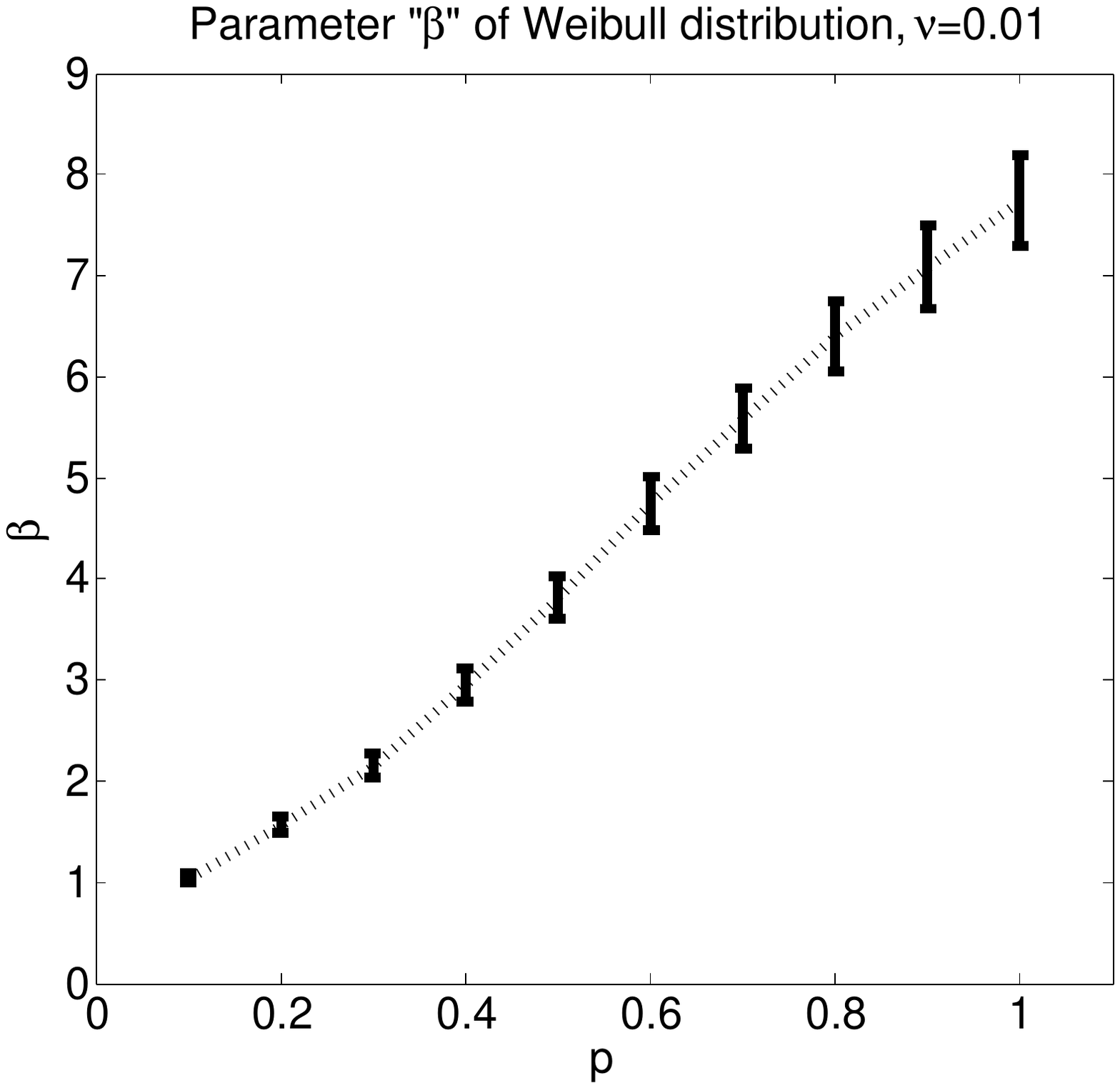}
\includegraphics[height=4cm,width=4cm]{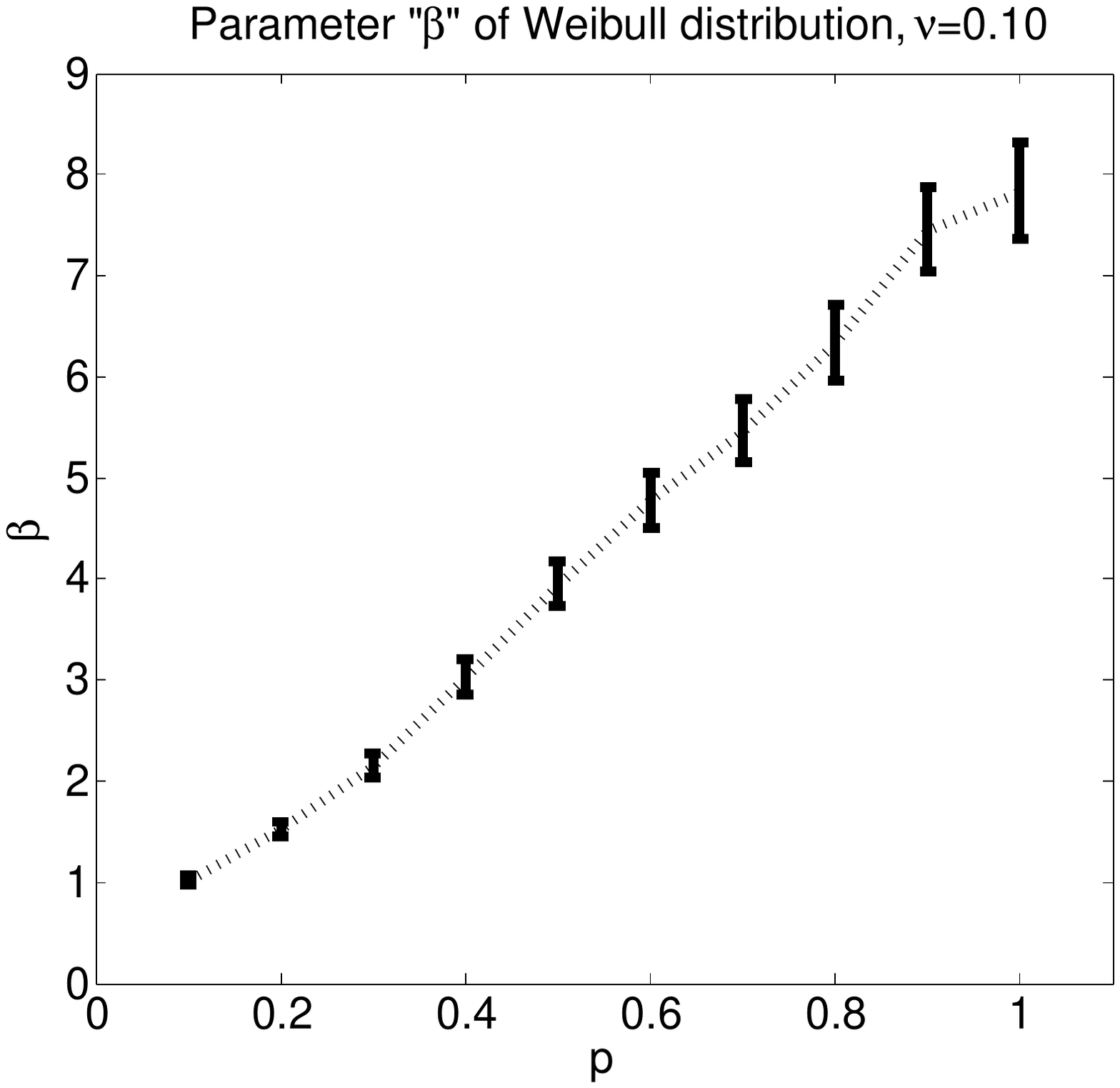}
\caption{(a) Parameter $\alpha$ of the Weibull fit as a function of $p$ for $\nu=0.01$, (b) parameter $\alpha$ of the Weibull fit as a function of $p$ for $\nu=0.10$, (c) parameter $\beta$ of the Weibull fit as a function of $p$ for $\nu=0.10$, (d) Parameter $\beta$ of the Weibull fit as a function of $p$ for $\nu=0.10$; the line between data points serves only to  emphasize some tendency }\label{fig3}
\end{figure} 

\begin{figure}
\centering
\includegraphics[height=5cm,width=4cm]{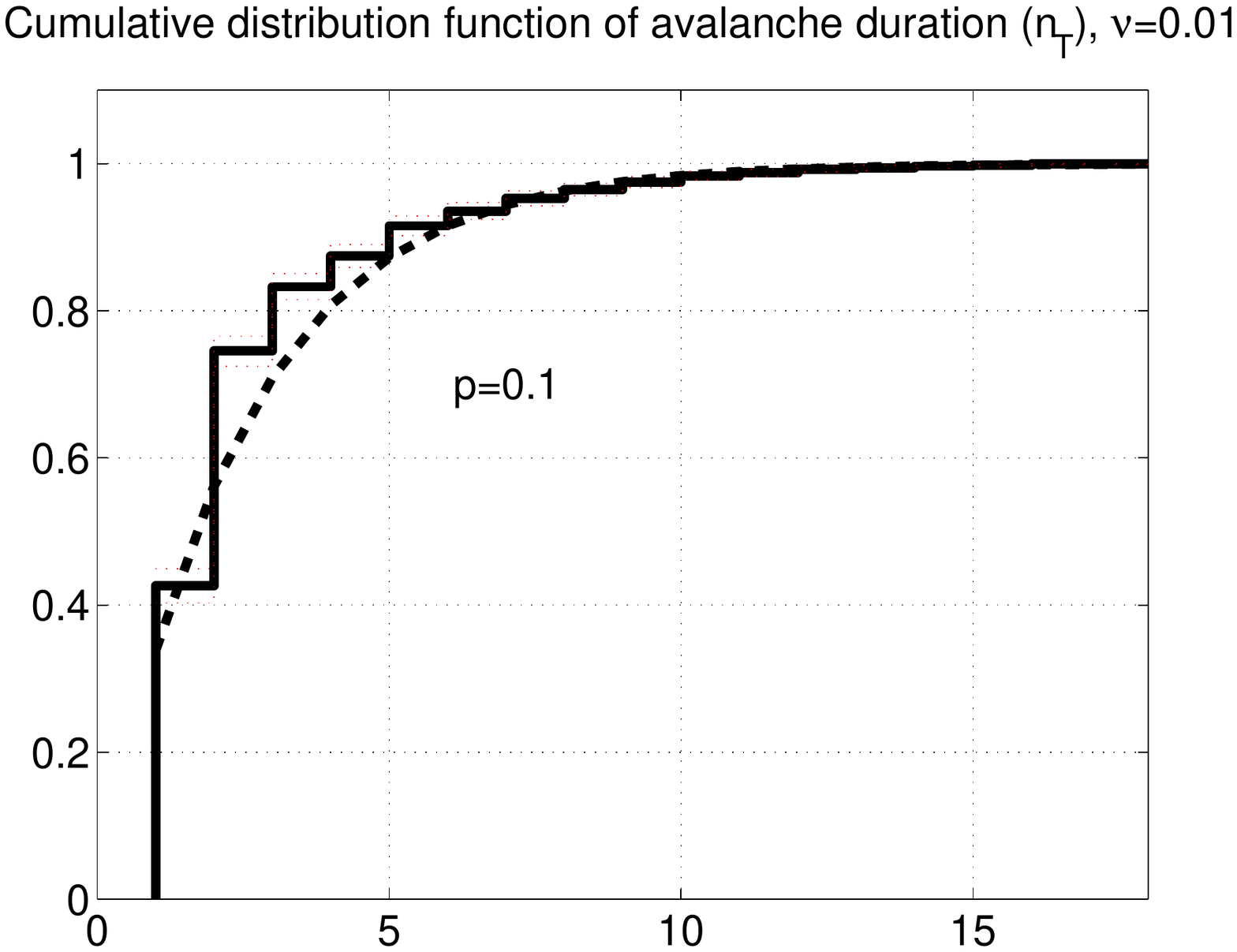}
\includegraphics[height=5cm,width=4cm]{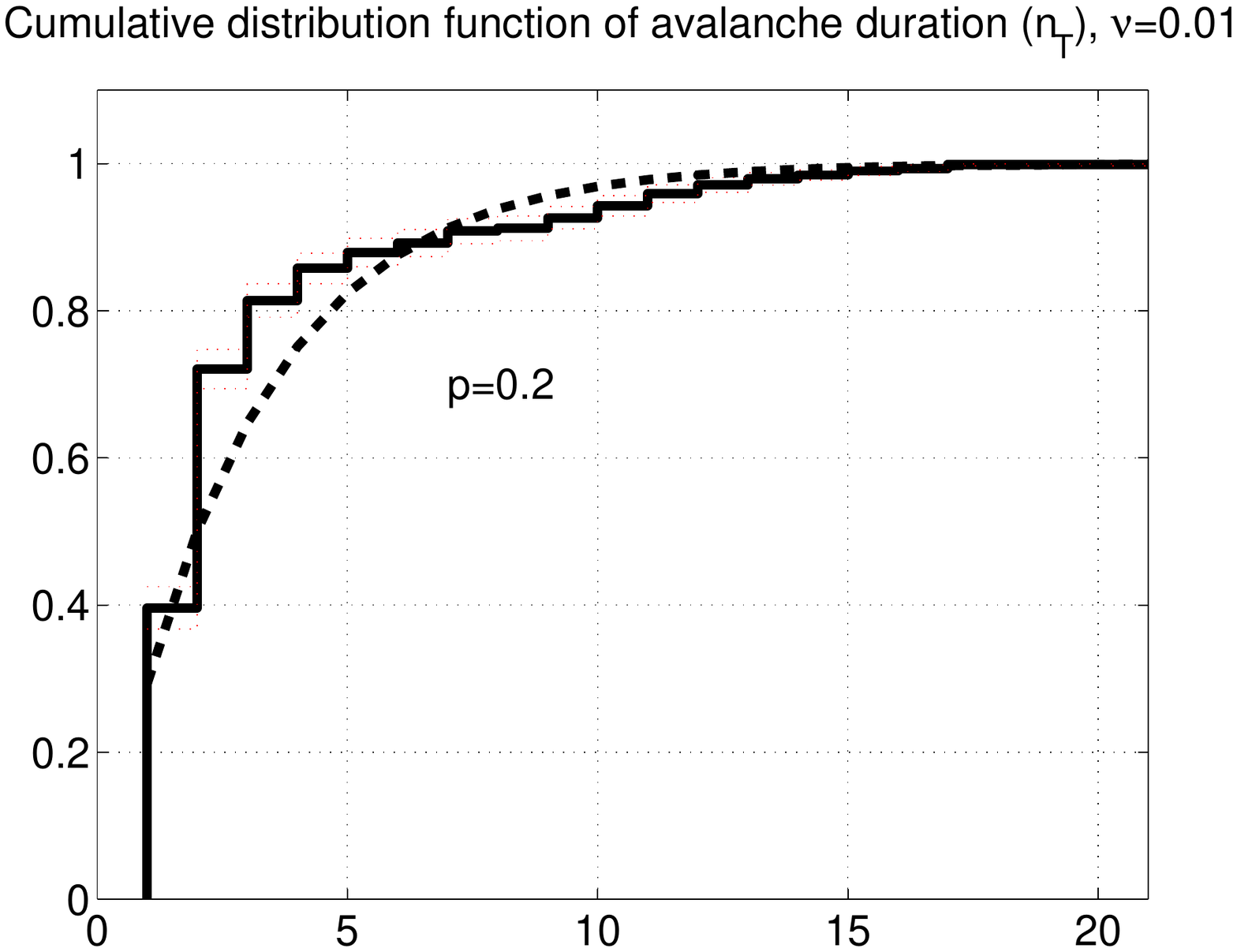}
\includegraphics[height=5cm,width=4cm]{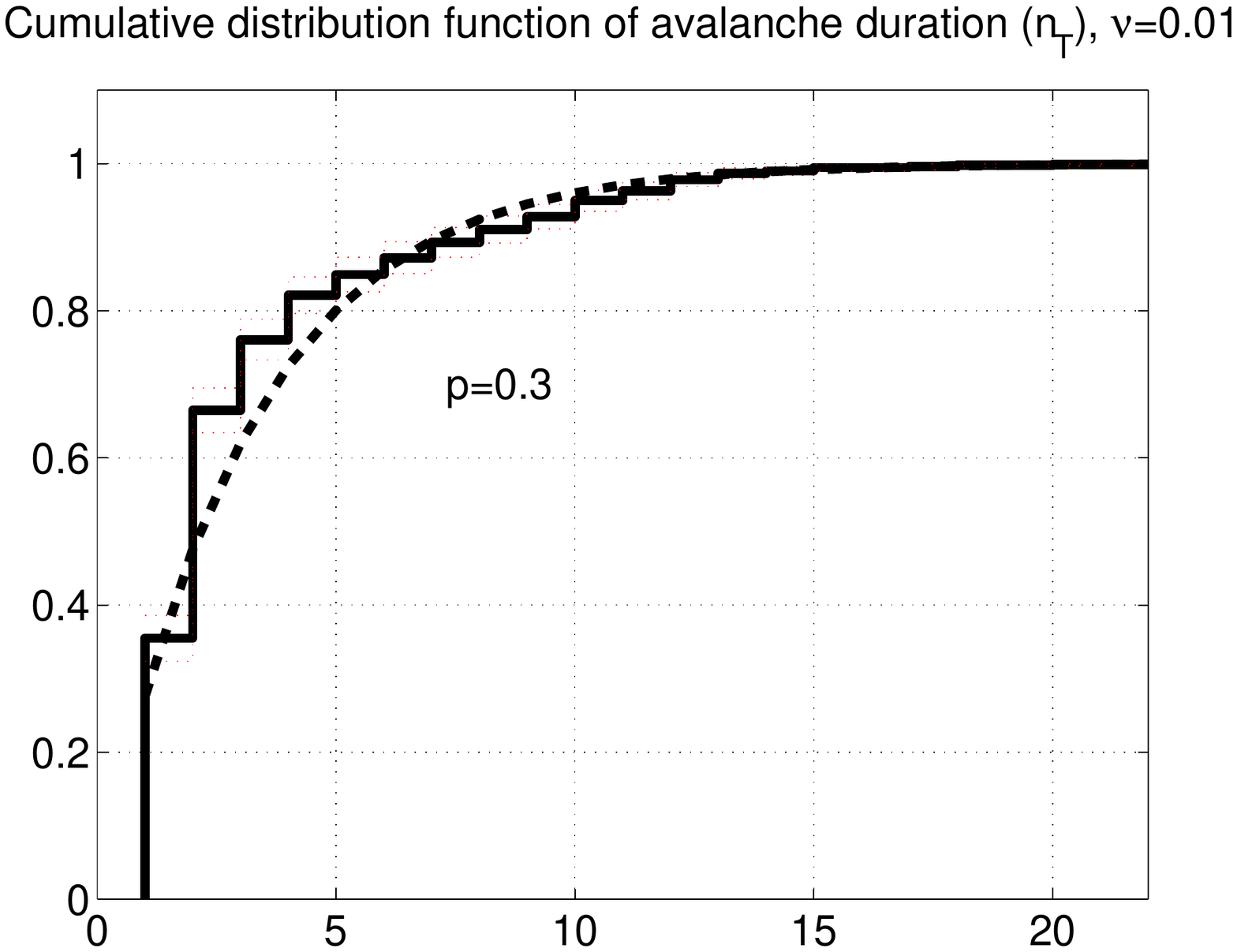}
\includegraphics[height=5cm,width=4cm]{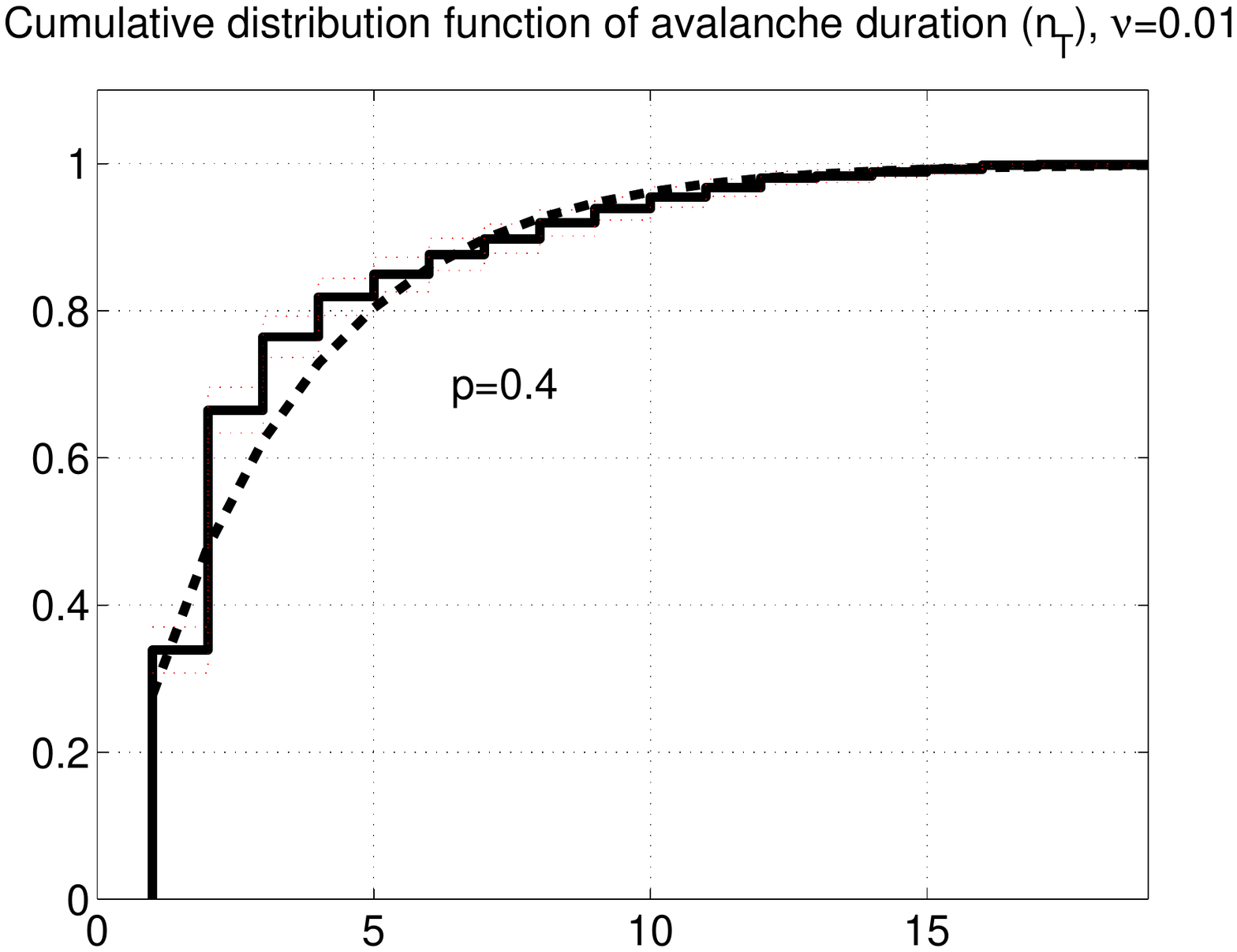}
\includegraphics[height=5cm,width=4cm]{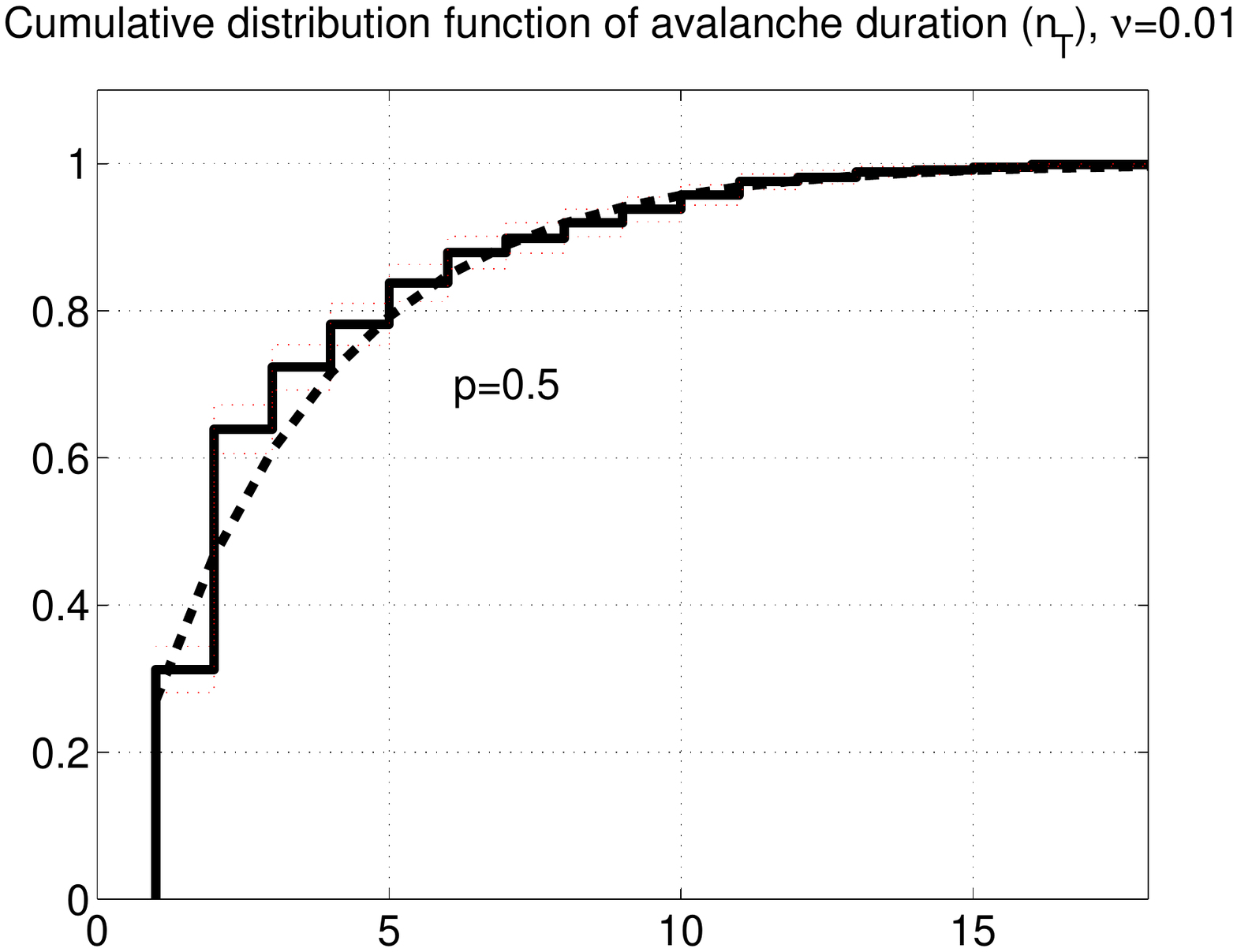}
\includegraphics[height=5cm,width=4cm]{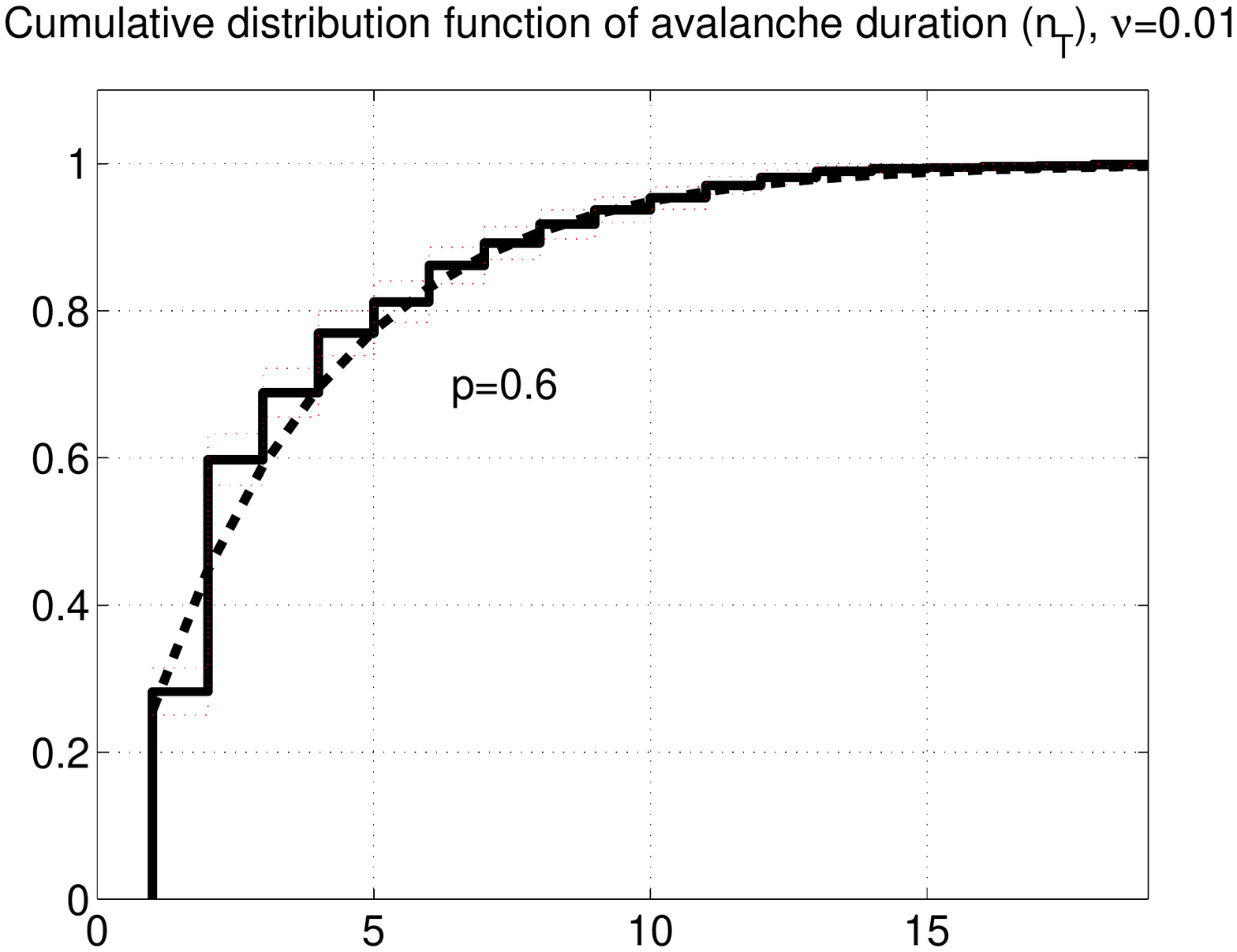}
\includegraphics[height=5cm,width=4cm]{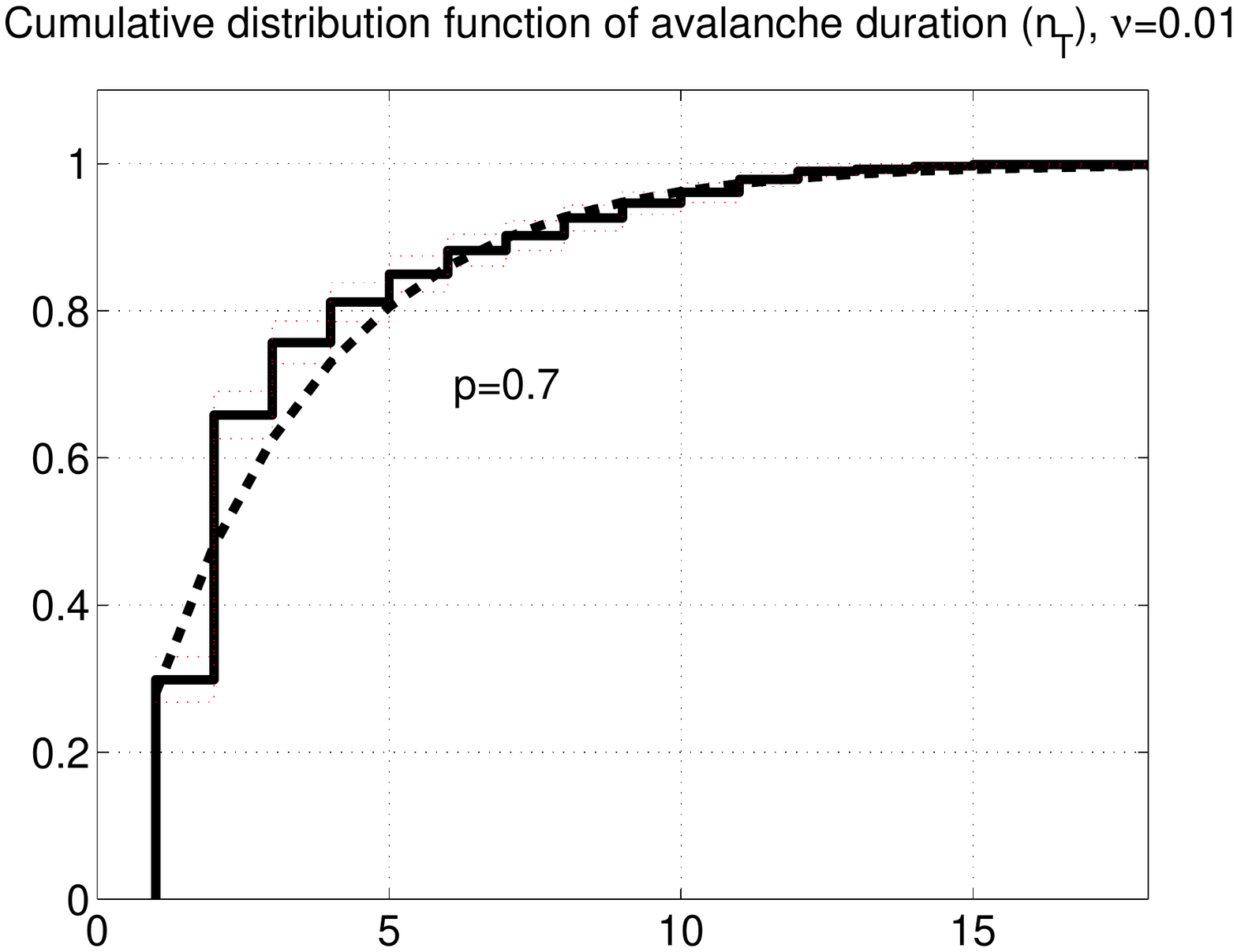}
\includegraphics[height=5cm,width=4cm]{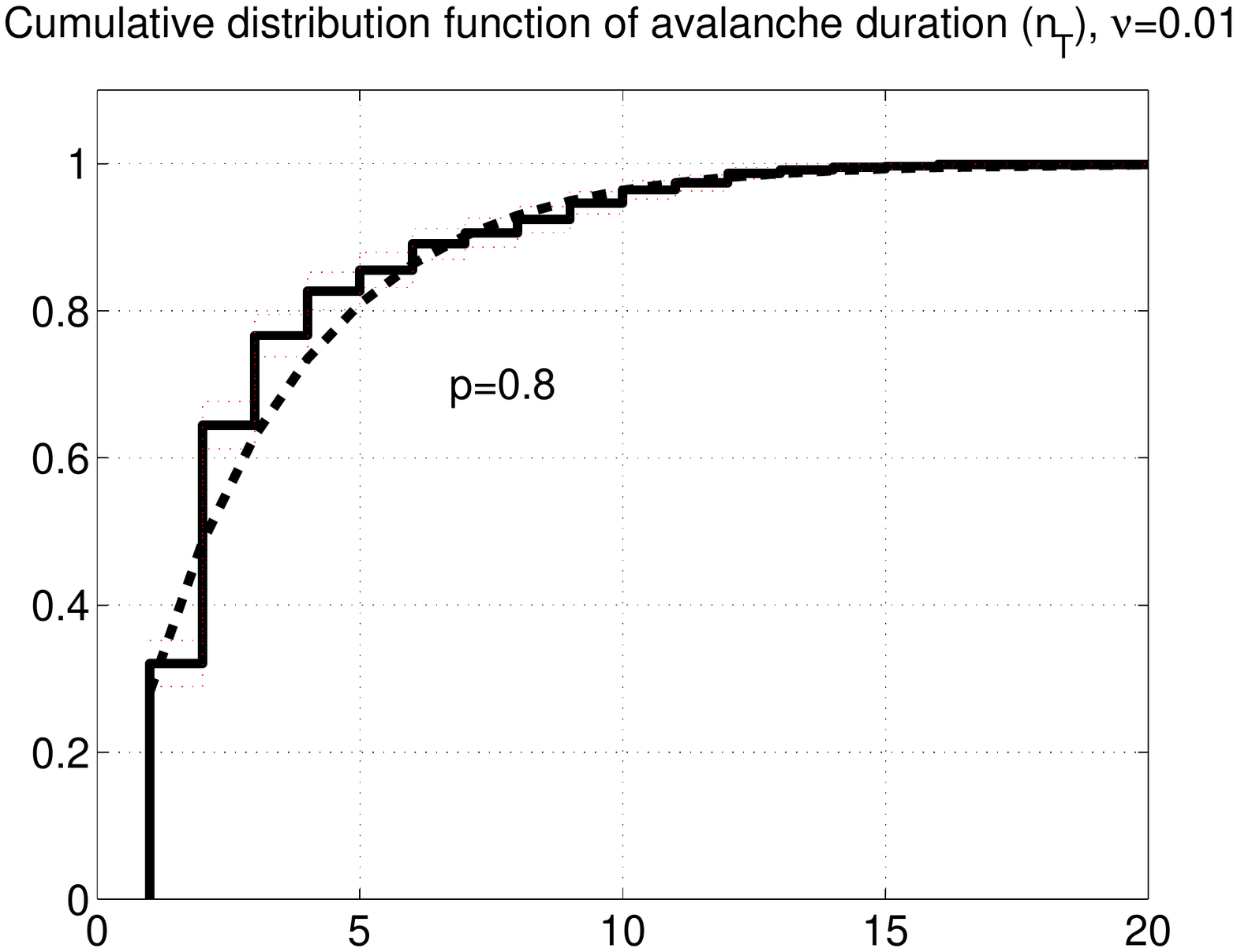}
\includegraphics[height=5cm,width=4cm]{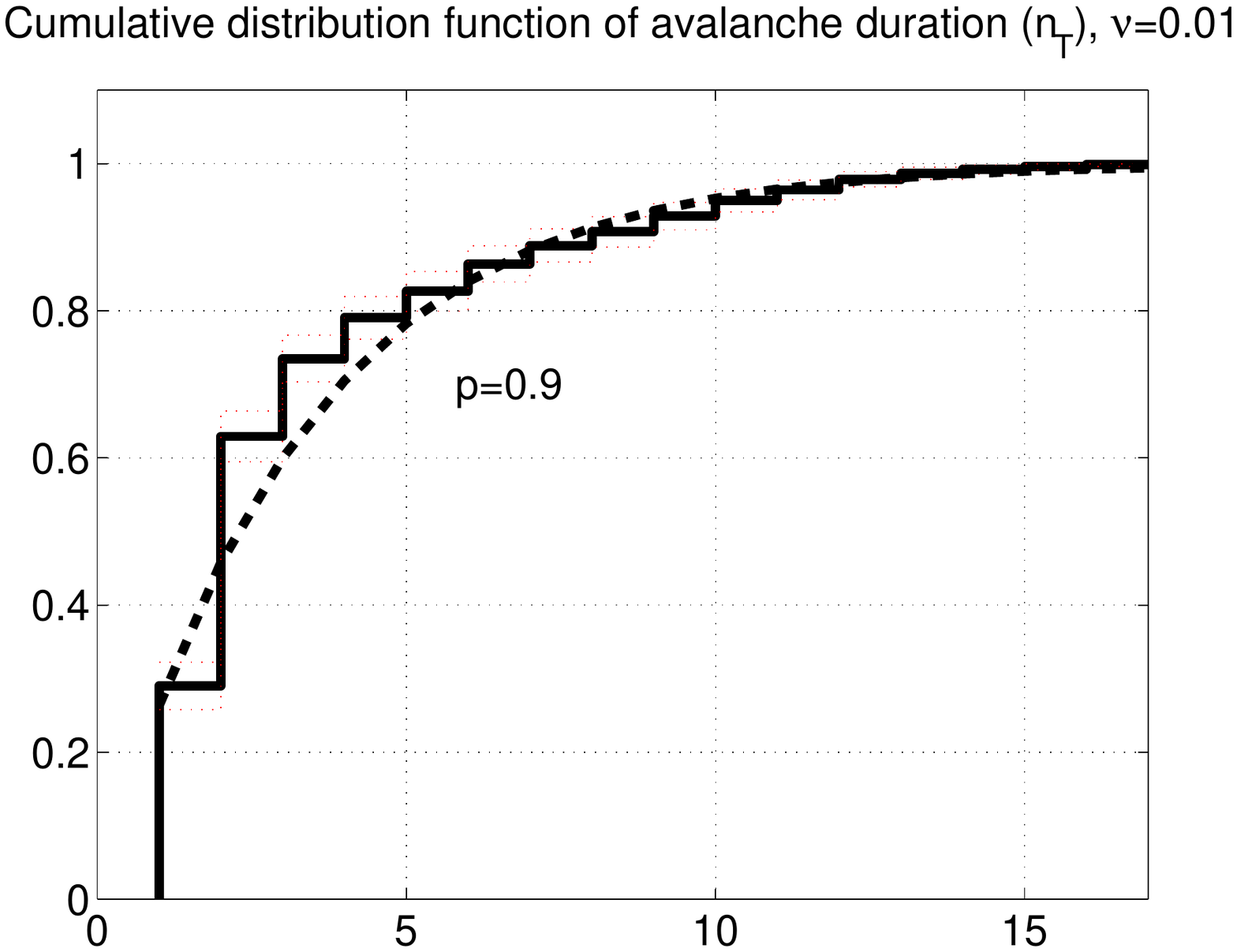}
\includegraphics[height=5cm,width=4cm]{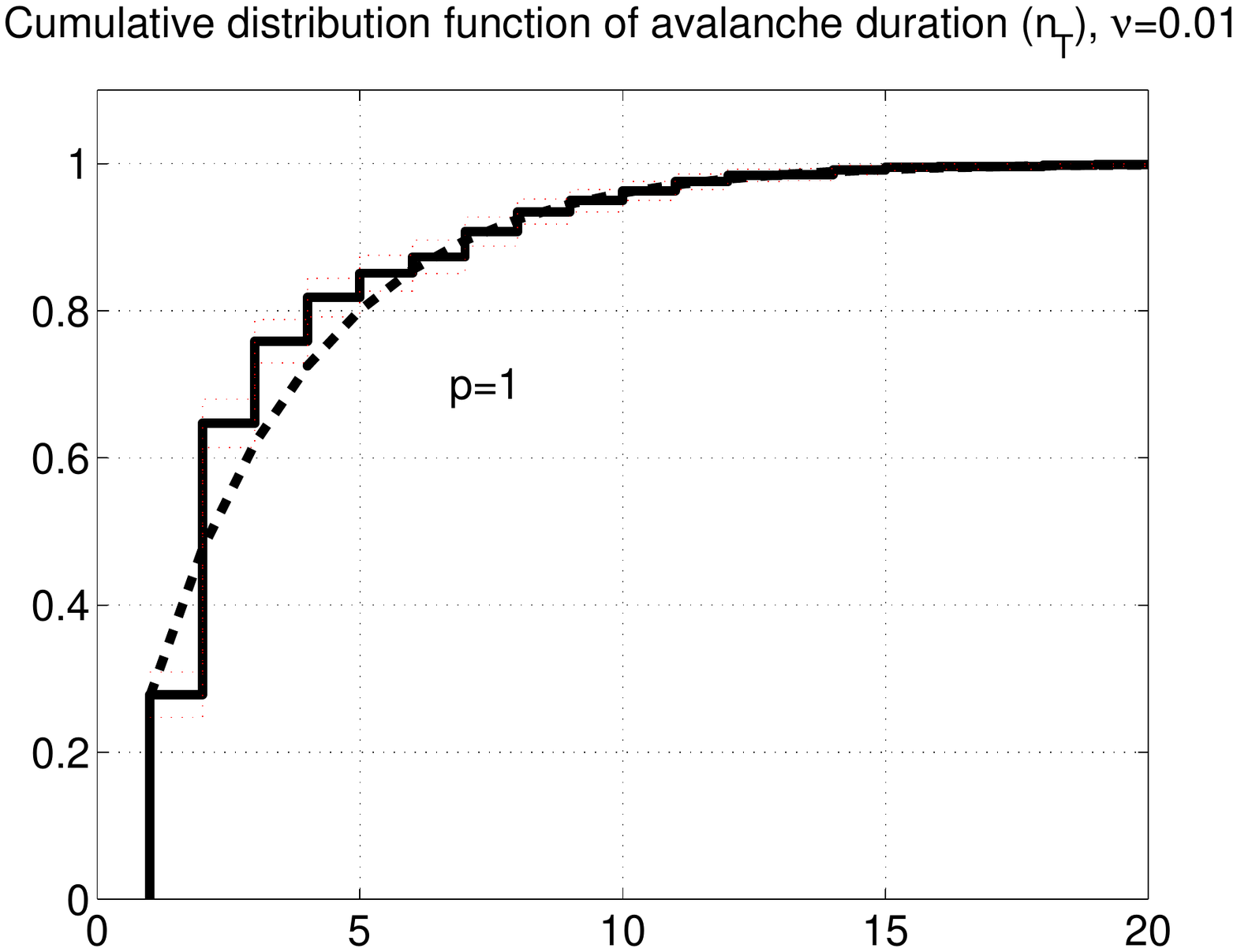}
\caption{Cumulative distribution function of the avalanche time duration fitted with an Exponential distribution. In all plots $\nu=0.01$ and for each plot the value of the connection probability $p$ is indicated}\label{fig2a}
\end{figure} 

\begin{figure}
\centering
\includegraphics[height=5cm,width=4cm]{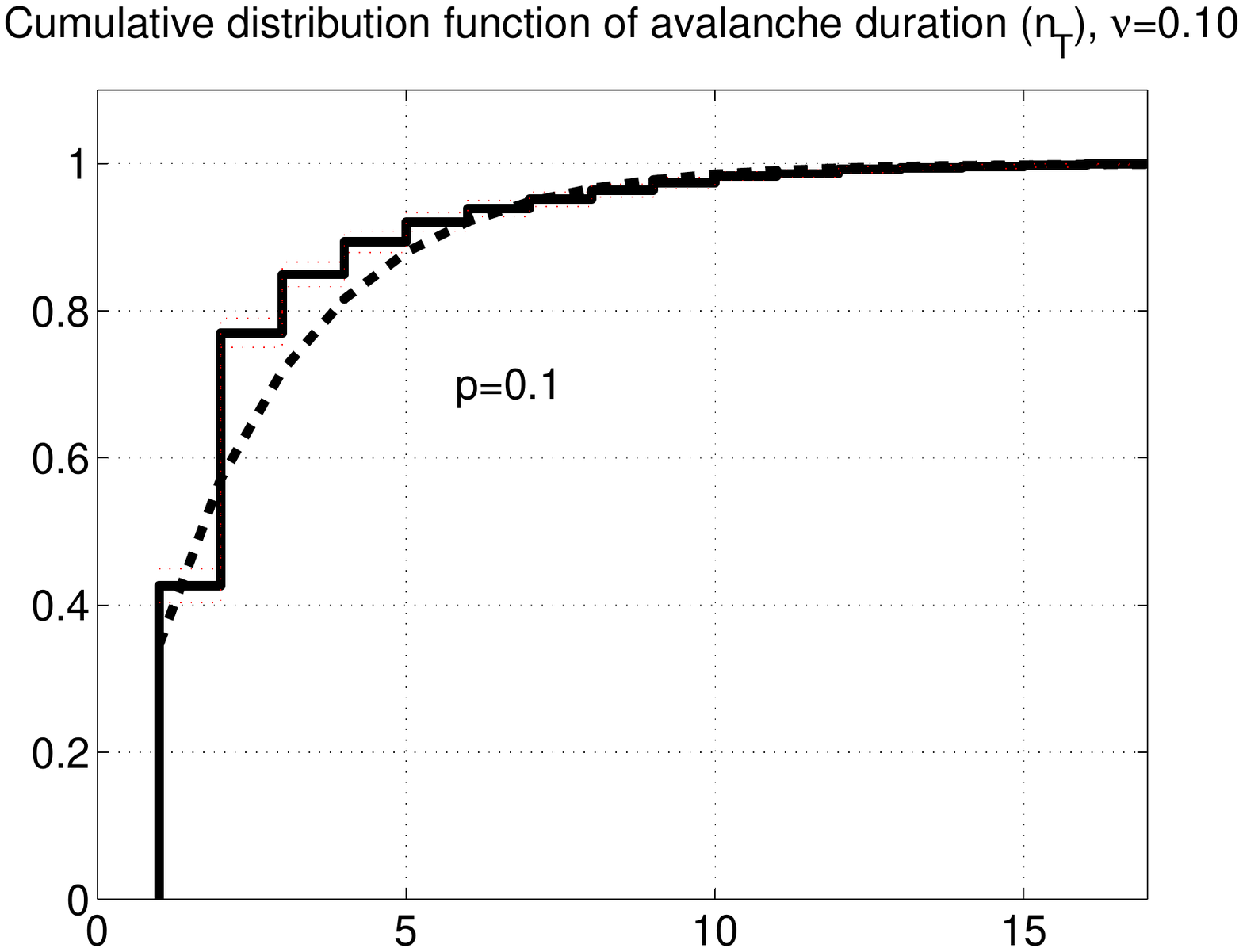}
\includegraphics[height=5cm,width=4cm]{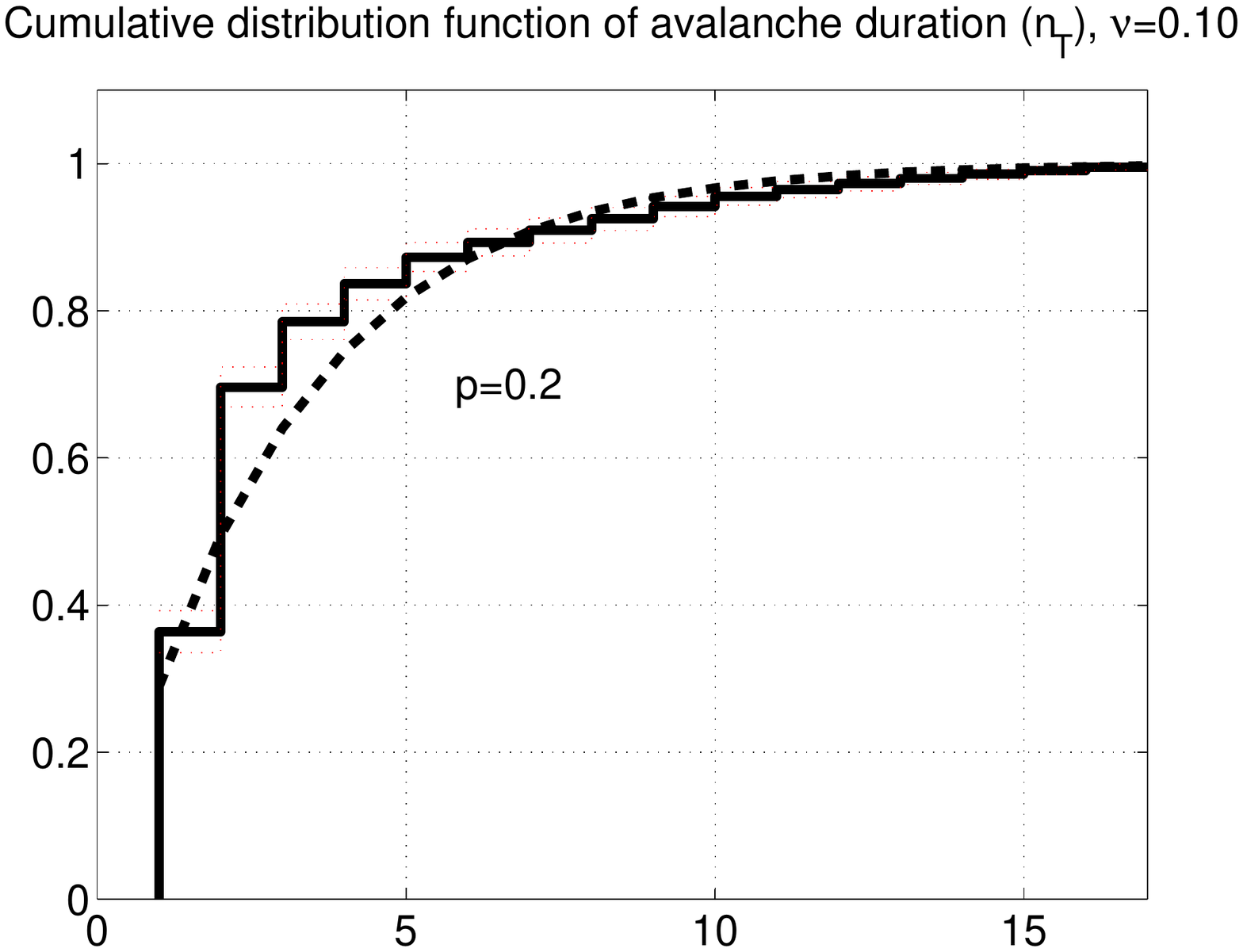}
\includegraphics[height=5cm,width=4cm]{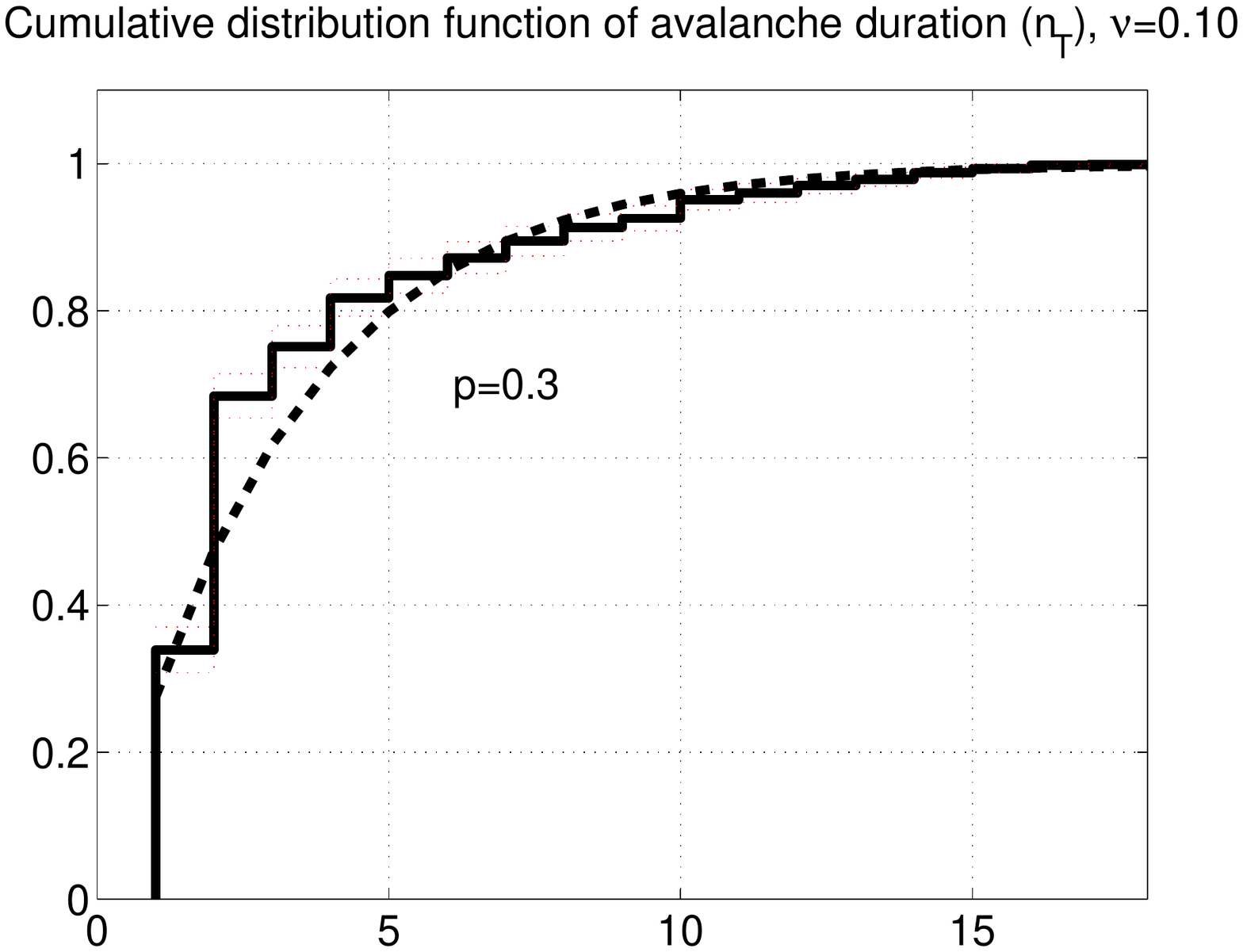}
\includegraphics[height=5cm,width=4cm]{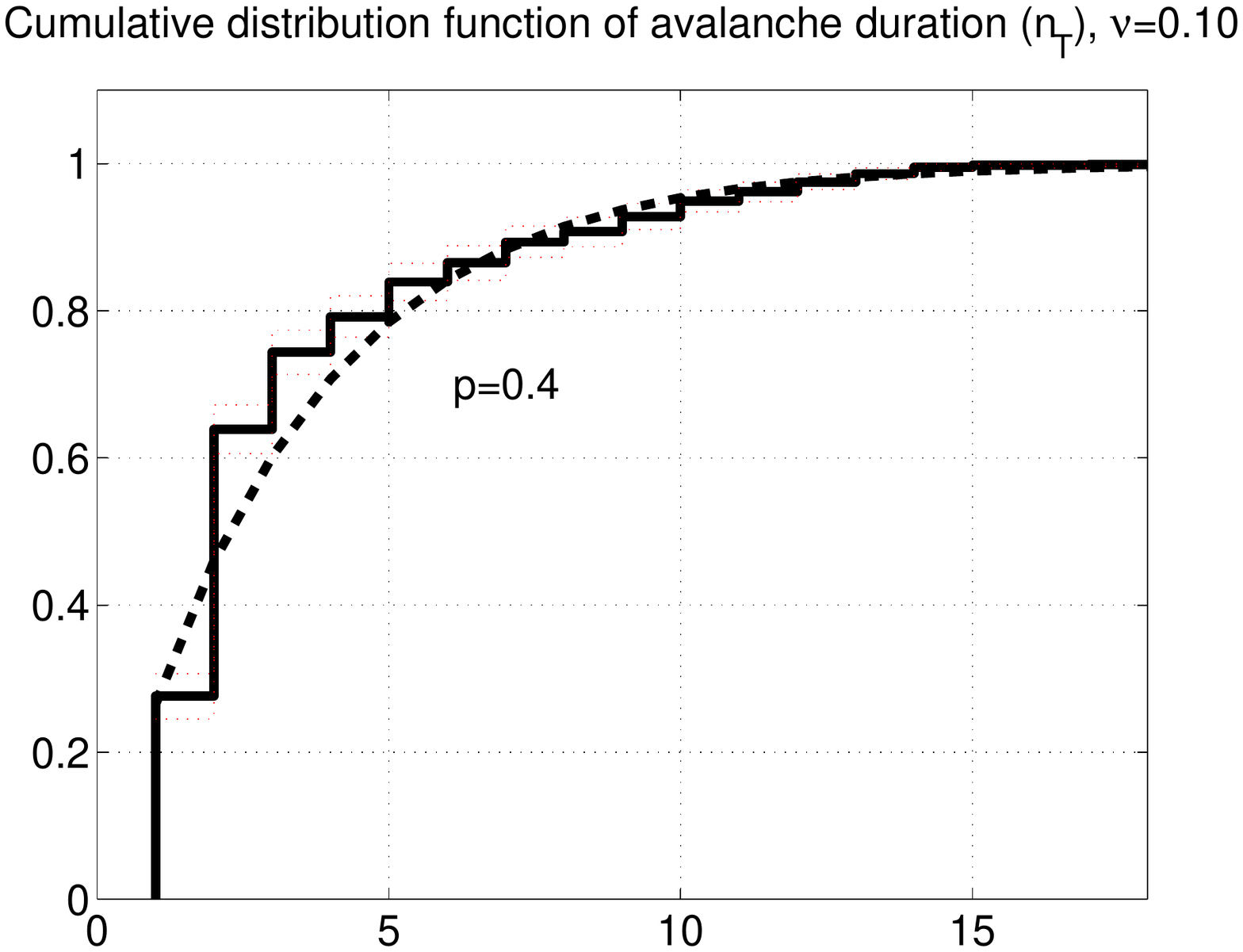}
\includegraphics[height=5cm,width=4cm]{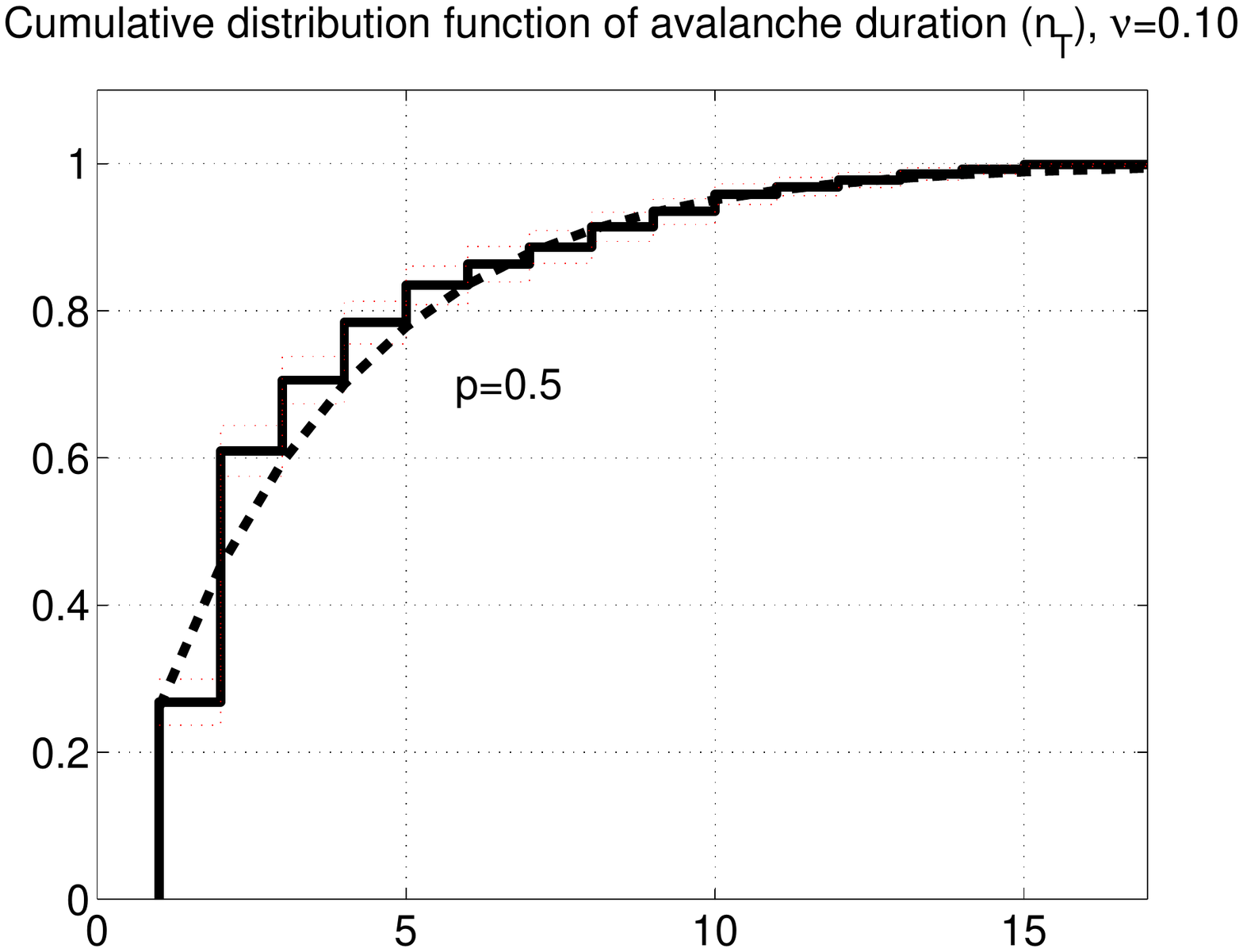}
\includegraphics[height=5cm,width=4cm]{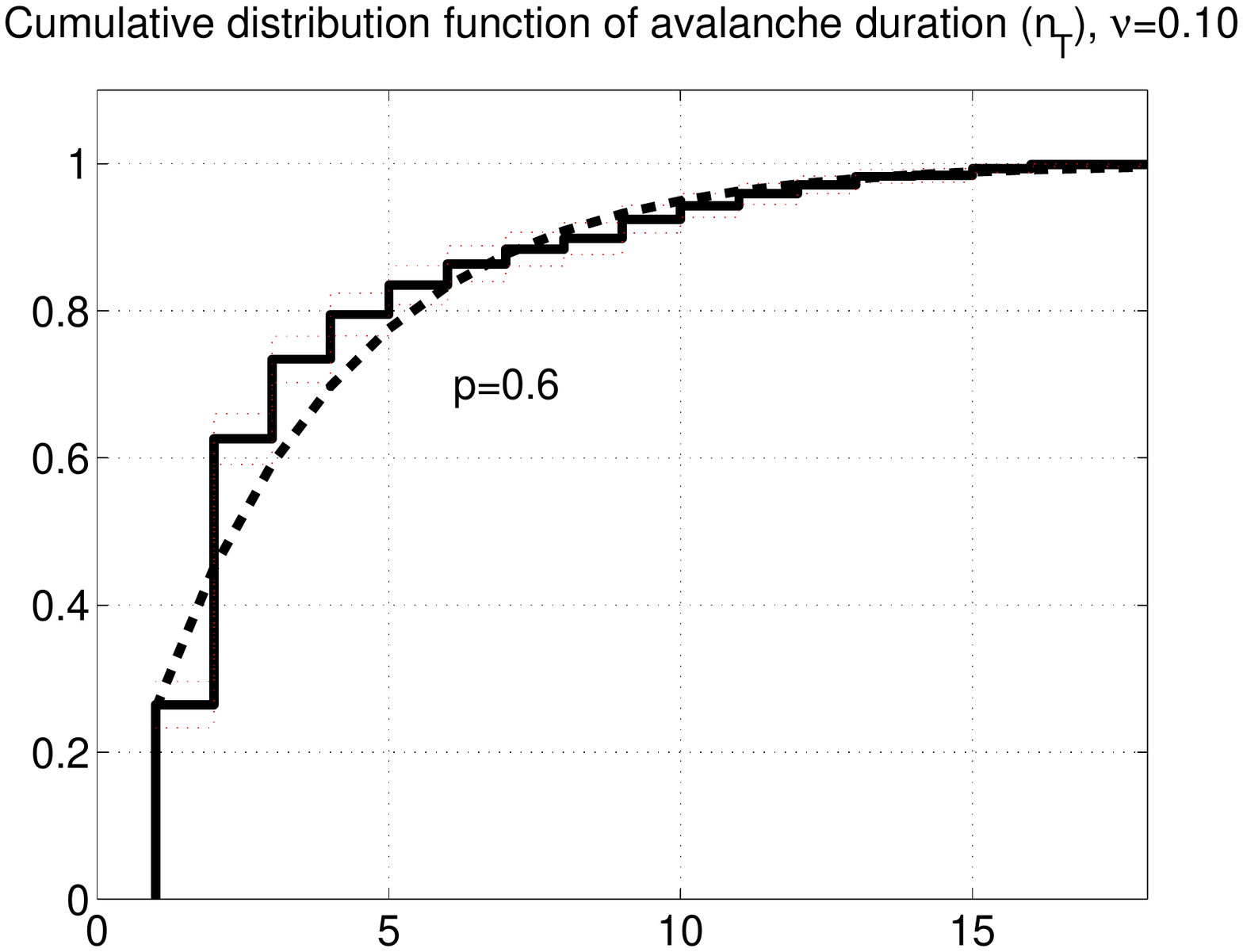}
\includegraphics[height=5cm,width=4cm]{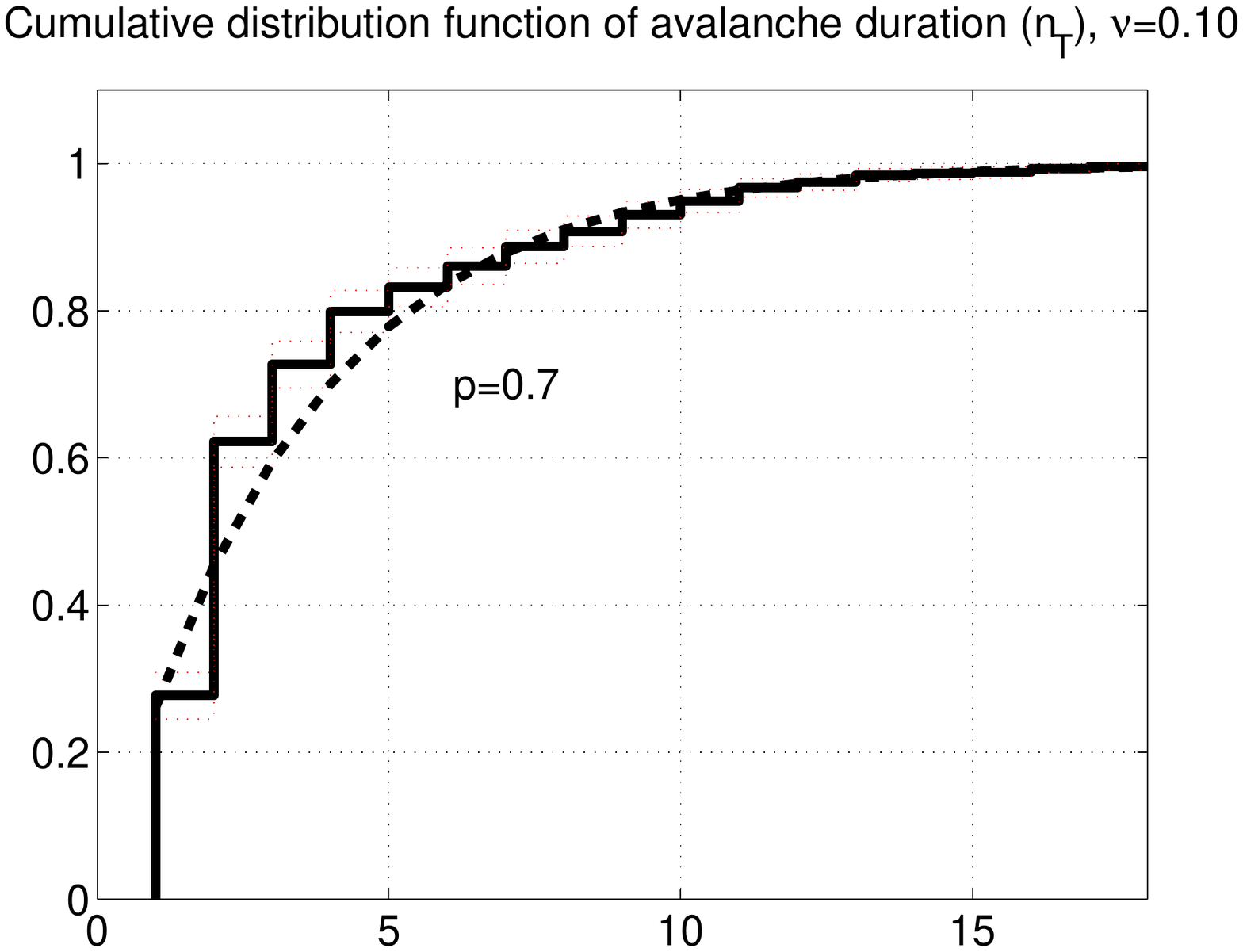}
\includegraphics[height=5cm,width=4cm]{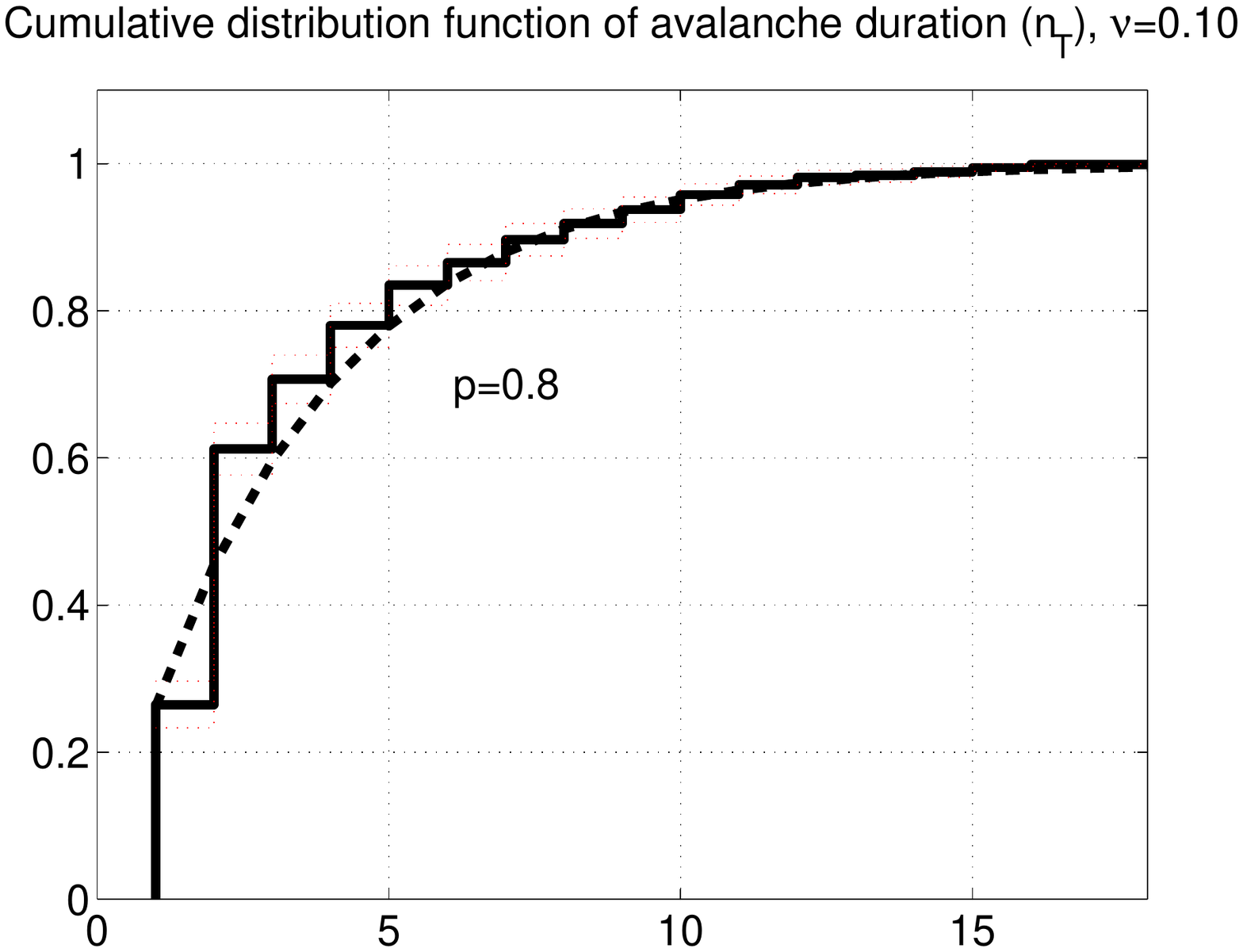}
\includegraphics[height=5cm,width=4cm]{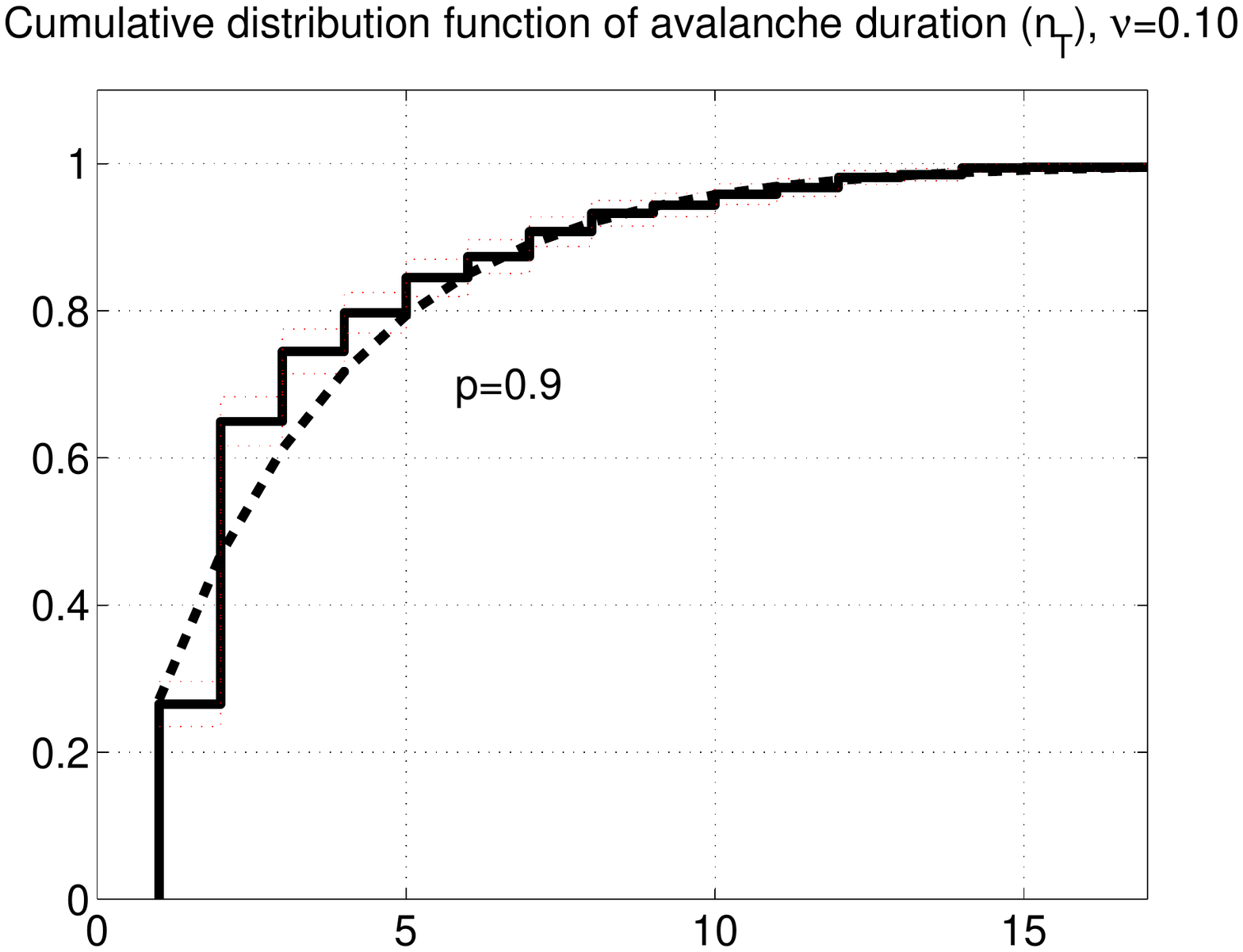}
\includegraphics[height=5cm,width=4cm]{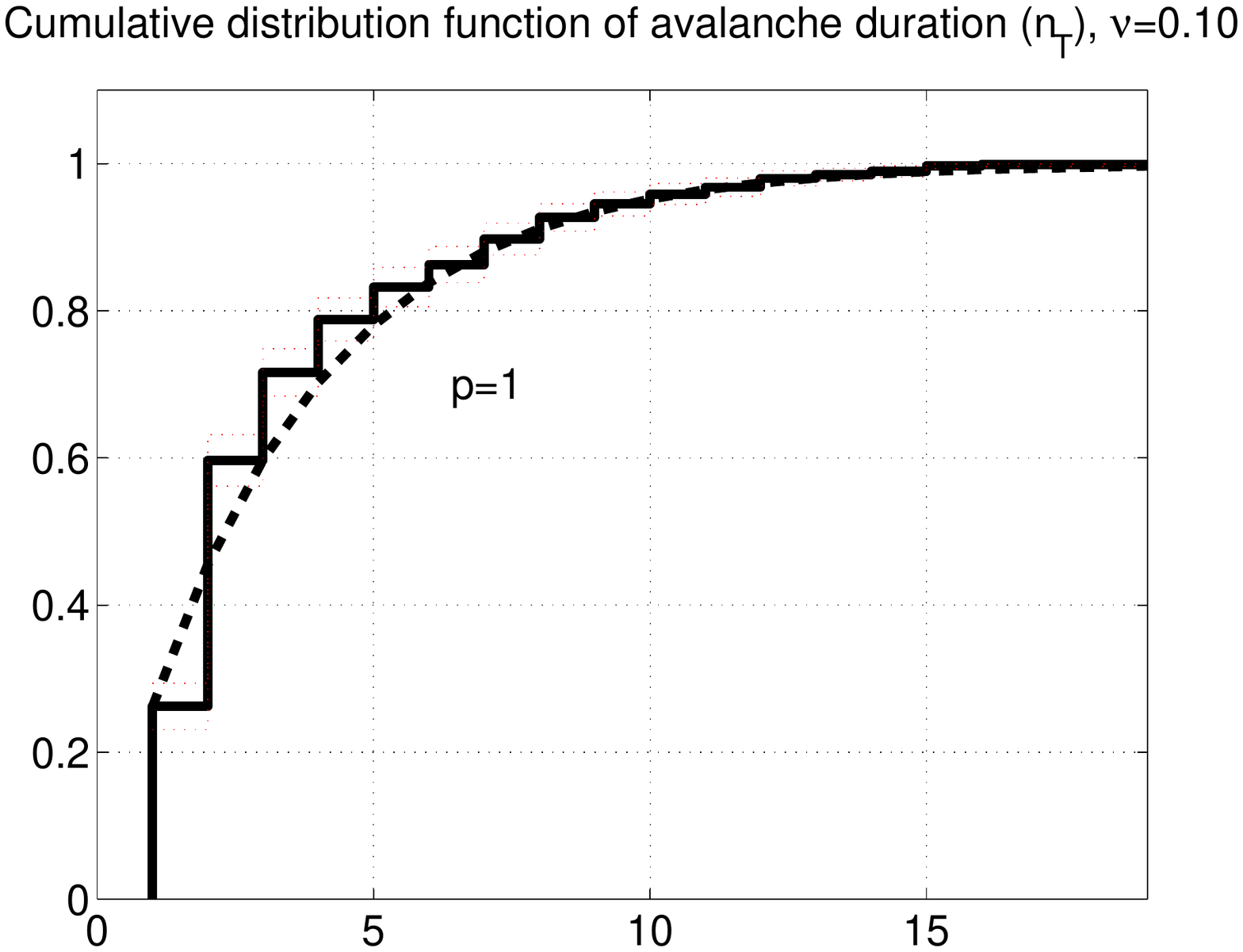}
\caption{Cumulative distribution function of the avalanche time duration fitted with an Exponential distribution. In all plots $\nu=0.10$ and for each plot the value of the connection probability $p$ is indicated}\label{fig2b}
\end{figure}

\begin{figure}
\centering
\includegraphics[height=8cm,width=4cm]{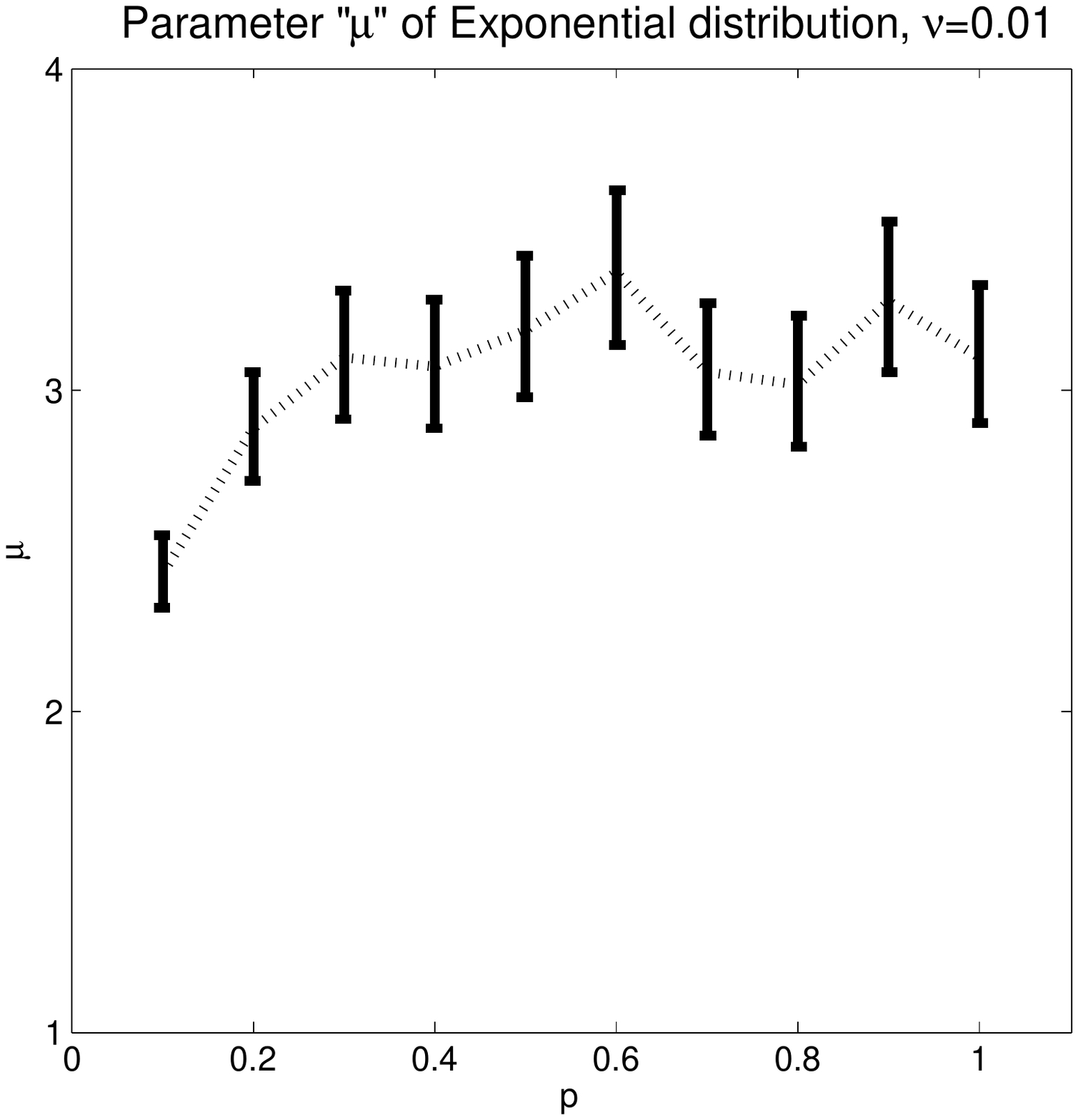}
\includegraphics[height=8cm,width=4cm]{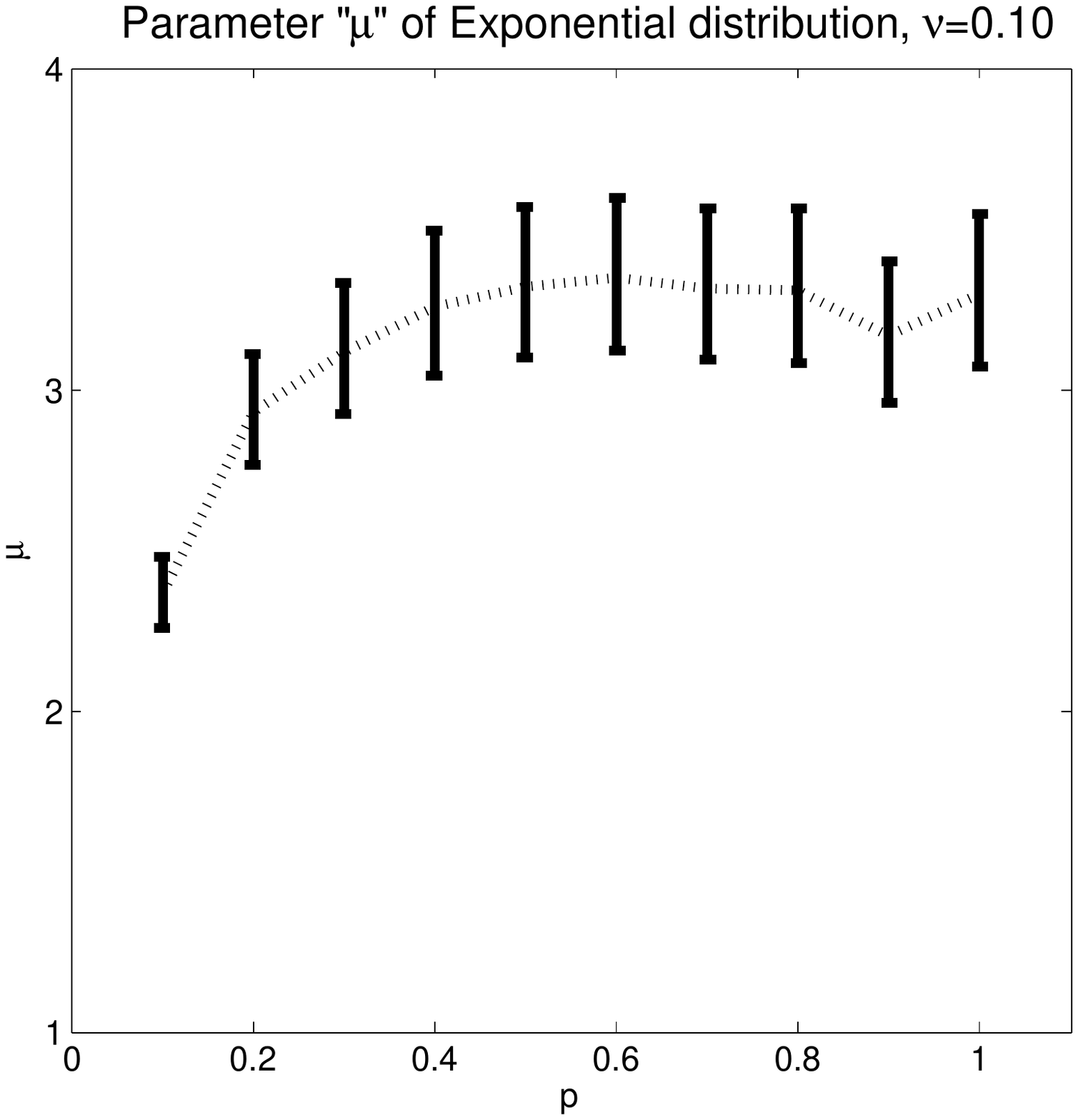}
\caption{(a) Parameter $\mu$ of the Exponential fit as a function of $p$ for $\nu=0.01$; (b) parameter $\mu$ of the Exponential fit as a function of $p$ for $\nu=0.10$; the line between data points serves only to  emphasize some tendency}\label{fig4}
\end{figure}

\section {Results}

In order to implement the above dynamics, we have here by chosen to examine the case of a finite and fixed size random network composed by $N$ nodes, and a connection probability $p$ varying between 0.10 and 1.0. We have examined  cases of $N$ less than 500; it was found that the size of the system can be reduced for answering the above questions, thus allowing some shortening of the computation time. Therefore we report results for $N=100$. We let $\nu$ varying between 0.01 and 0.15;  however for reasons which will become evident, here we show only results for two cases $\nu=0.01$ and $\nu=0.10$. In all cases,  the displayed results are obtained after averaging over 100 simulations.

In Figs. \ref{fig1a} and \ref{fig1b}, we show the cumulative distribution functions (cdf) for the avalanche size, normalized to the number of links $L$, for each $p$ and for two values of the external field $\nu=0.01$ and  $\nu=0.10$. Each cdf  is fitted with the corresponding  cdf of a Weibull distribution  that can be written 
\begin{equation}
F(x)=1-e^{-({x/\alpha})^\beta}
\end{equation}

The behavior of the parameters $\alpha$ and $\beta$ as function of the connection probability $p$ is shown in Fig. \ref{fig3}   for  both field values. One can notice that the parameters do not depend on the external field value $\nu$ but only on the probability $p$. The percolation transition of a  Erd\"os-R\'enyi network, i.e. when a connected network occurs \cite{luczac}  is adequately seen near $p_0 = (1/N) ln (N)$ $\sim 0.05$.
 
In Figs. \ref{fig2a} and \ref{fig2b} we show the cdf of the time duration of the avalanches. Also in this case we show the cdf  for each studied $p$ and for two values of $\nu=0.01$ and  $\nu=0.10$. In this case  each cdf is fitted with an exponential cdf written as
\begin{equation}
G(x)=1-e^{-{x/ \mu}}.
\end{equation}

The behavior of the parameter  $\mu$ as a function of $p$ and for both values of $\nu$ is shown in figure \ref{fig4}. Also in this case there is no  field dependence of the parameter $\mu$ and the dependence on $p$ is not strong. 
 
  Furthermore, in Fig. 7,  the number of avalanches $N_A$ before all nodes are fully aware is shown  as a function of $p$ for $\nu=0.01$ and for $\nu=0.10$. One finds independently of $\nu$:  $N_A \simeq 8 + 0.1\; p^2 $.
 Moreover, in  order to see how long it takes before a  network is fully aware it is of interest to display the number $ N_h$ of sequences ($h$) before all nodes are aware as function of $p$; this is shown in Fig. 8 for $\nu=0.01$ and   $=0.10$.  One finds $ N_h \simeq (1/\nu)(6+0.02 p^{-5/2})$. 

 The number of $hot$ nodes, for example with $a$ between $\phi-0.10\phi$ and $\phi$  as displayed in Figs. 9-10 as a function of $h$ for $\nu=0.01$  and $\nu=0.10$  respectively.  In each plot the value of the connection probability $p$ is indicated. The curve is fitted with a three parameter  logistic function written as
\begin{equation}
f(h)=\frac{A}{1+B\;e^{Kh }}.
\end{equation}
The parameters $A$ and $B$ are used respectively for the normalization of the logistic function. Their values (e.g. $A \sim 10$; $B\sim 0.025 $) are easily understood, in view of the type of studied network. 
It is interesting, as a complement to the above, to calculate the number of nodes with $a$ still below $\phi$ as function of $h$.  This is shown for both $\nu $  values  In Figs. 11-12;   for each plot the value of the connection probability $p$ is indicated.
The data can be fitted with a three parameter logistic function; the same $K$ value, i.e.  $K \cong 0.005$ when $\nu=0.01$ and  $K\cong 0.05$ when $\nu=0.10$,   are found for the number of hot nodes and for the number of not yet fully aware nodes. However the parameters $A$ and $B$ are increased by a nuumerical factor, i.e.  $A \sim 100$; $B\sim  0.065$. 

From the preceding  illustrations, Figs. (9-12) one also observes that the (x-) ``time'' scale is  an order of magnitude different,  as easily understood due to the value  of $\nu$, while the y-axis scale  for the number of hot nodes is an order of magnitude smaller than for the total number of not yet fully aware nodes, due to the assumed initial conditions on the $a_i$ distribution, the type of network and the model dynamics. The important parameter for the model is $K$, playing the role of an inverse  relaxation time, more exactly here a growth rate;  we emphasize  that $K$ does not depend on the value of $p$, but one has a universal relation:  $K = \nu/2$. 

\begin{figure}
\centering
\includegraphics[height=8cm,width=4cm]{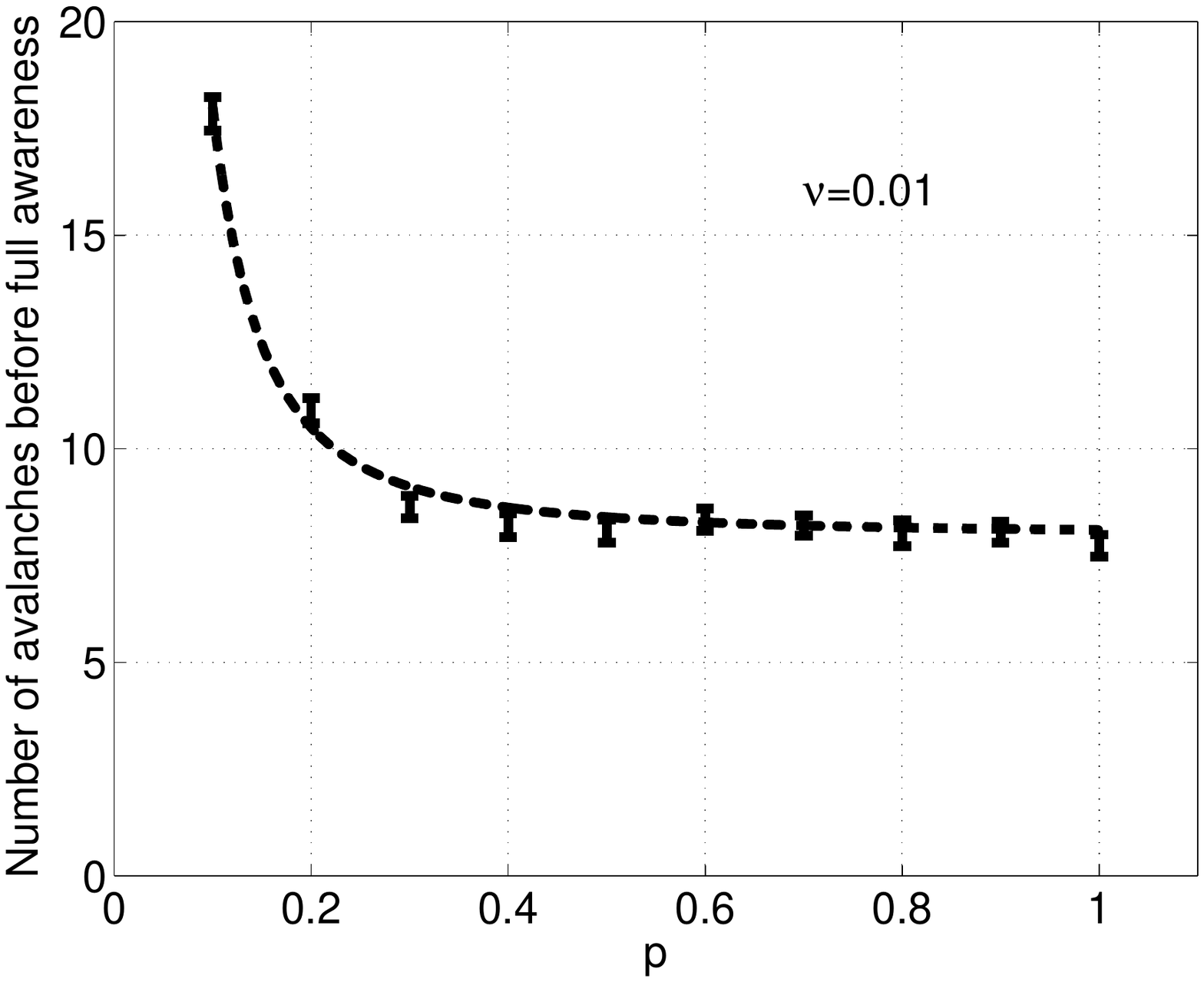}
\includegraphics[height=8cm,width=4cm]{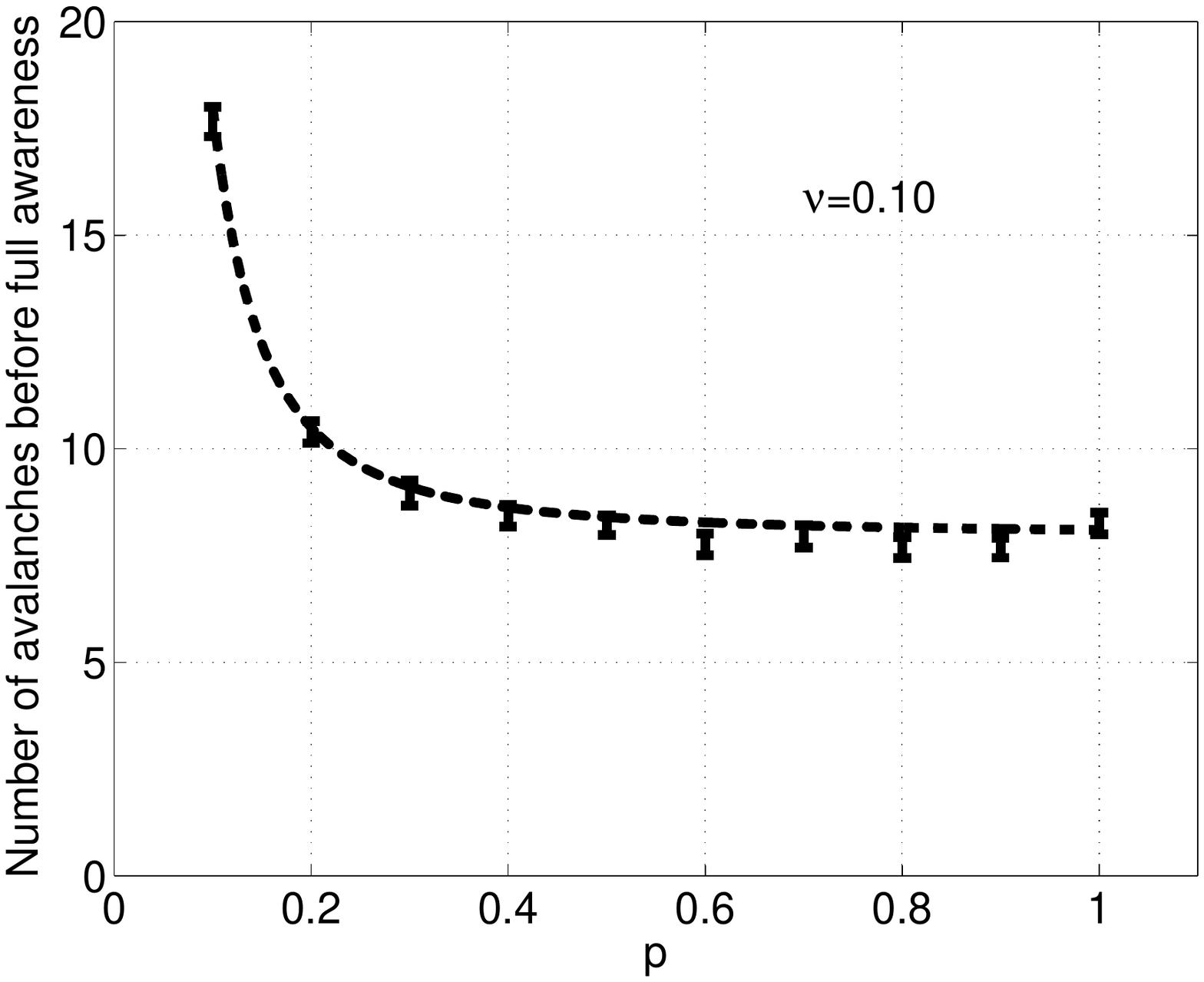}
\caption{ Number of avalanches before all nodes are aware as function of $p$ : (a) for $\nu=0.01$, (b) for $\nu=0.10$; the line between data points has an analytical form given in the main text}\label{figavalnumber}
\end{figure}

\begin{figure}
\centering
\includegraphics[height=8cm,width=4cm]{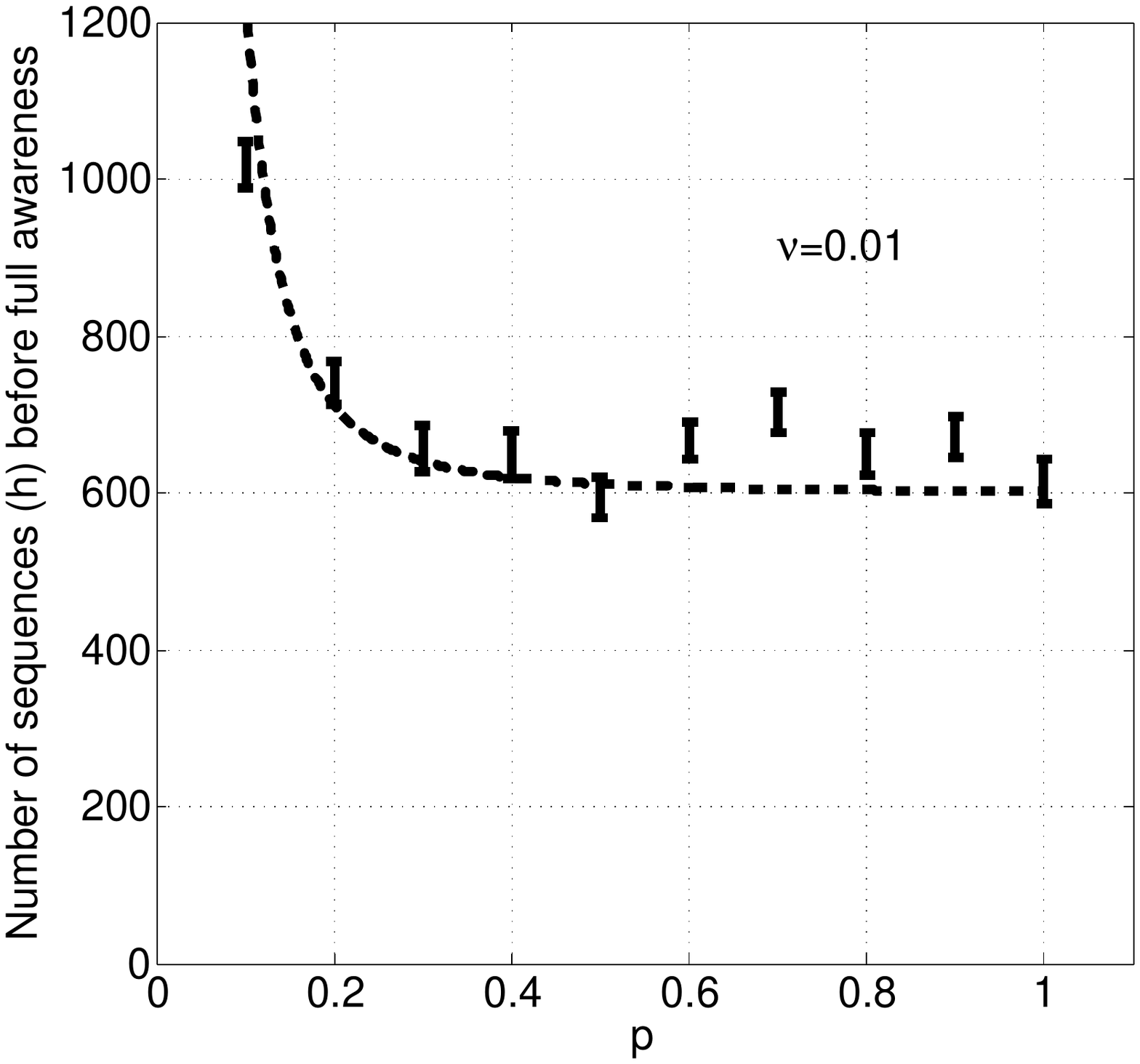}
\includegraphics[height=8cm,width=4cm]{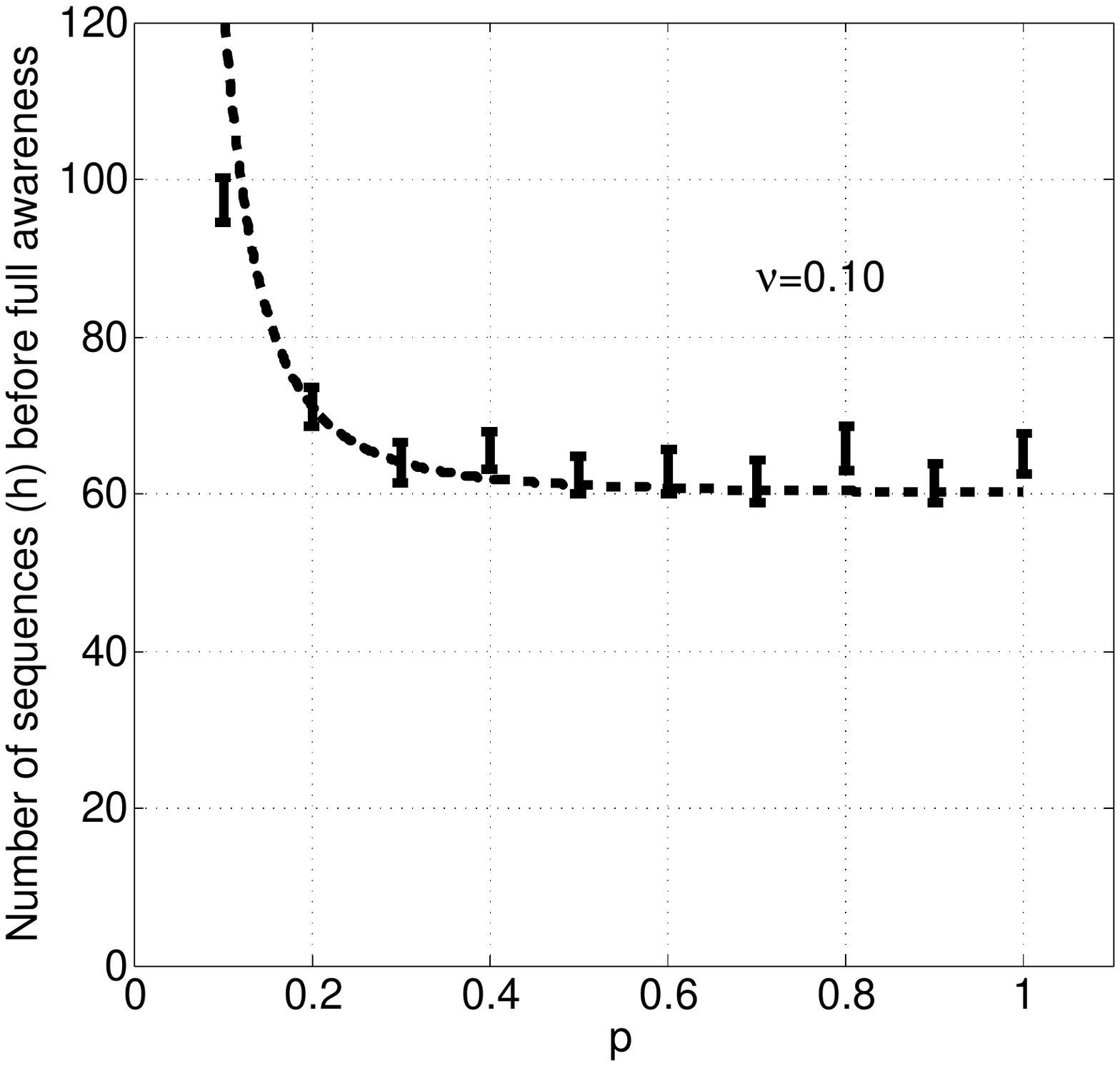}
\caption{Number of sequences ($h$) before all nodes are aware as function of $p$: (a)  for $\nu=0.01$, (b) for $\nu=0.10$; the line between data points has an analytical form given in the main text} \label{figseqnumber}
\end{figure}

\begin{figure}
\centering
\includegraphics[height=5cm,width=4cm]{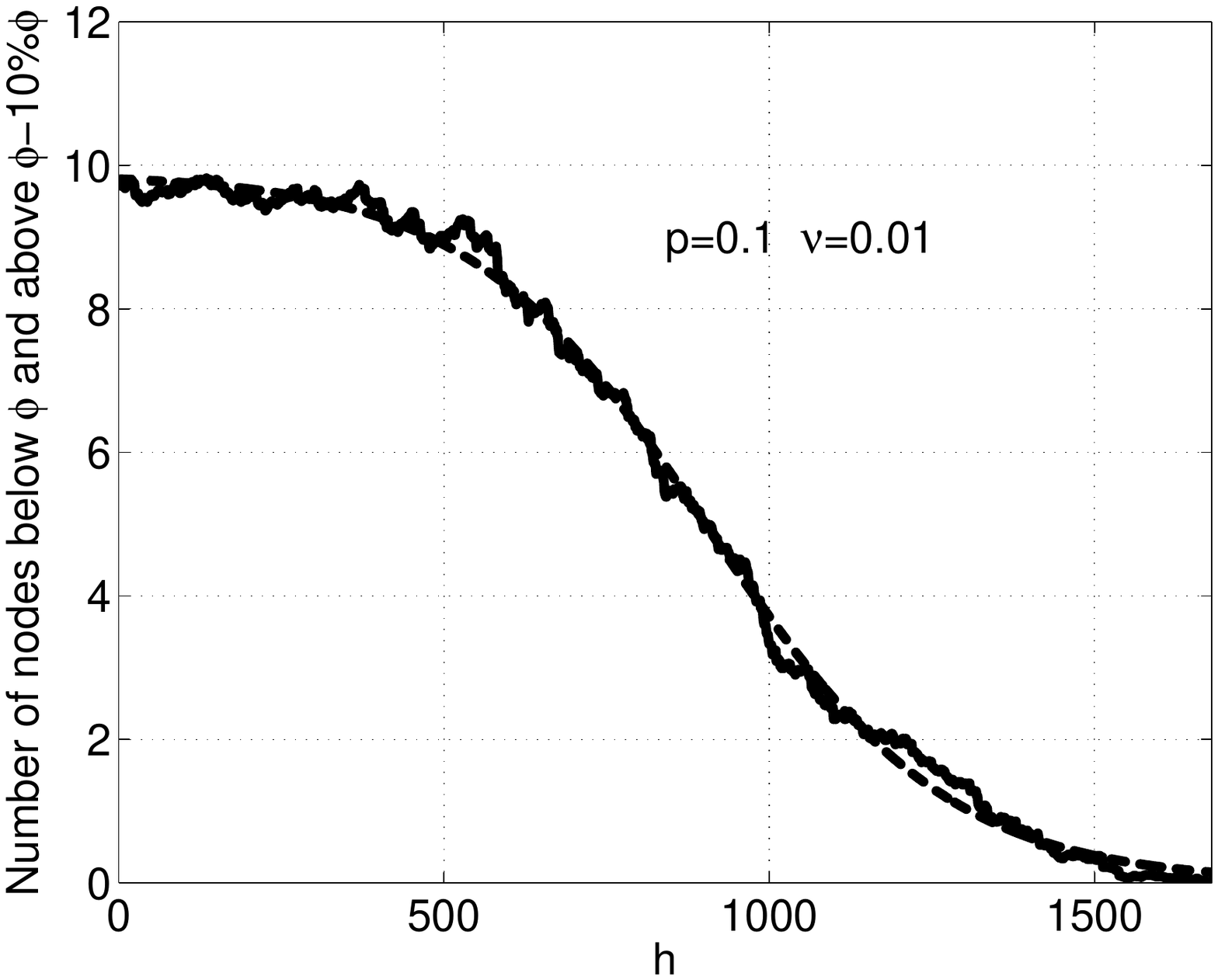}
\includegraphics[height=5cm,width=4cm]{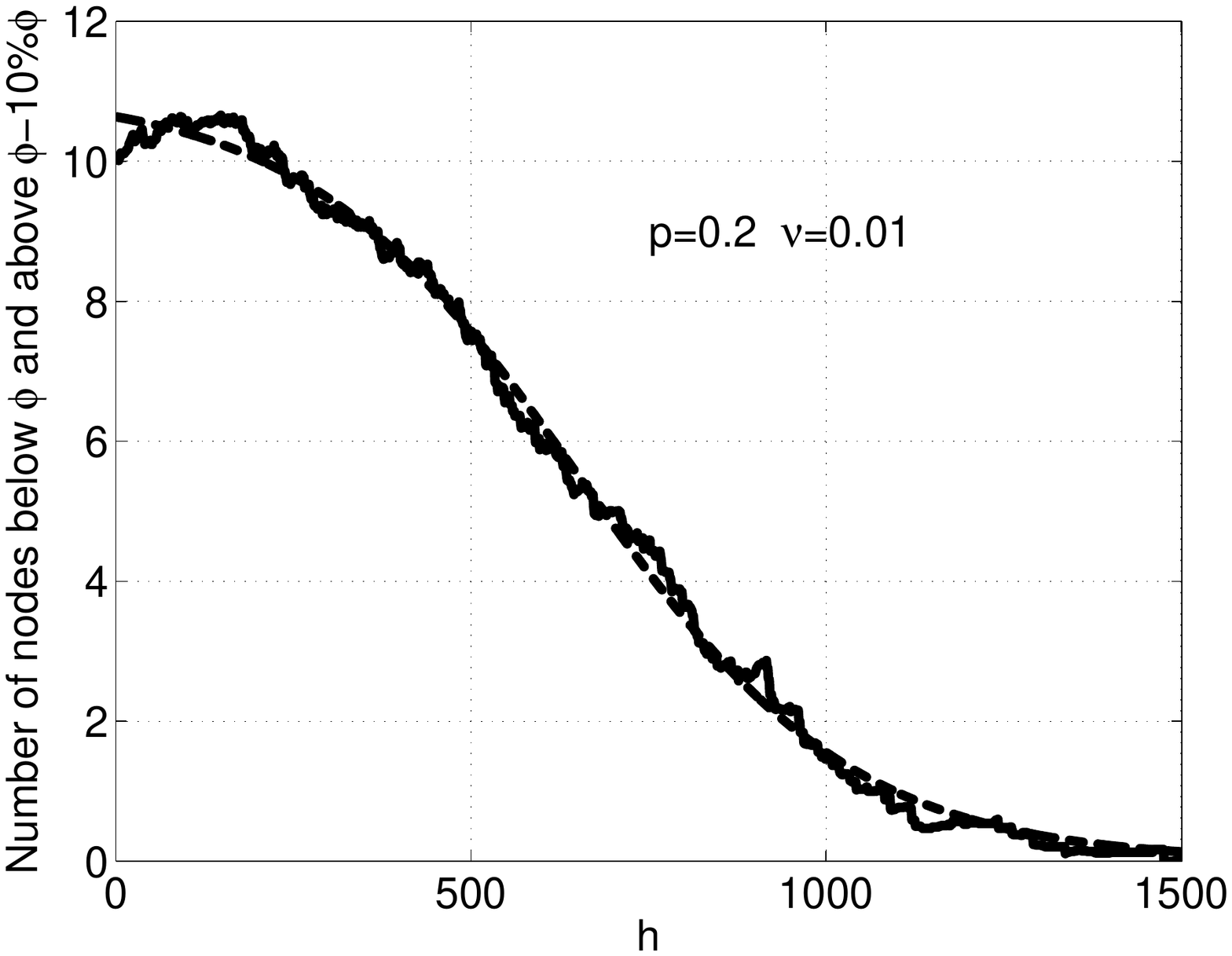}
\includegraphics[height=5cm,width=4cm]{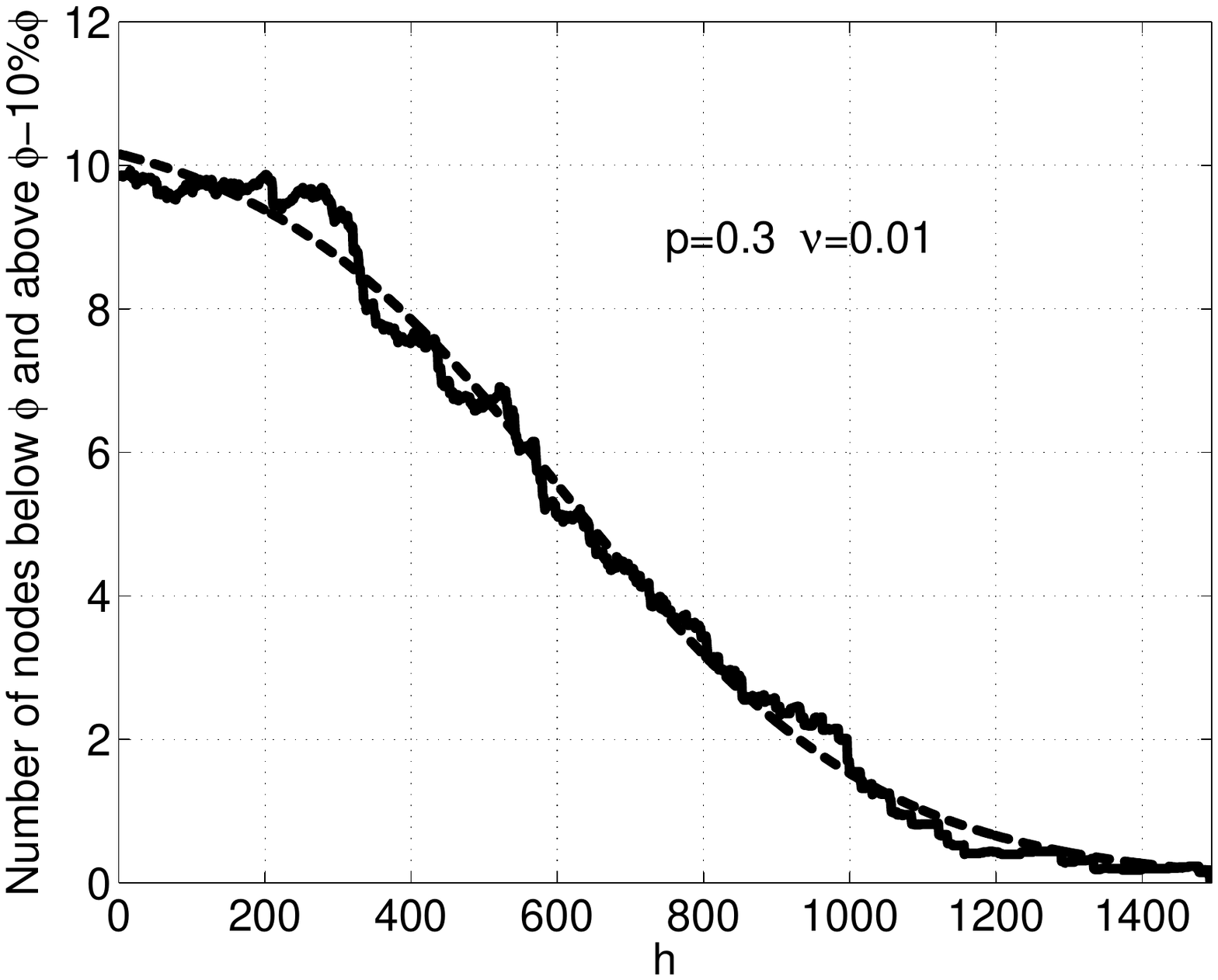}
\includegraphics[height=5cm,width=4cm]{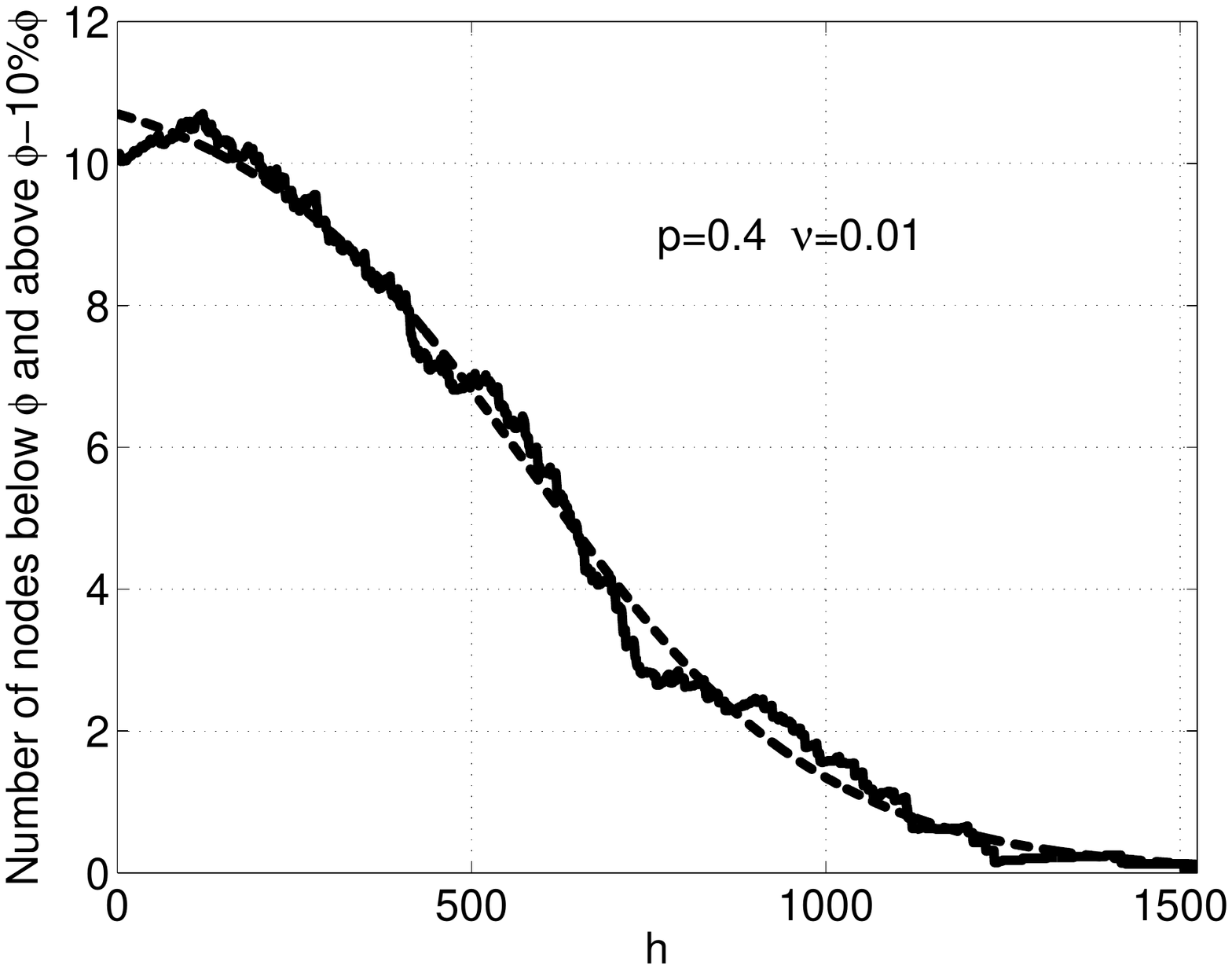}
\includegraphics[height=5cm,width=4cm]{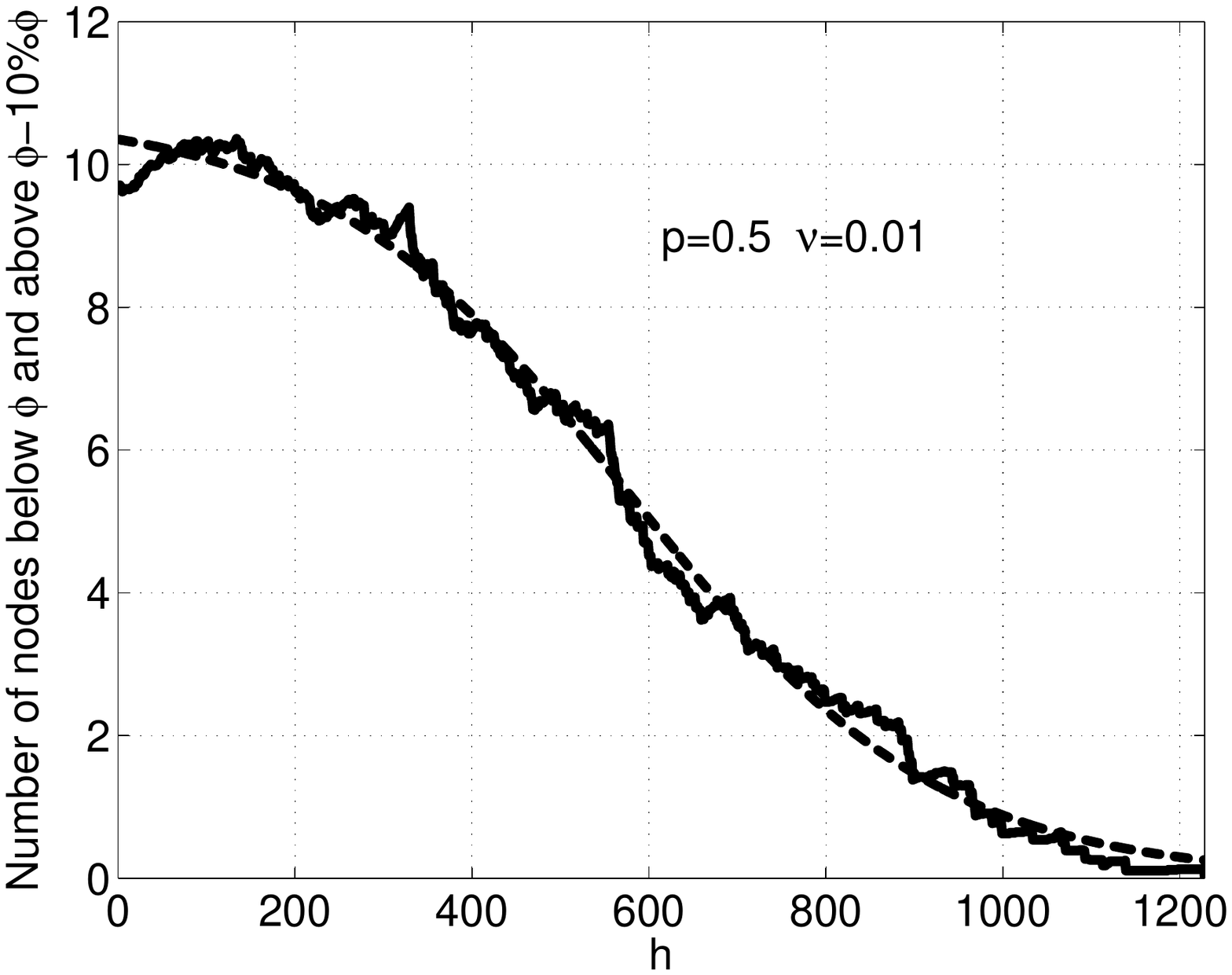}
\includegraphics[height=5cm,width=4cm]{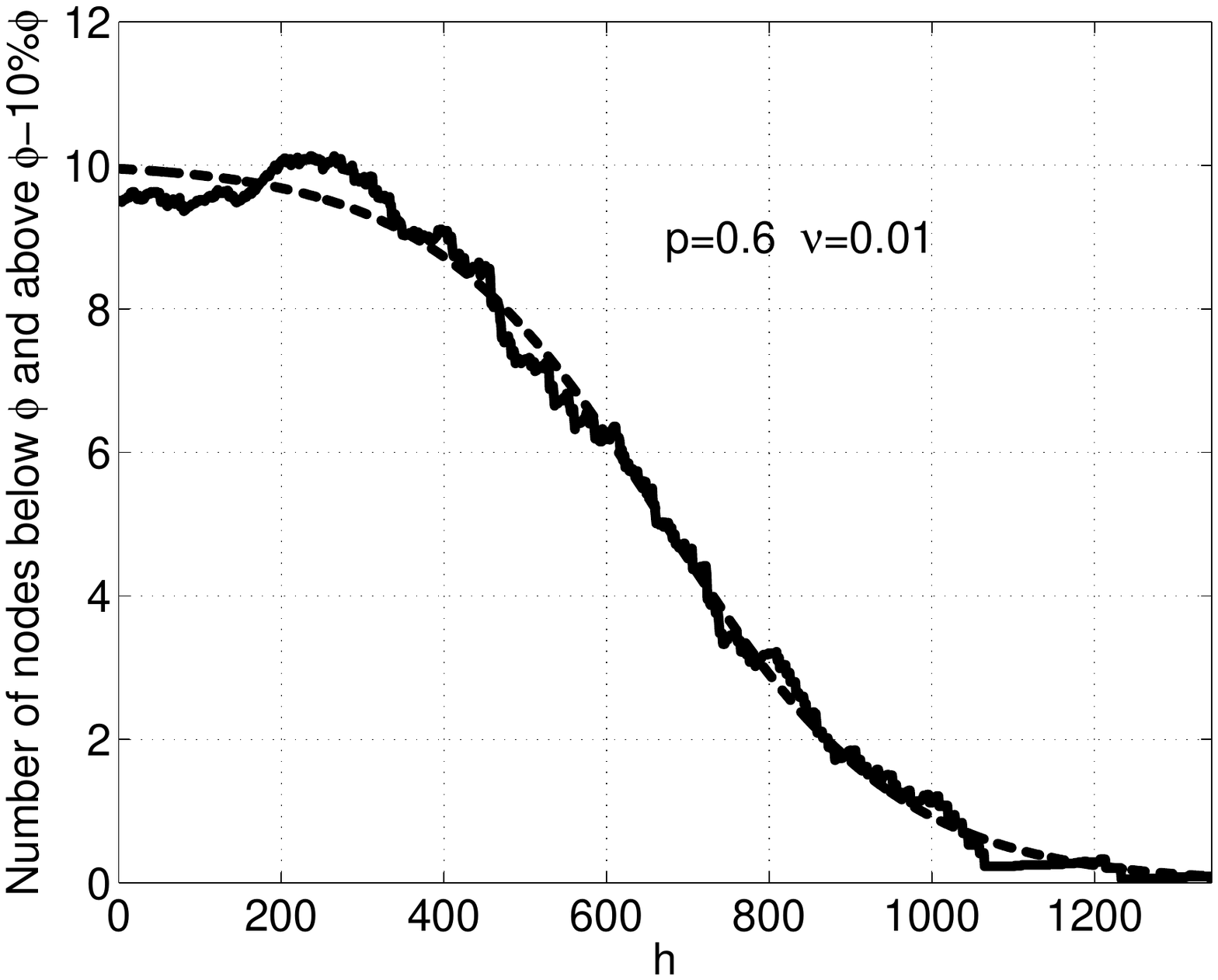}
\includegraphics[height=5cm,width=4cm]{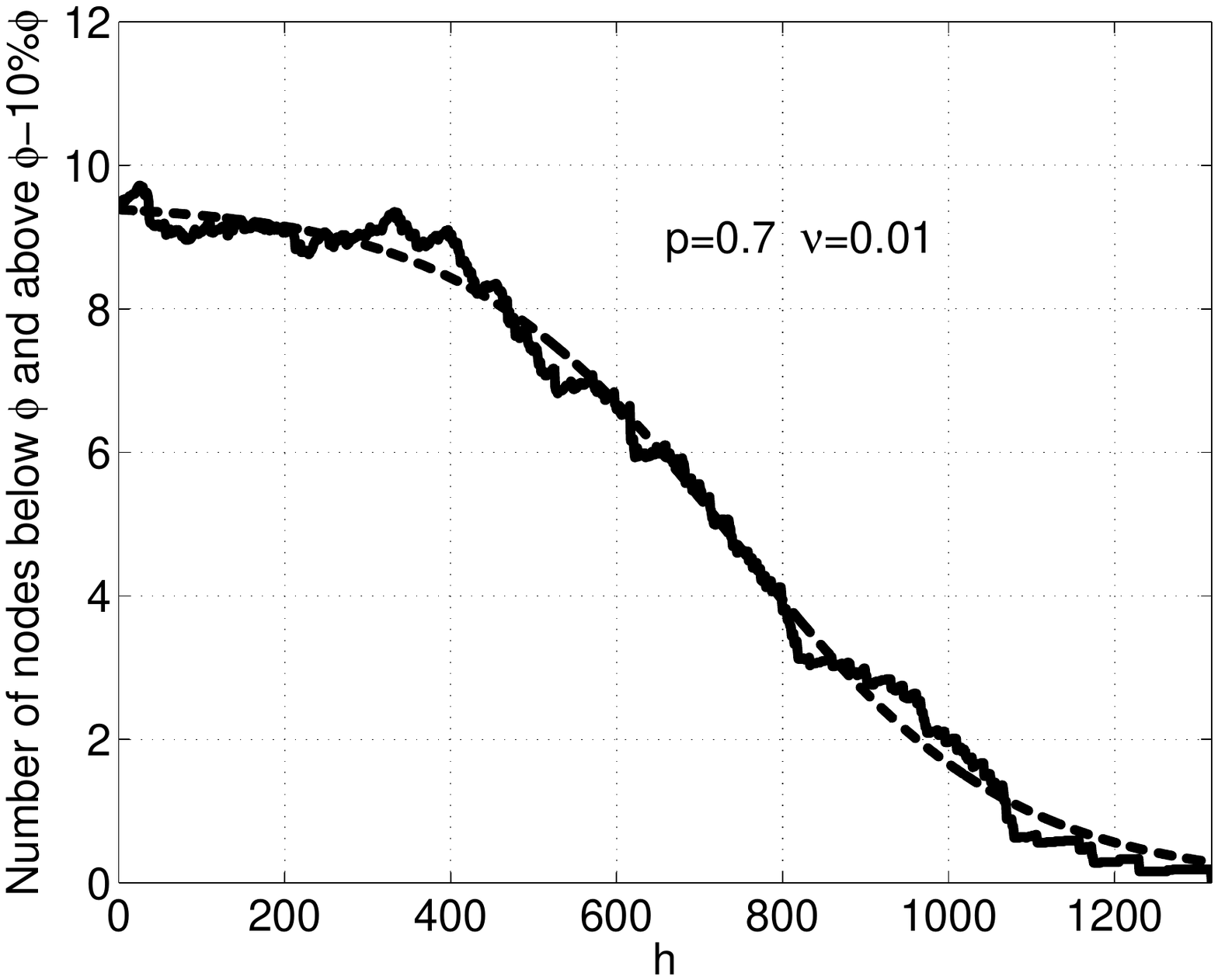}
\includegraphics[height=5cm,width=4cm]{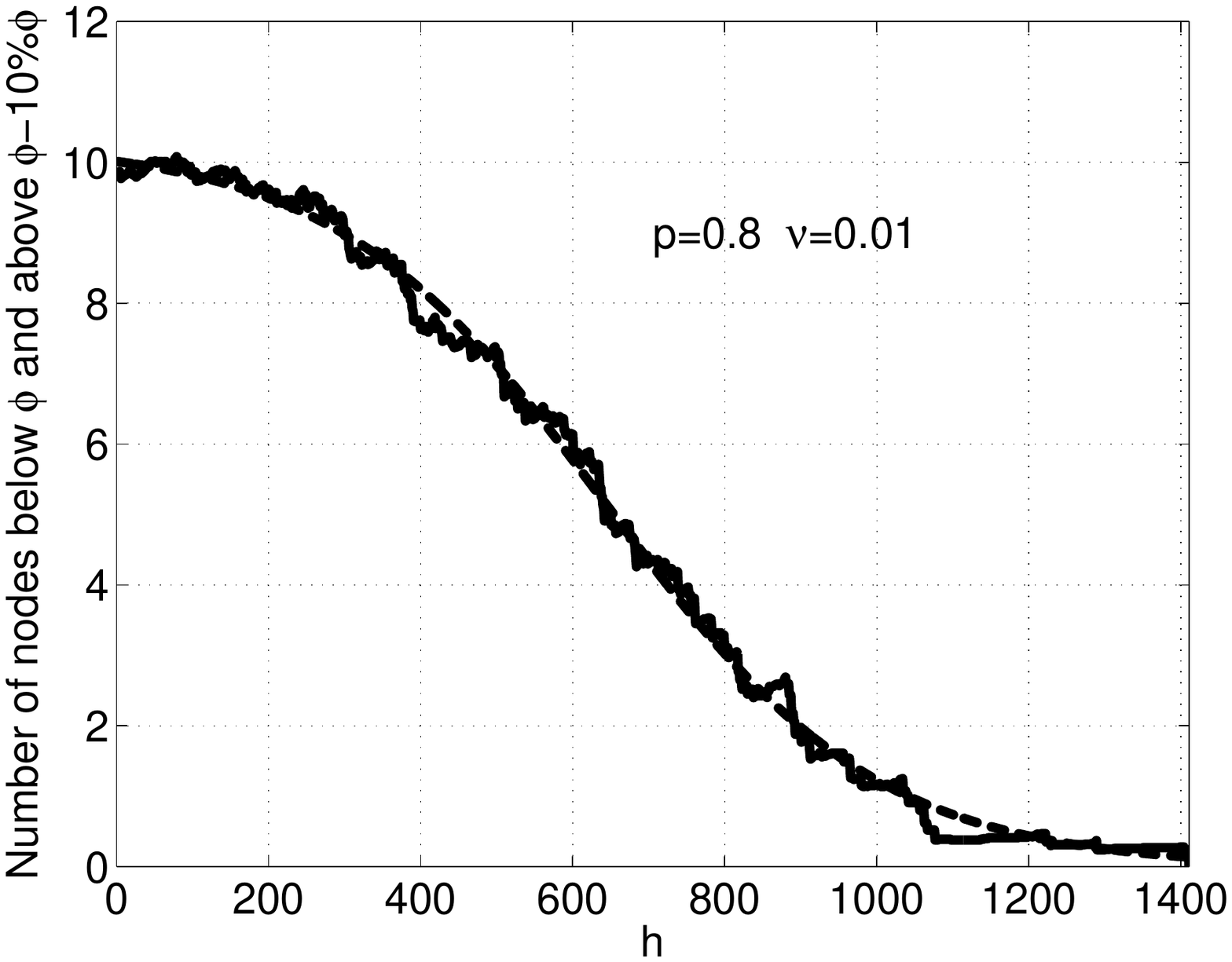}
\includegraphics[height=5cm,width=4cm]{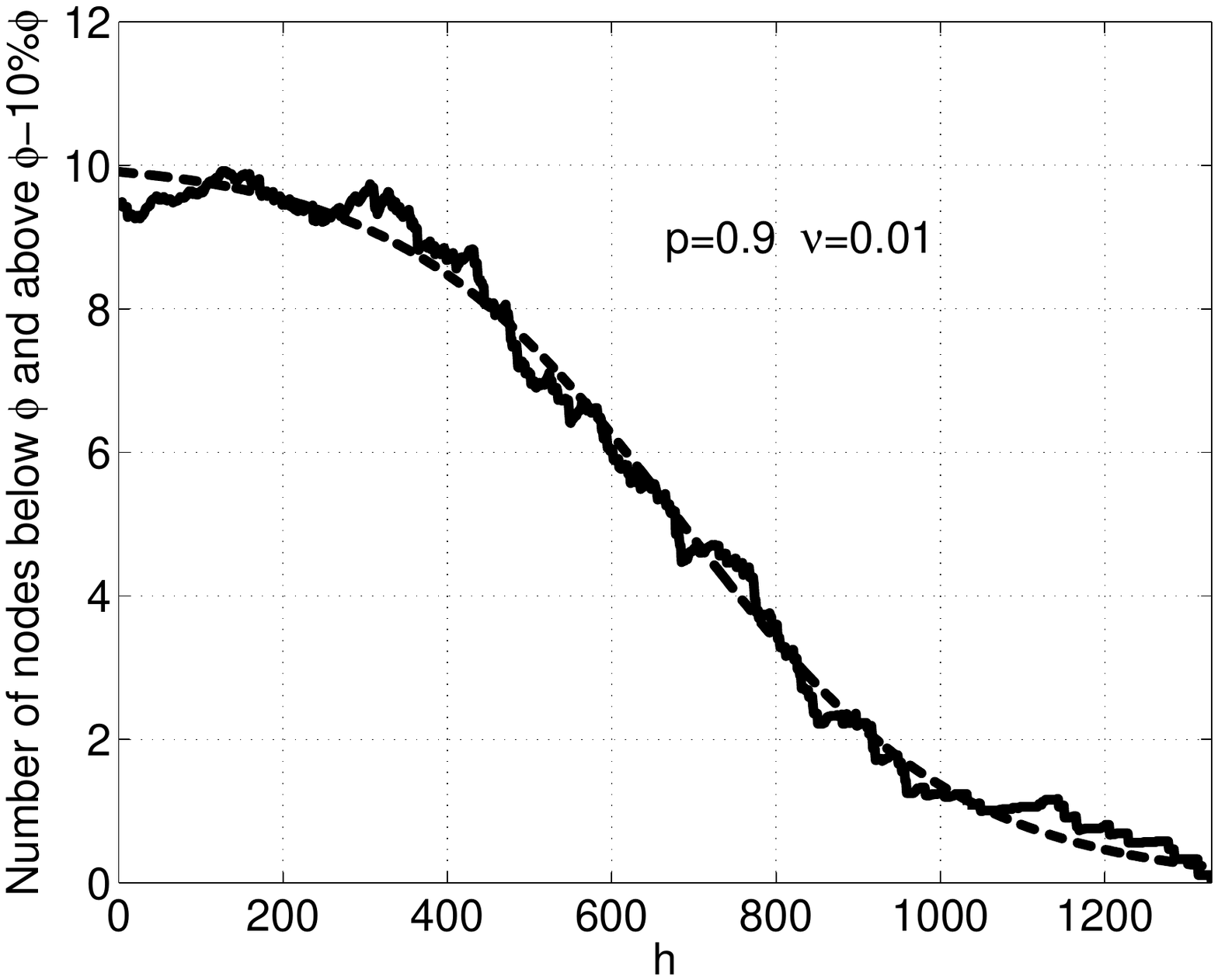}
\includegraphics[height=5cm,width=4cm]{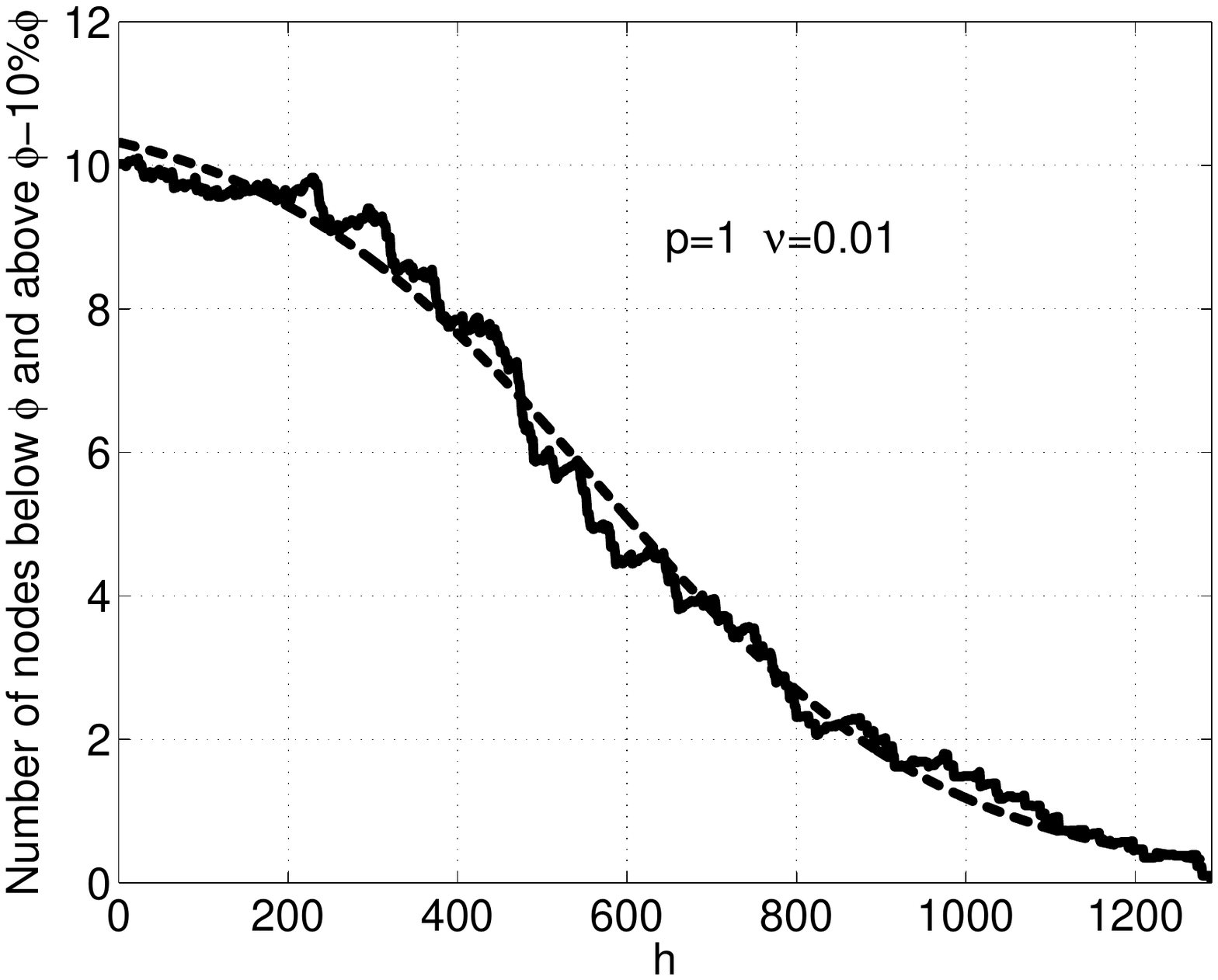}
\caption{Number of nodes with $a$ between $\phi-0.10\phi$ and $\phi$ as function of $h$. In all plots $\nu=0.01$ and for each plot the value of the connection probability $p$ is indicated. The curve is fitted with a logistic function as explained in the text.} \label{fighotnodenumber001}
\end{figure}

\begin{figure}
\centering
\includegraphics[height=5cm,width=4cm]{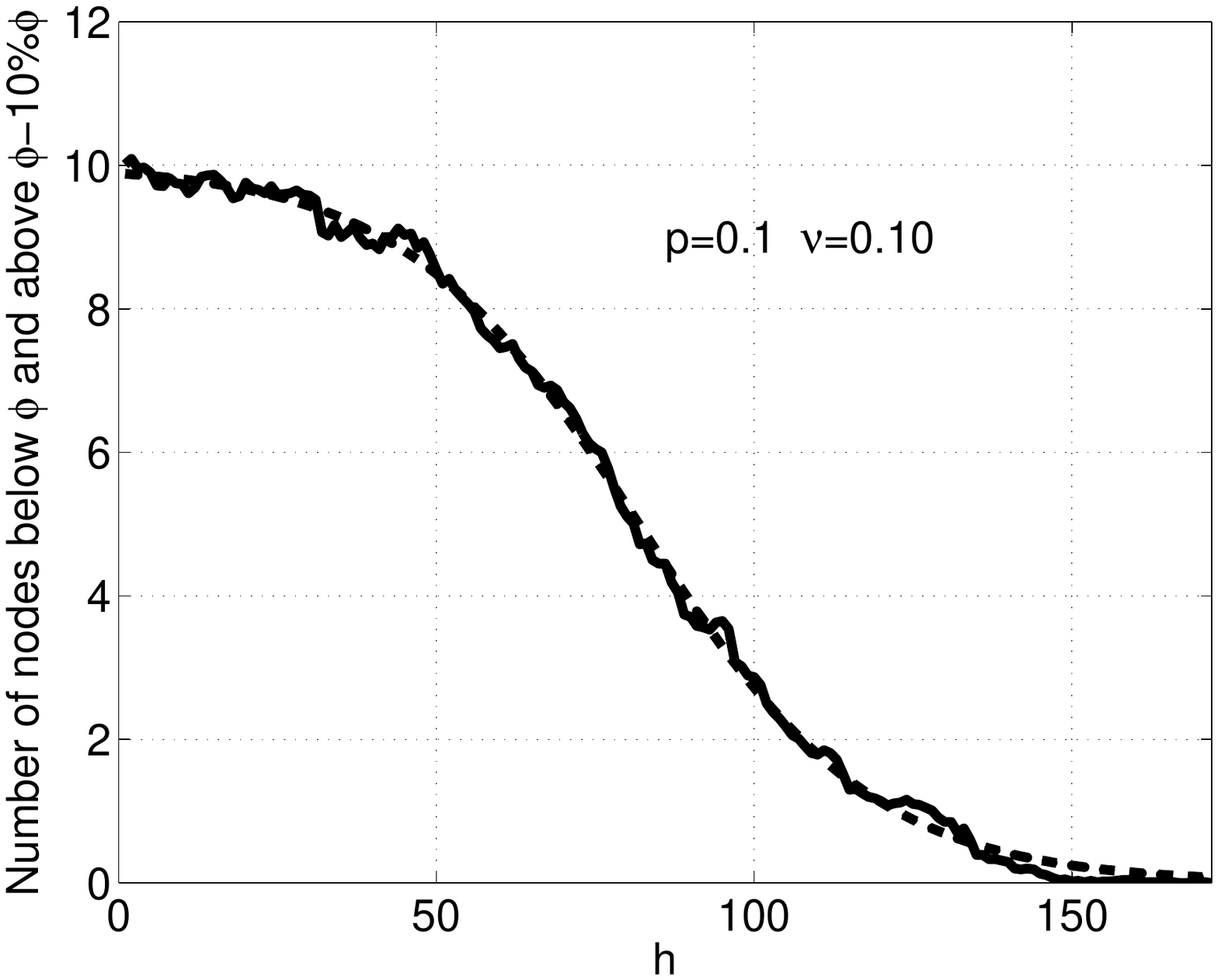}
\includegraphics[height=5cm,width=4cm]{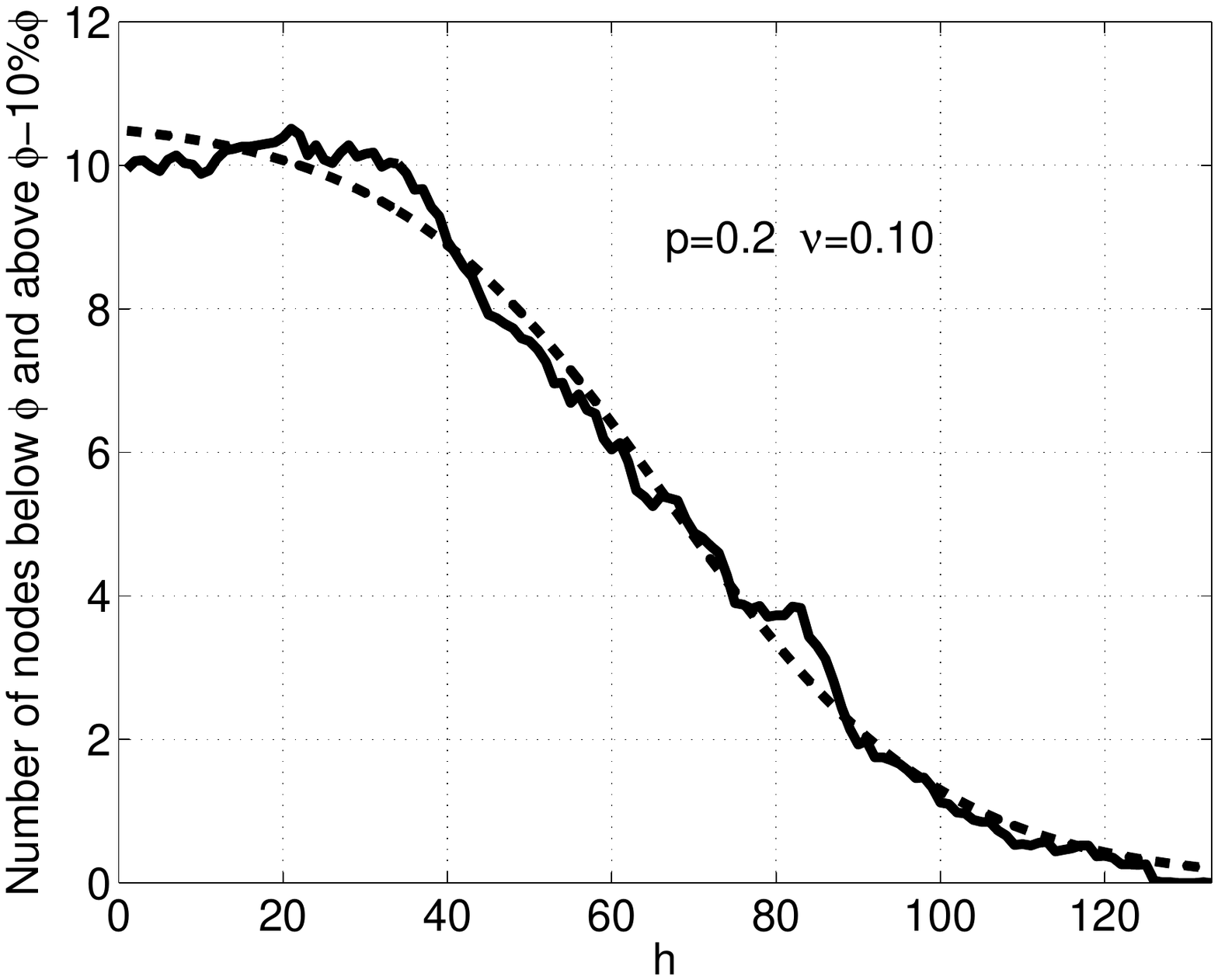}
\includegraphics[height=5cm,width=4cm]{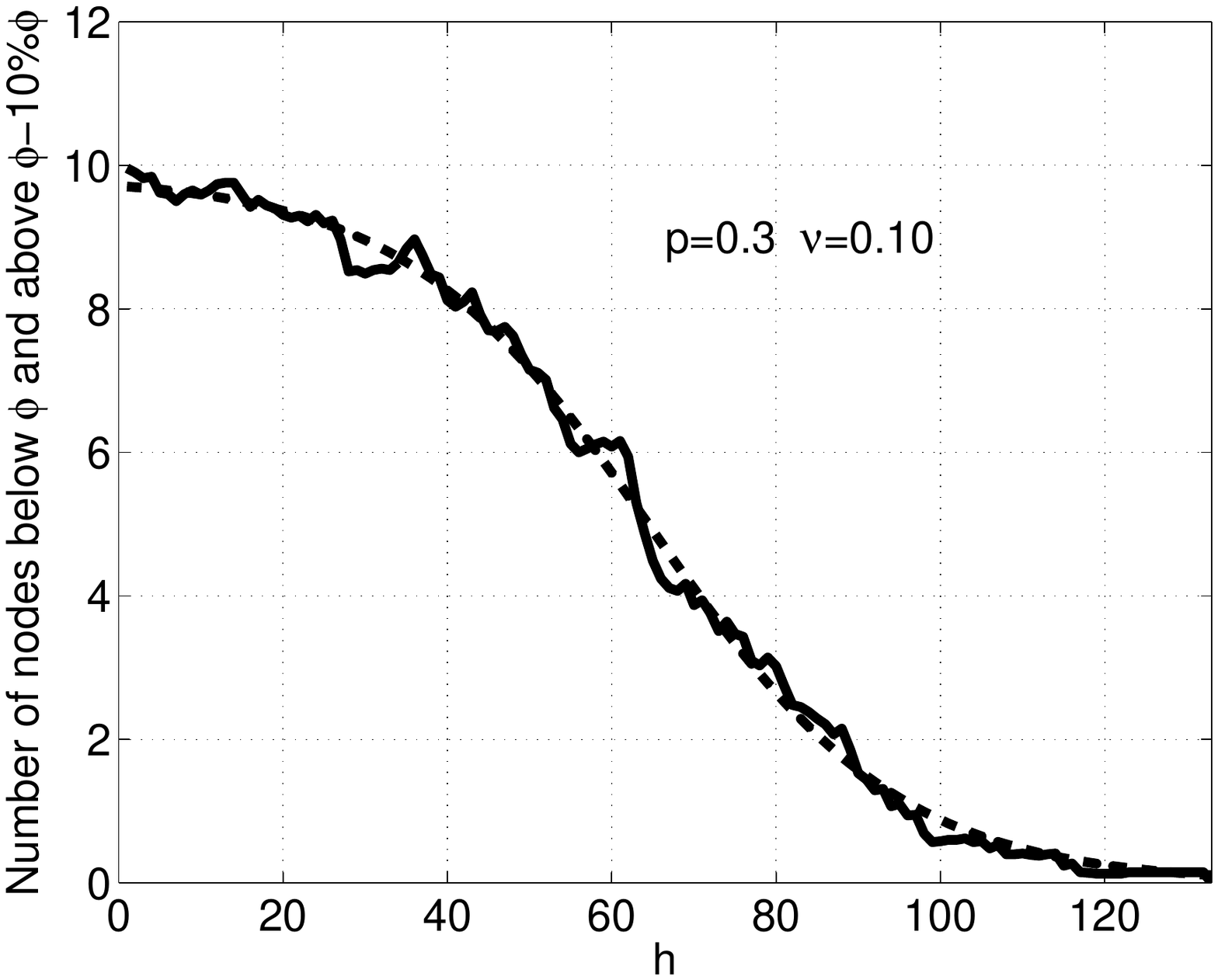}
\includegraphics[height=5cm,width=4cm]{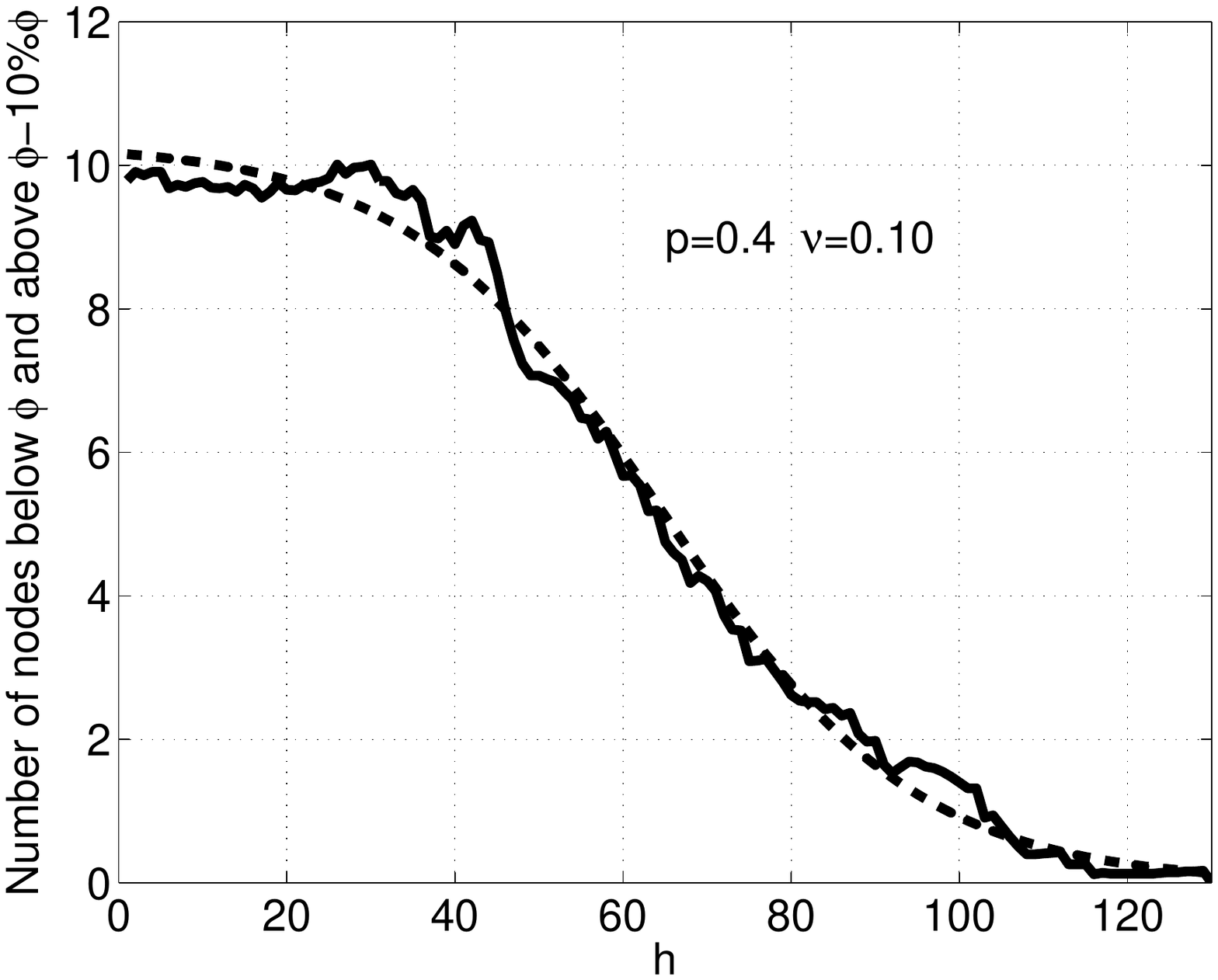}
\includegraphics[height=5cm,width=4cm]{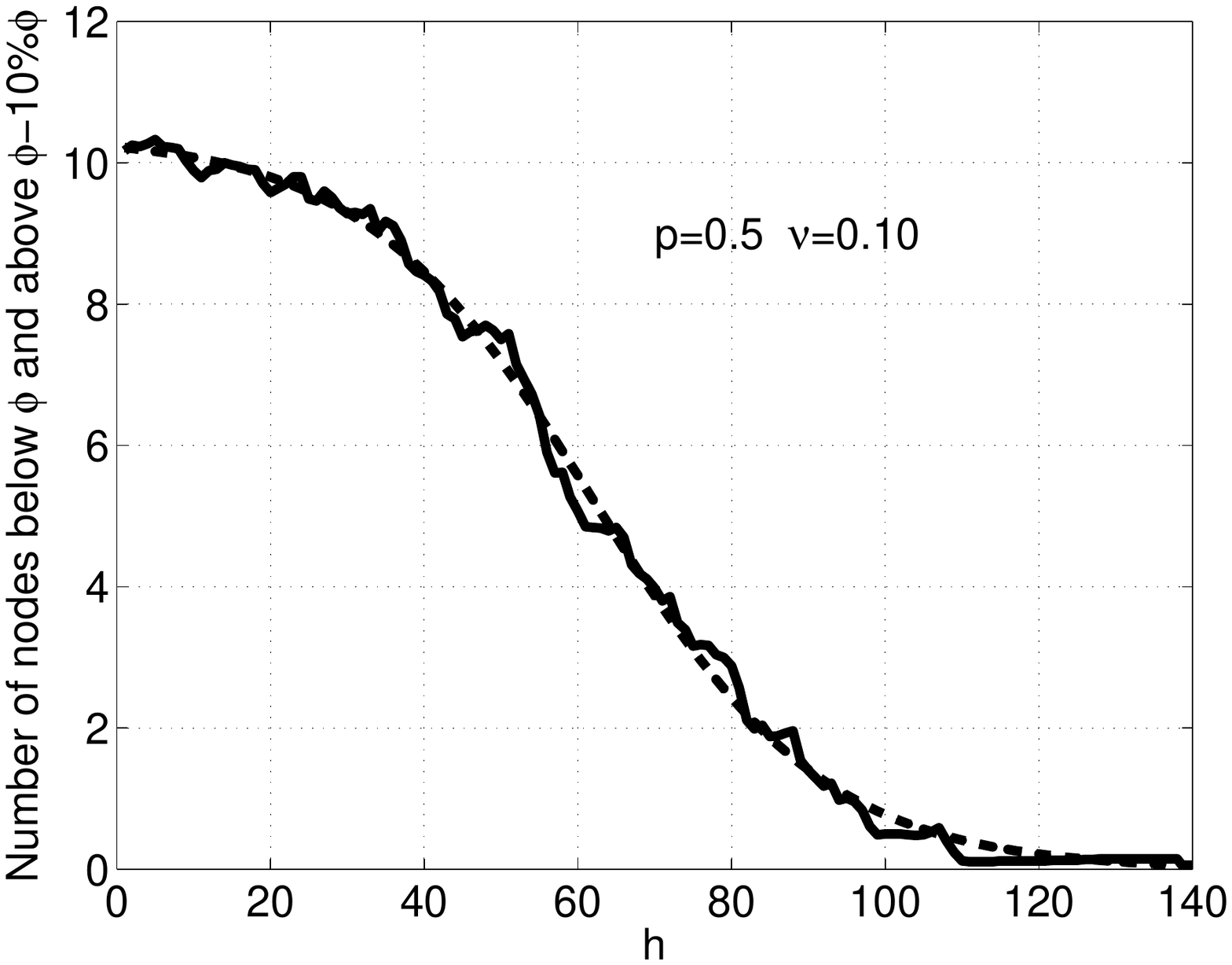}
\includegraphics[height=5cm,width=4cm]{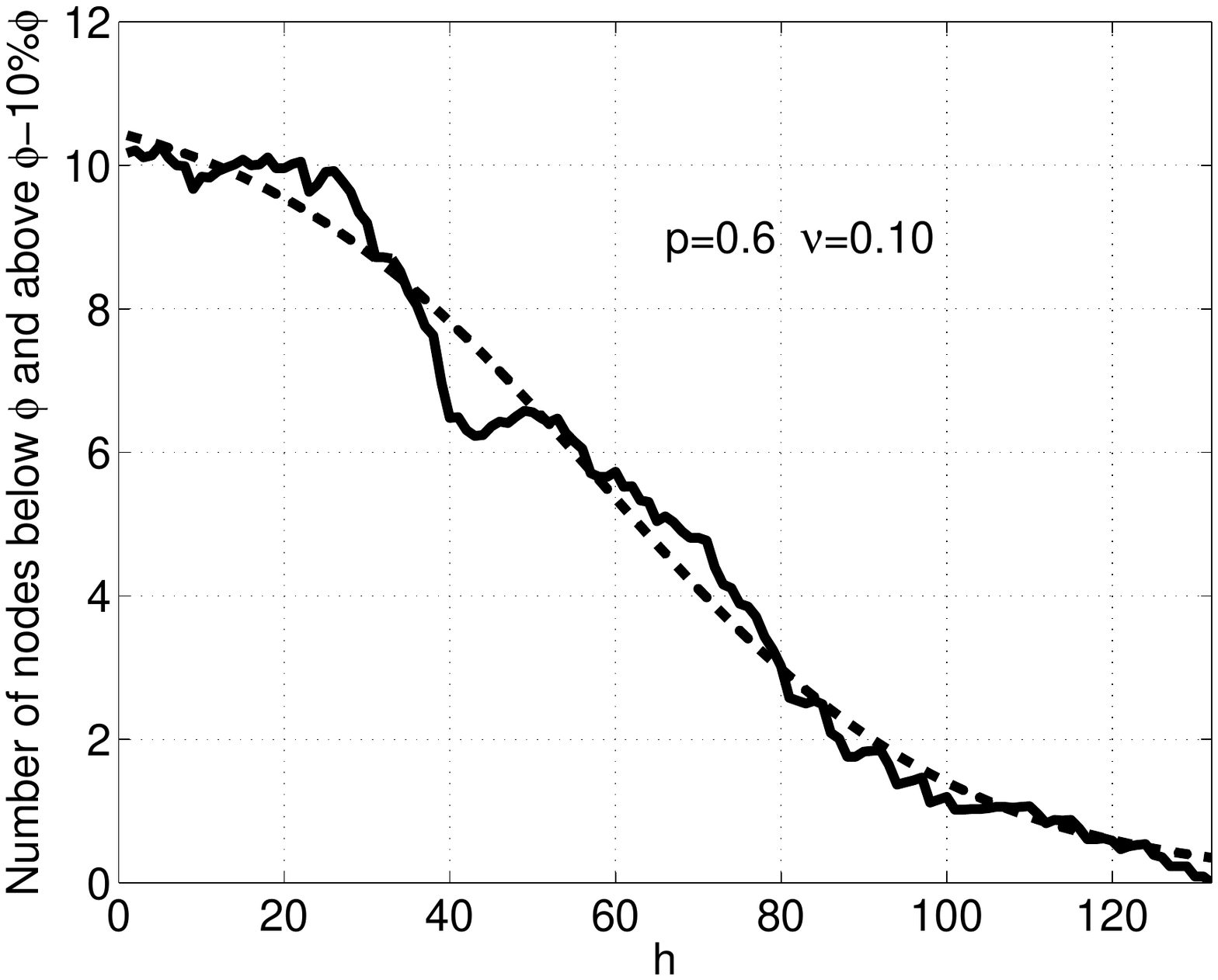}
\includegraphics[height=5cm,width=4cm]{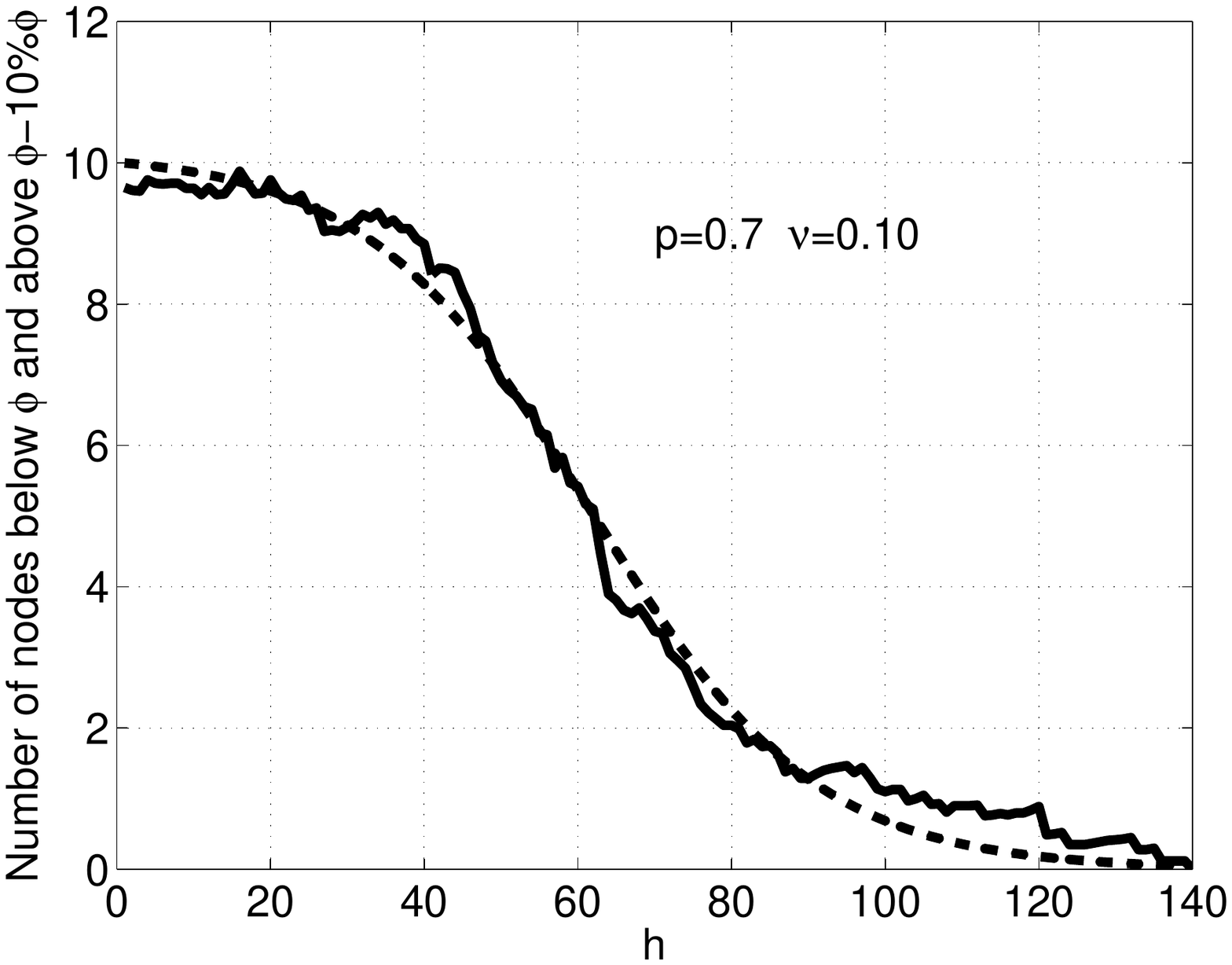}
\includegraphics[height=5cm,width=4cm]{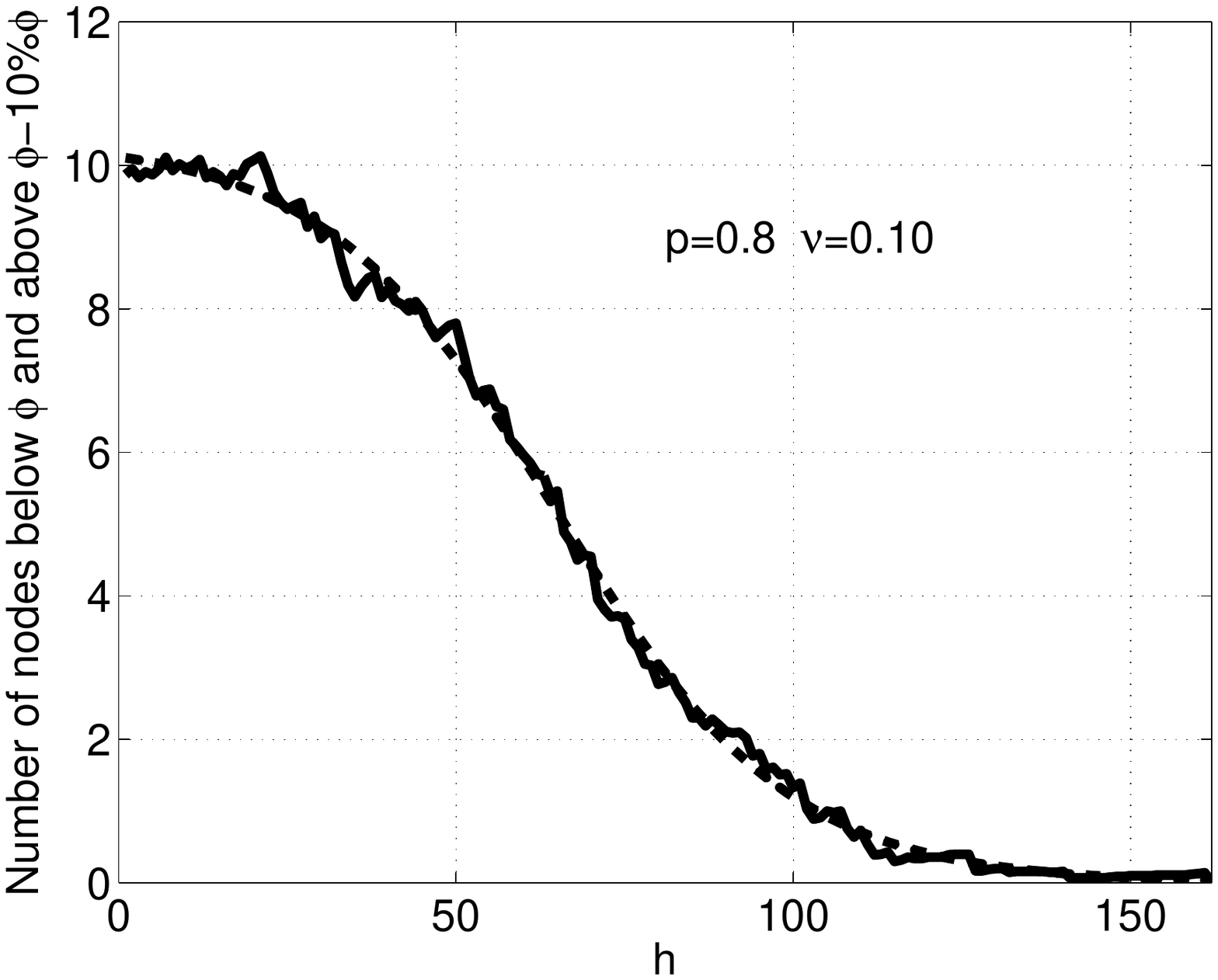}
\includegraphics[height=5cm,width=4cm]{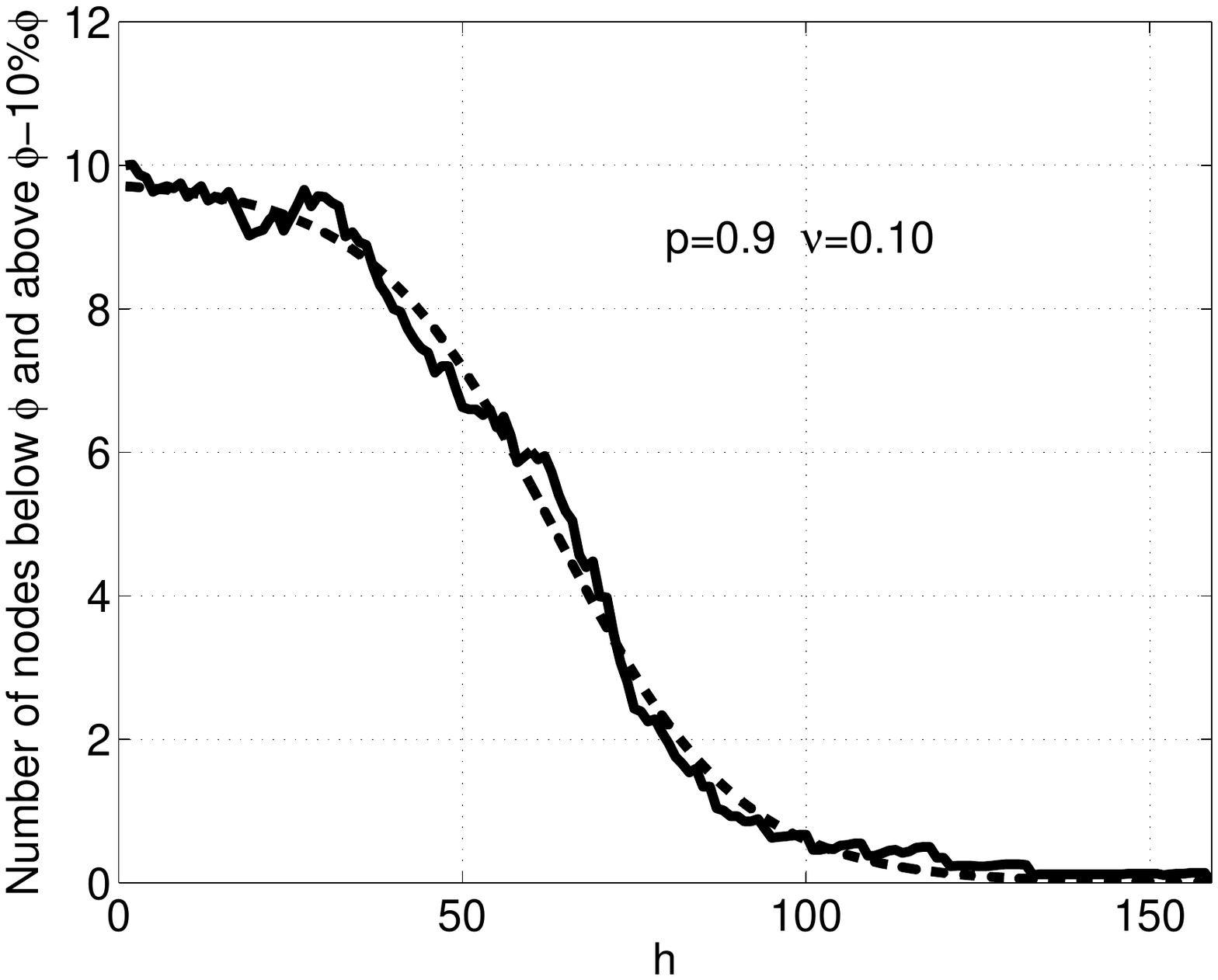}
\includegraphics[height=5cm,width=4cm]{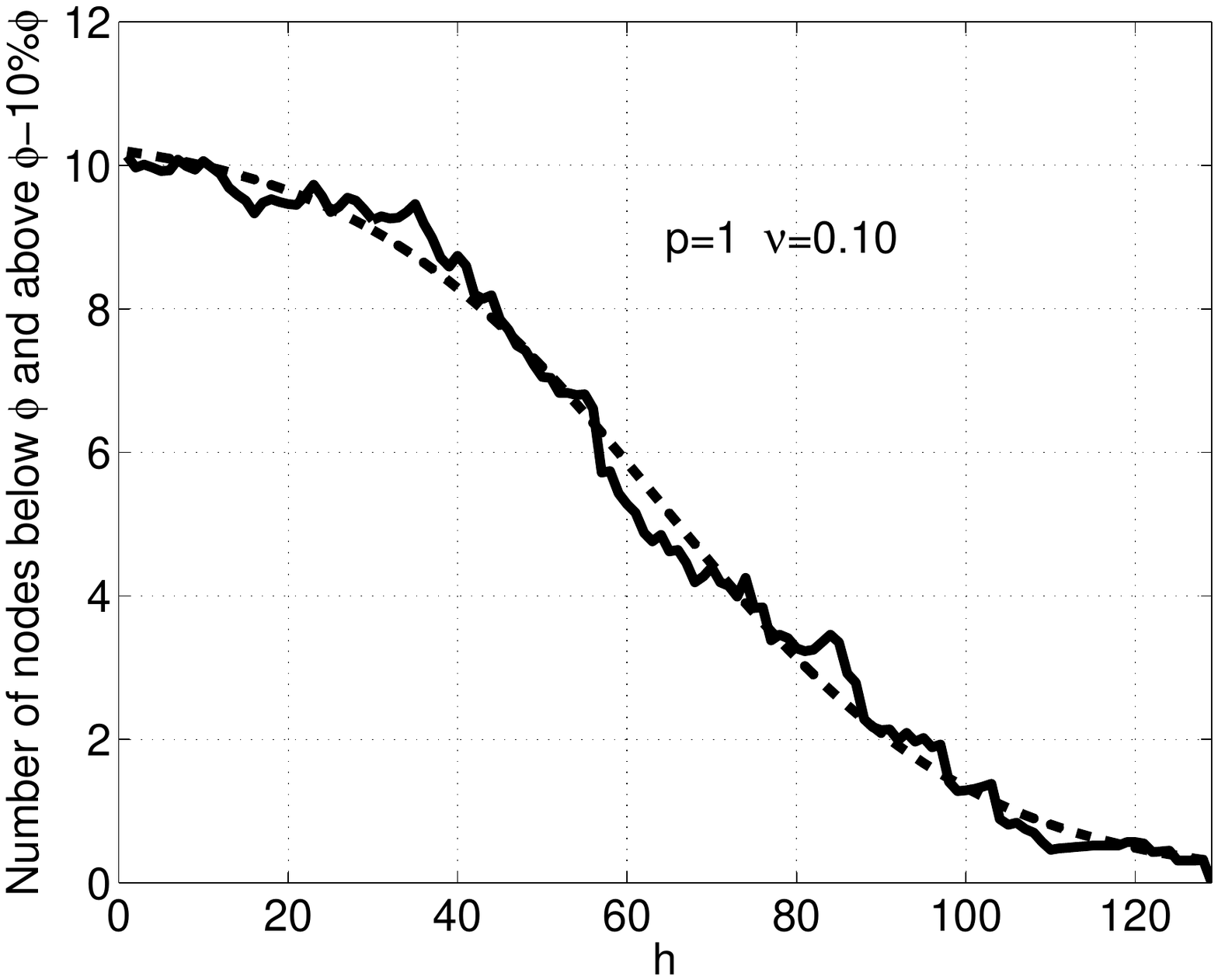}
\caption{Number of nodes with $a$ between $\phi-0.10\phi$ and $\phi$ as function of $h$. In all plots $\nu=0.10$ and for each plot the value of the connection probability $p$ is indicated. The curve is fitted with a logistic function as explained in the text.}\label{fighotnodenumber010}
\end{figure}

\begin{figure}
\centering
\includegraphics[height=5cm,width=4cm]{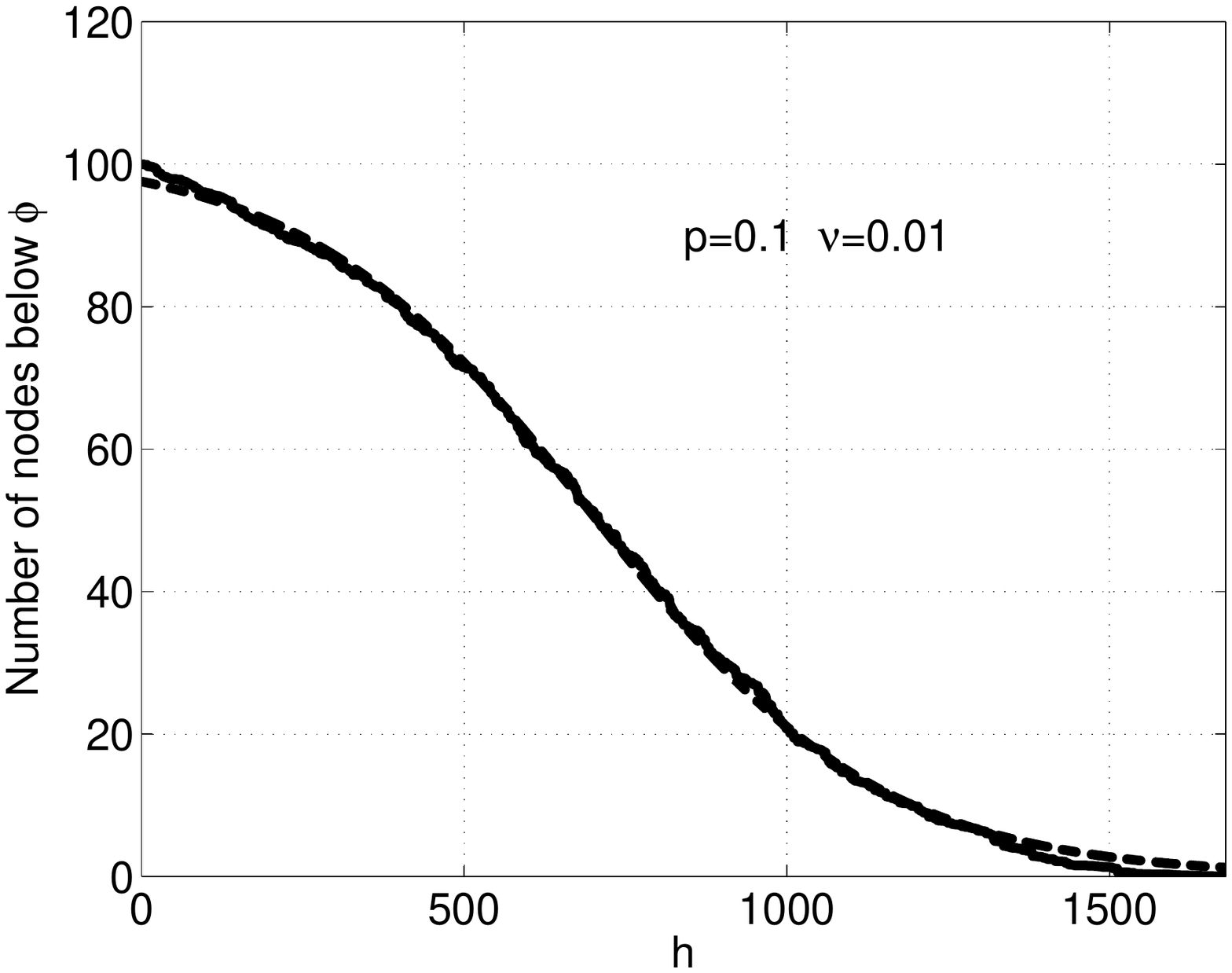}
\includegraphics[height=5cm,width=4cm]{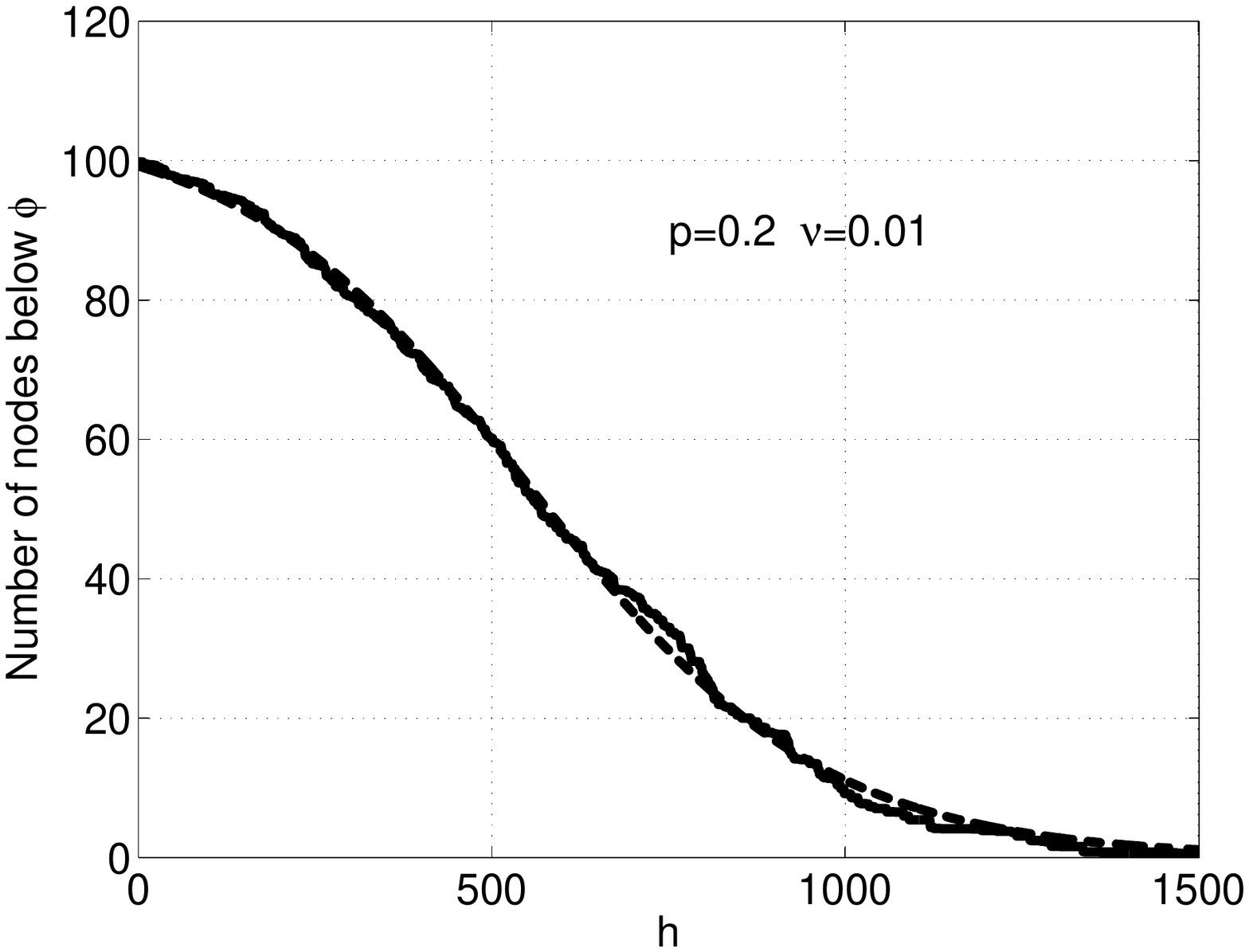}
\includegraphics[height=5cm,width=4cm]{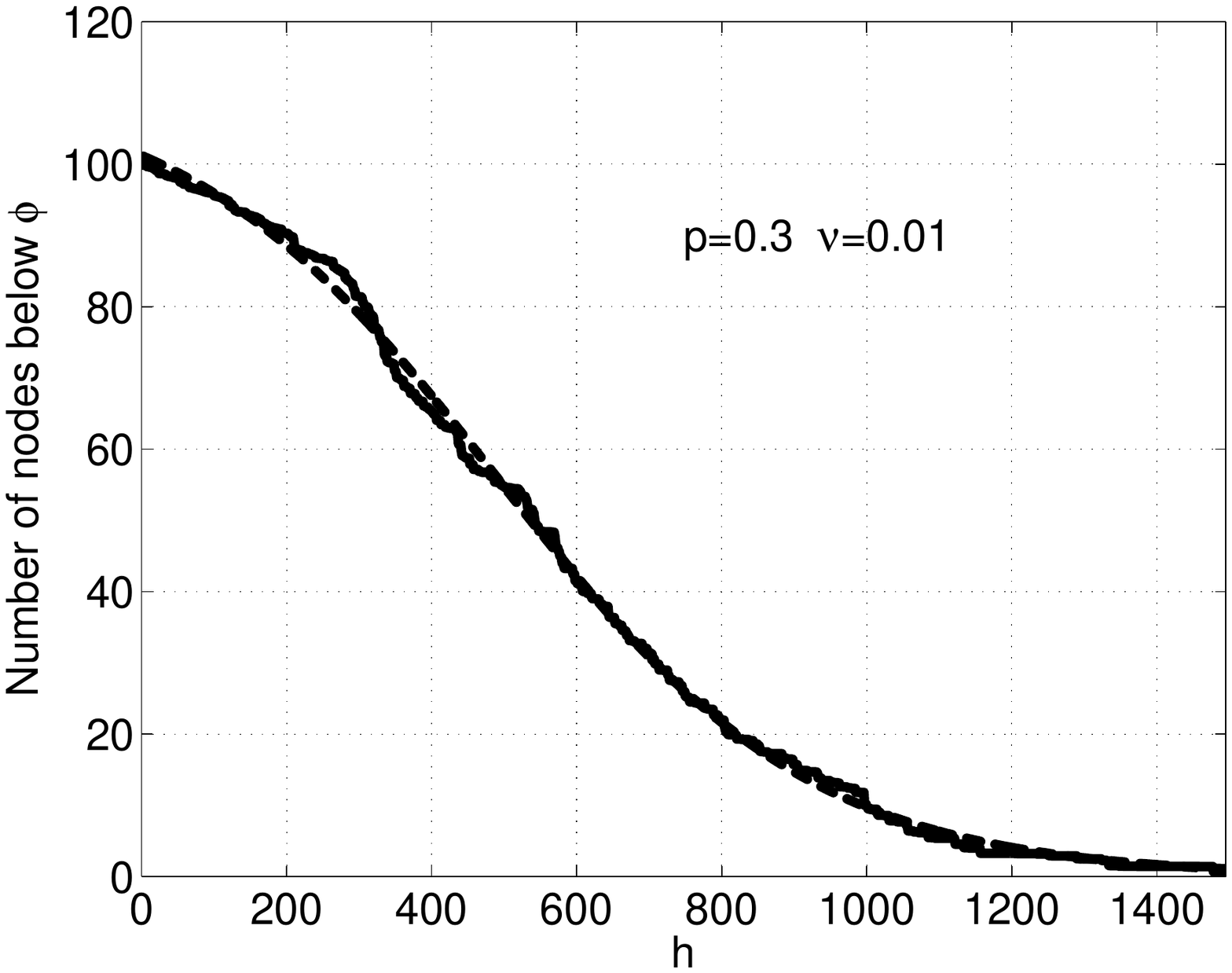}
\includegraphics[height=5cm,width=4cm]{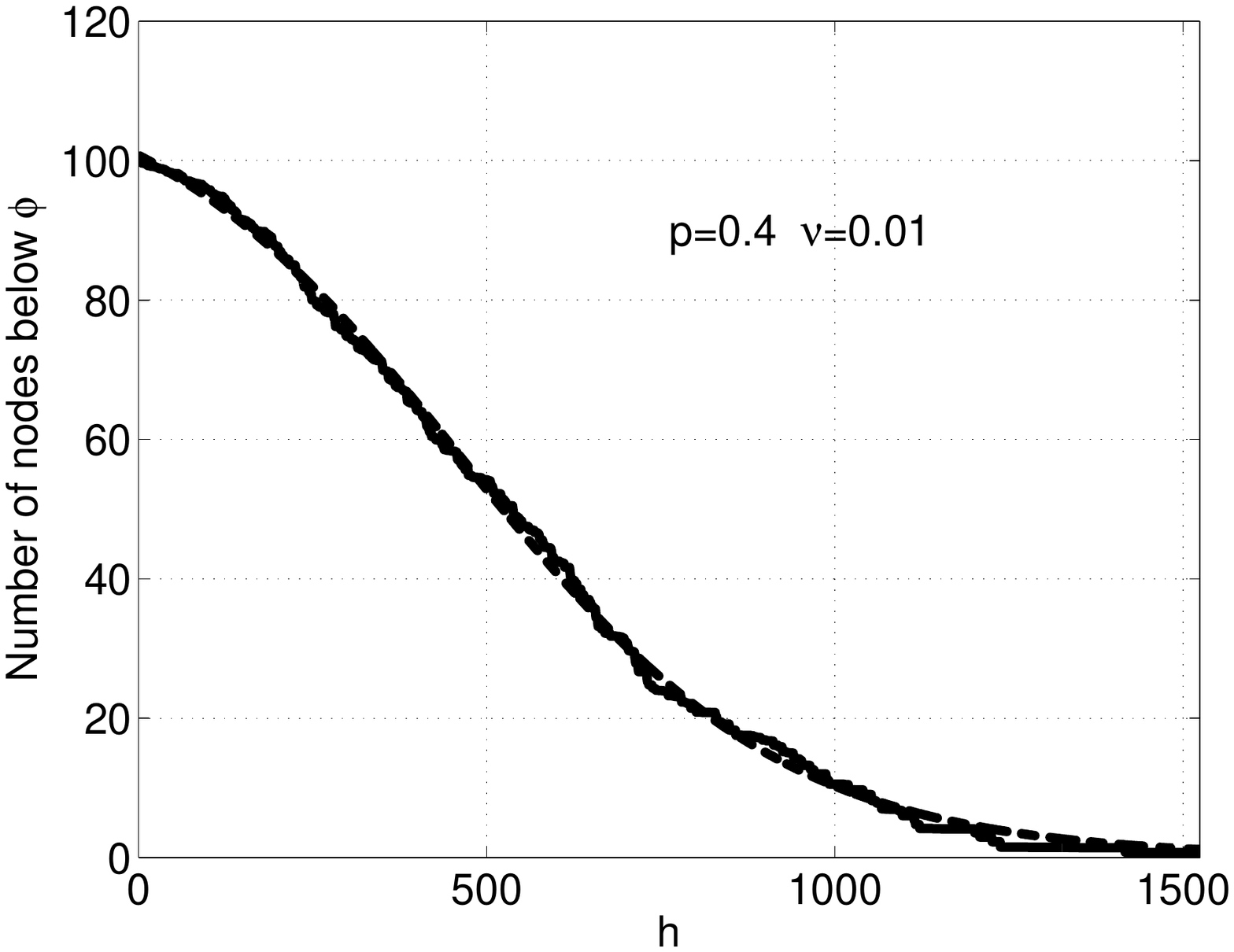}
\includegraphics[height=5cm,width=4cm]{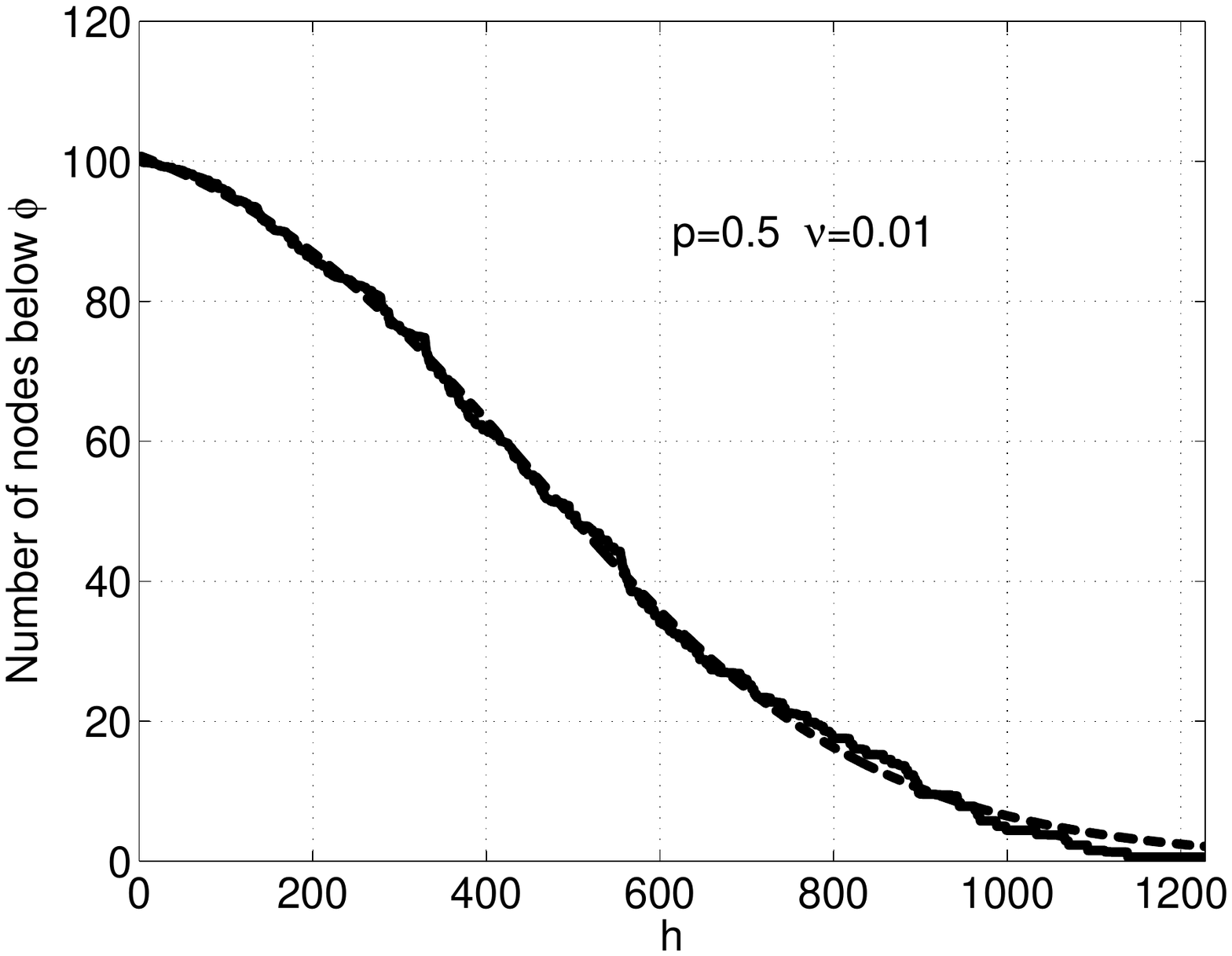}
\includegraphics[height=5cm,width=4cm]{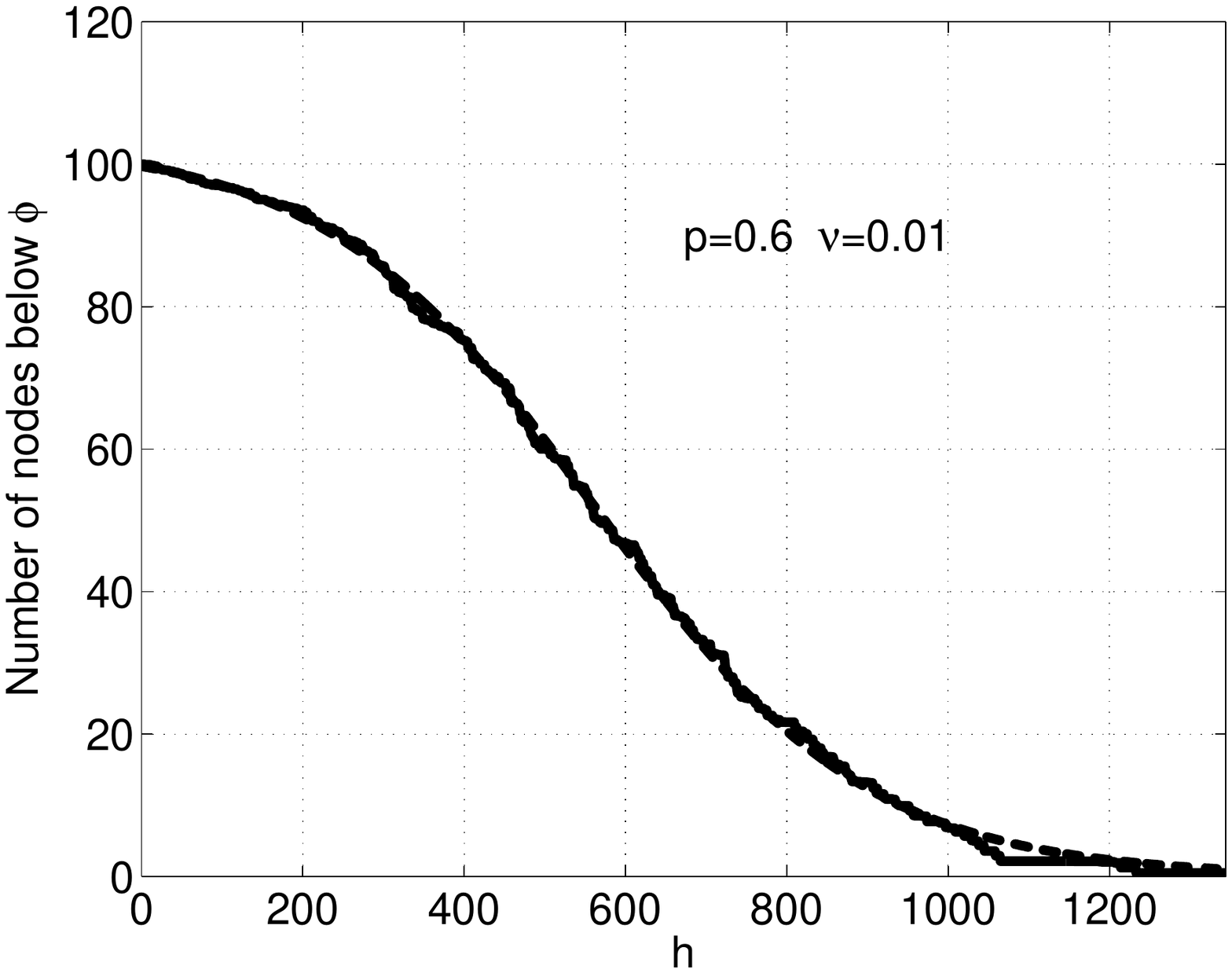}
\includegraphics[height=5cm,width=4cm]{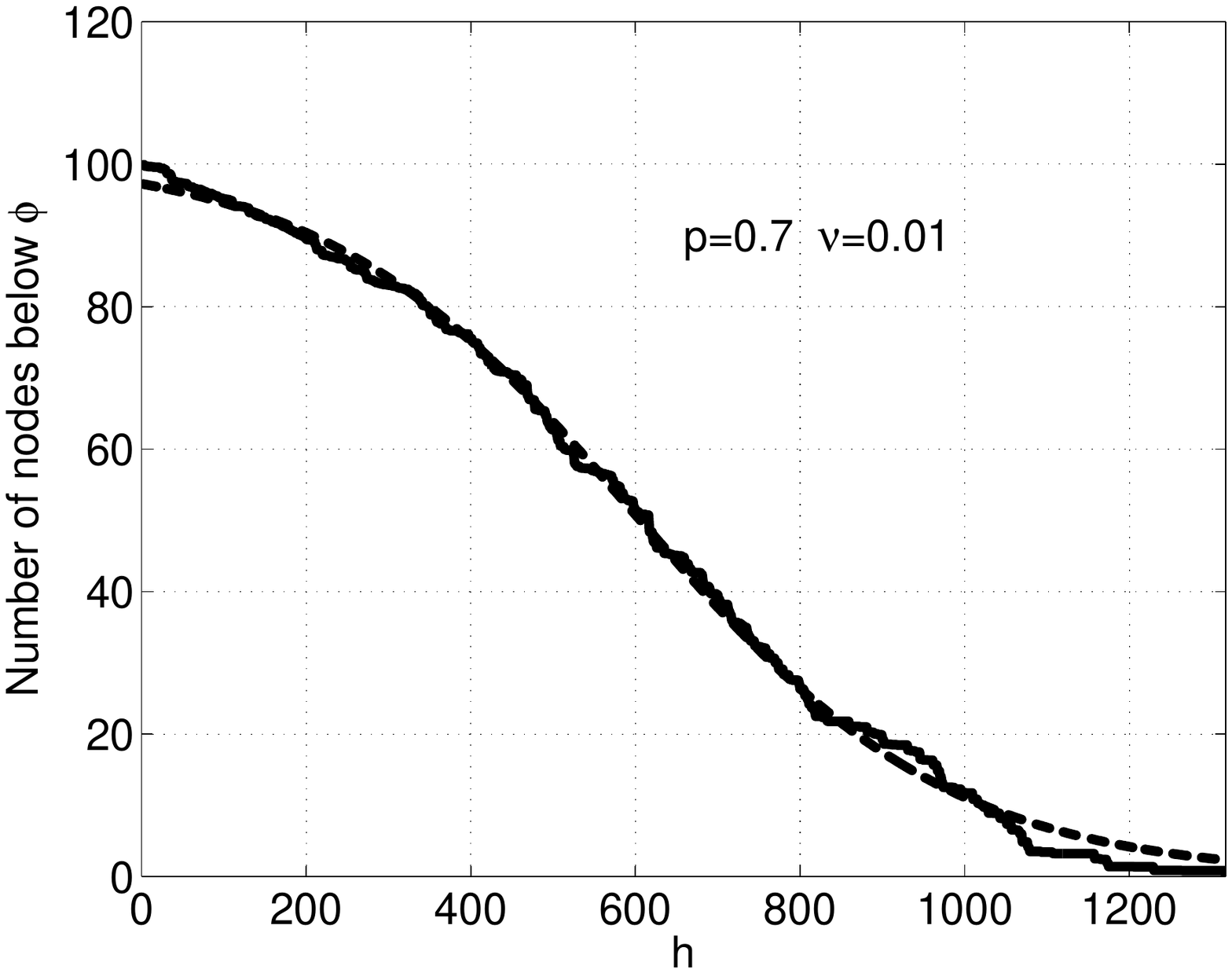}
\includegraphics[height=5cm,width=4cm]{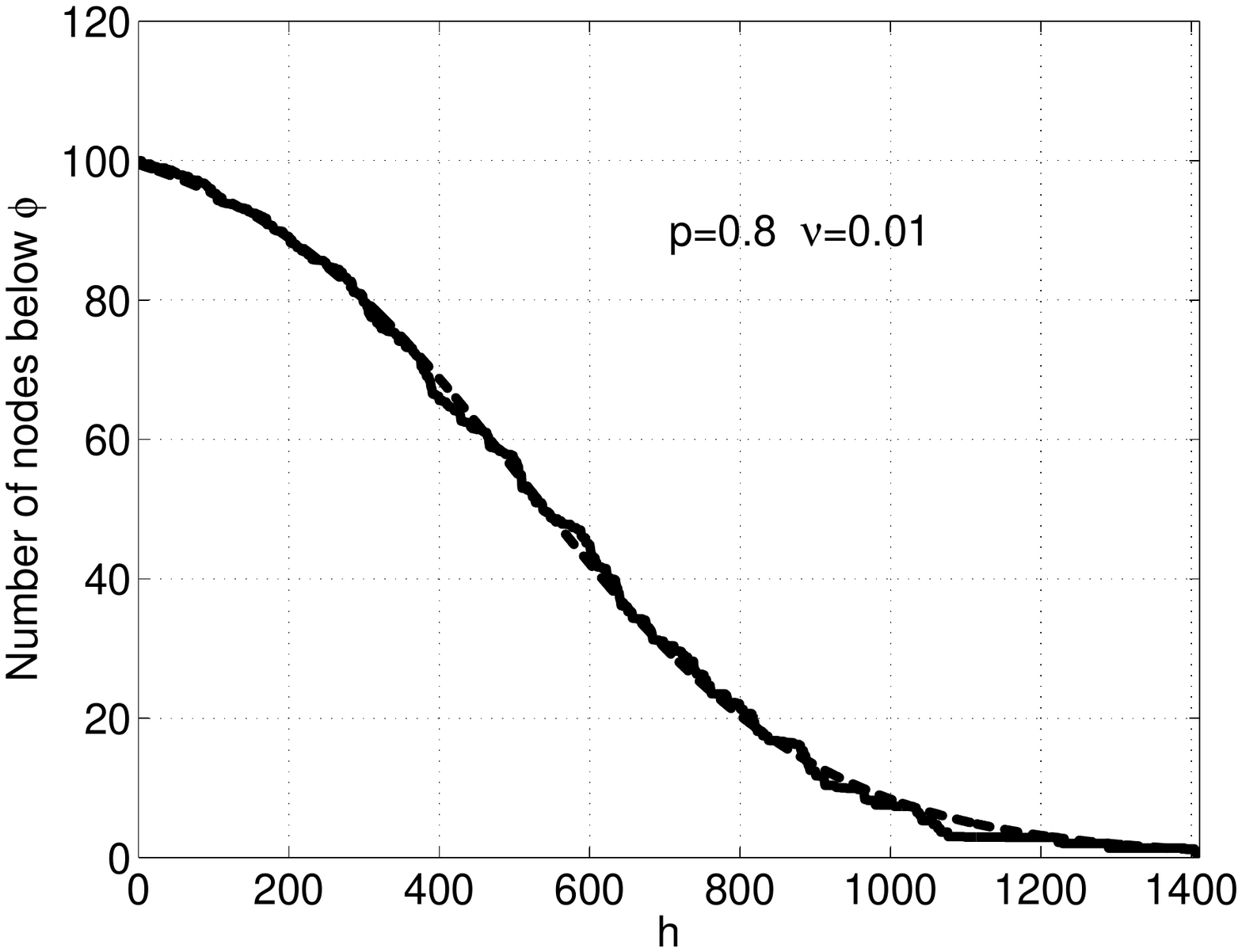}
\includegraphics[height=5cm,width=4cm]{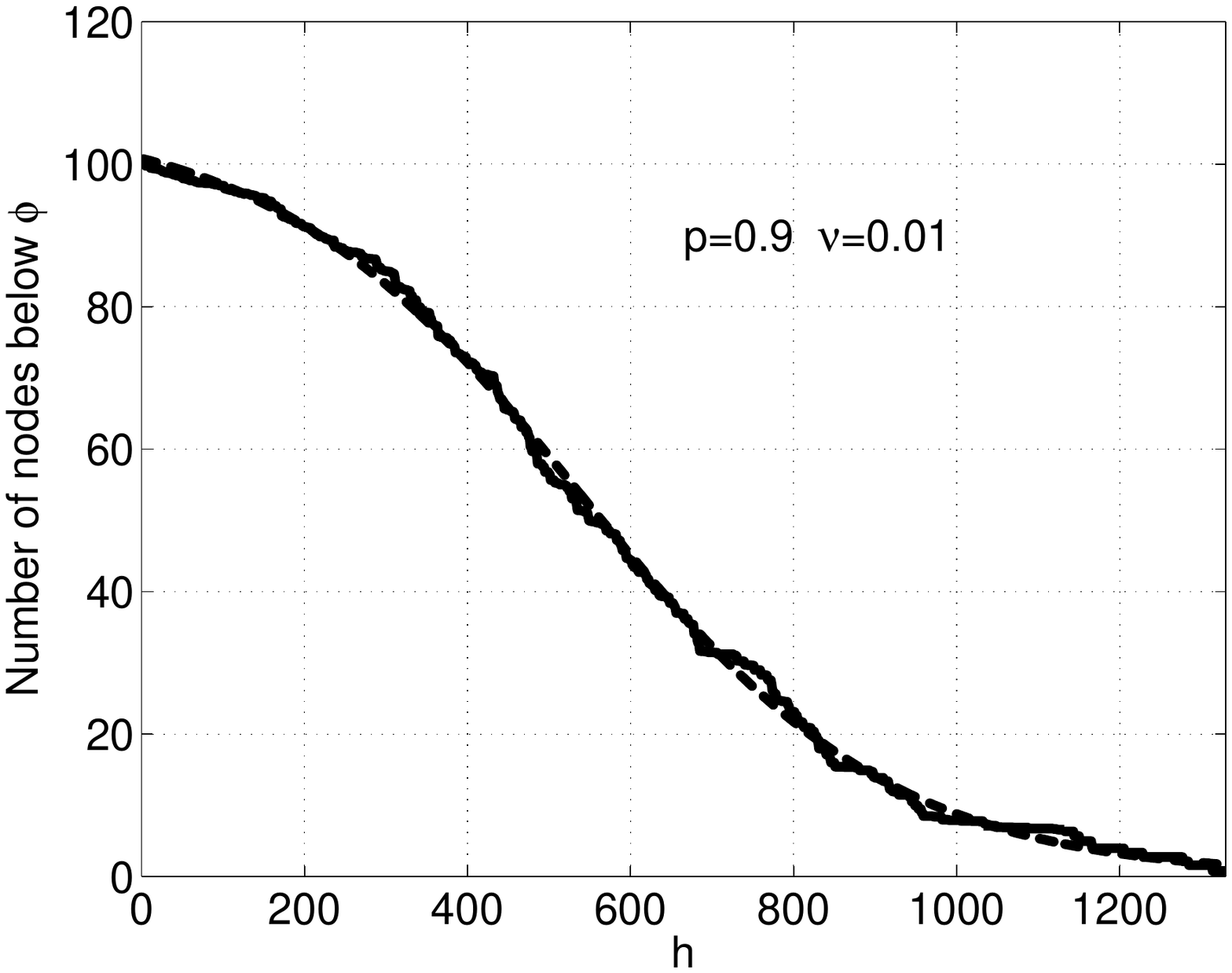}
\includegraphics[height=5cm,width=4cm]{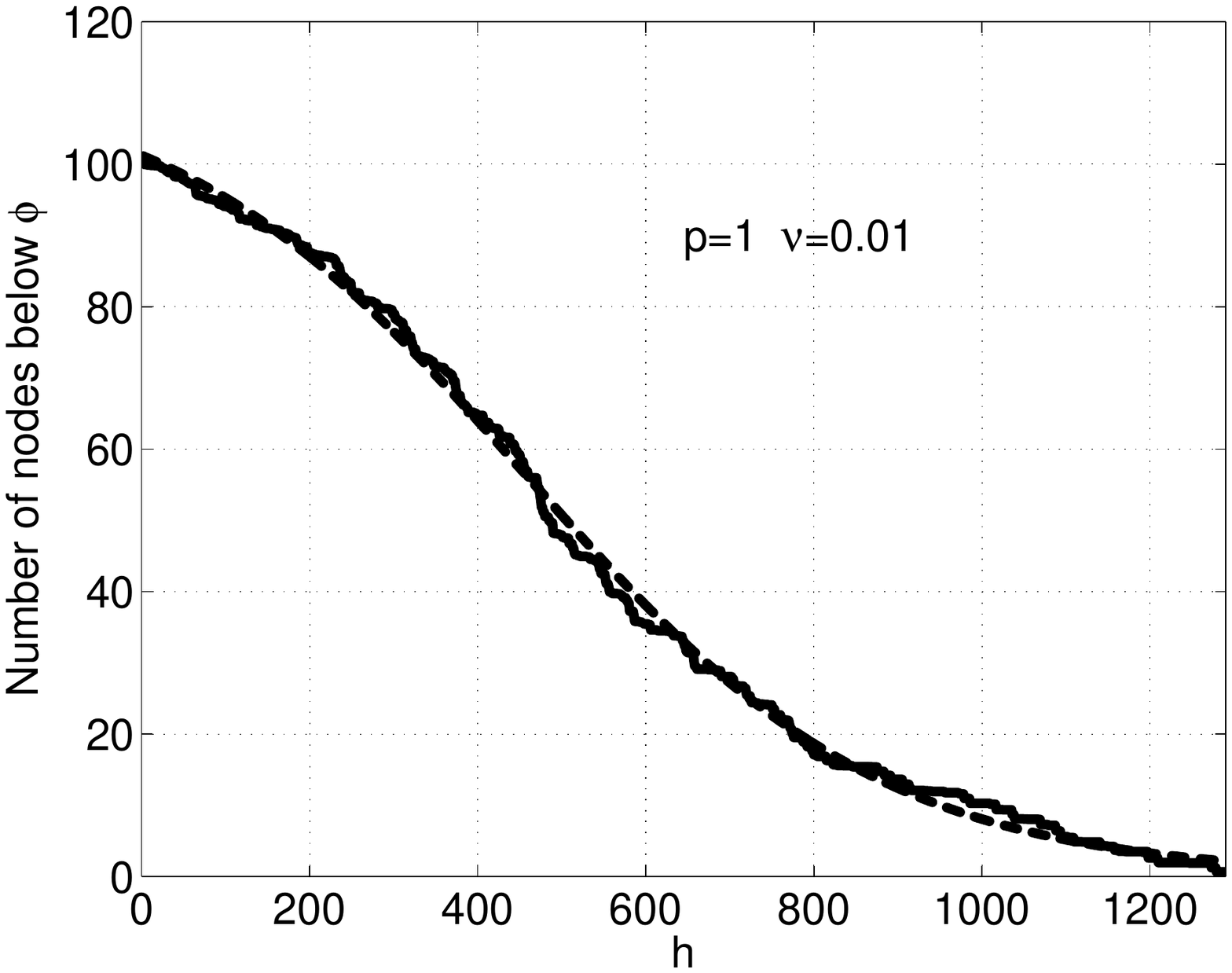}
\caption{Number of nodes with $a$ below $\phi$ as function of $h$. In all plots $\nu=0.01$ and for each plot the value of the connection probability $p$ is indicated. The curve is fitted with a logistic function as explained in the text.}\label{fignodenumber001}
\end{figure}

\begin{figure}
\centering
\includegraphics[height=5cm,width=4cm]{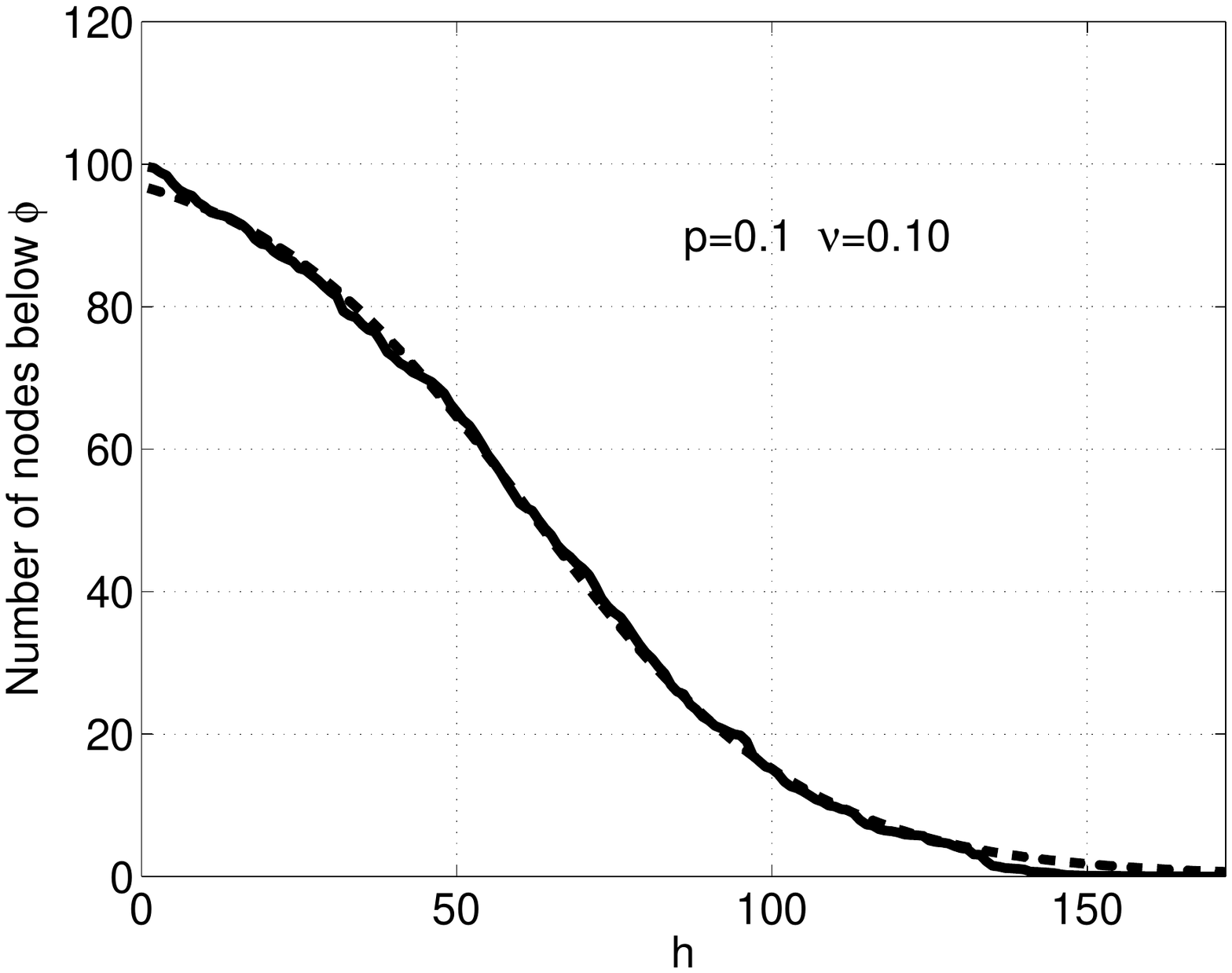}
\includegraphics[height=5cm,width=4cm]{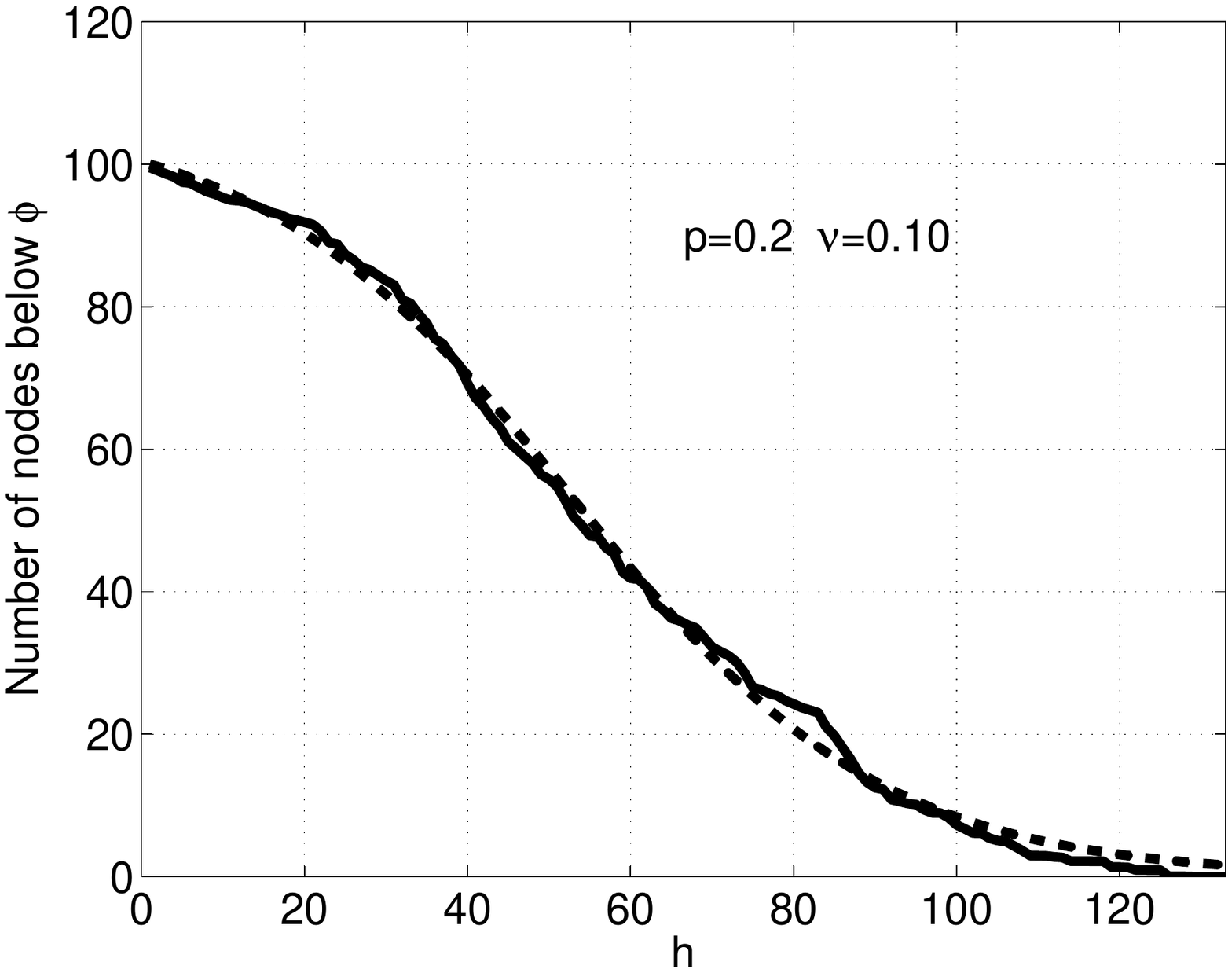}
\includegraphics[height=5cm,width=4cm]{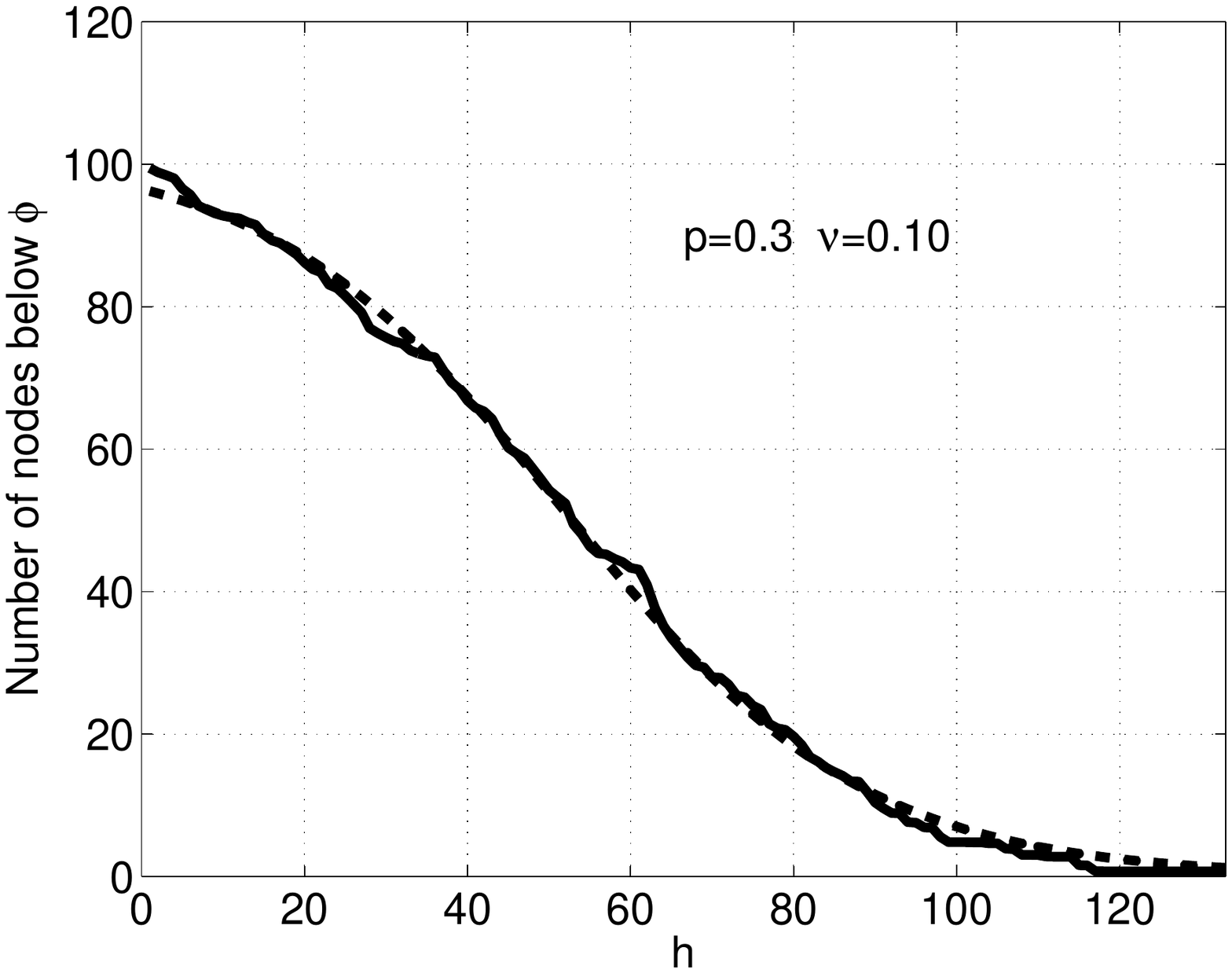}
\includegraphics[height=5cm,width=4cm]{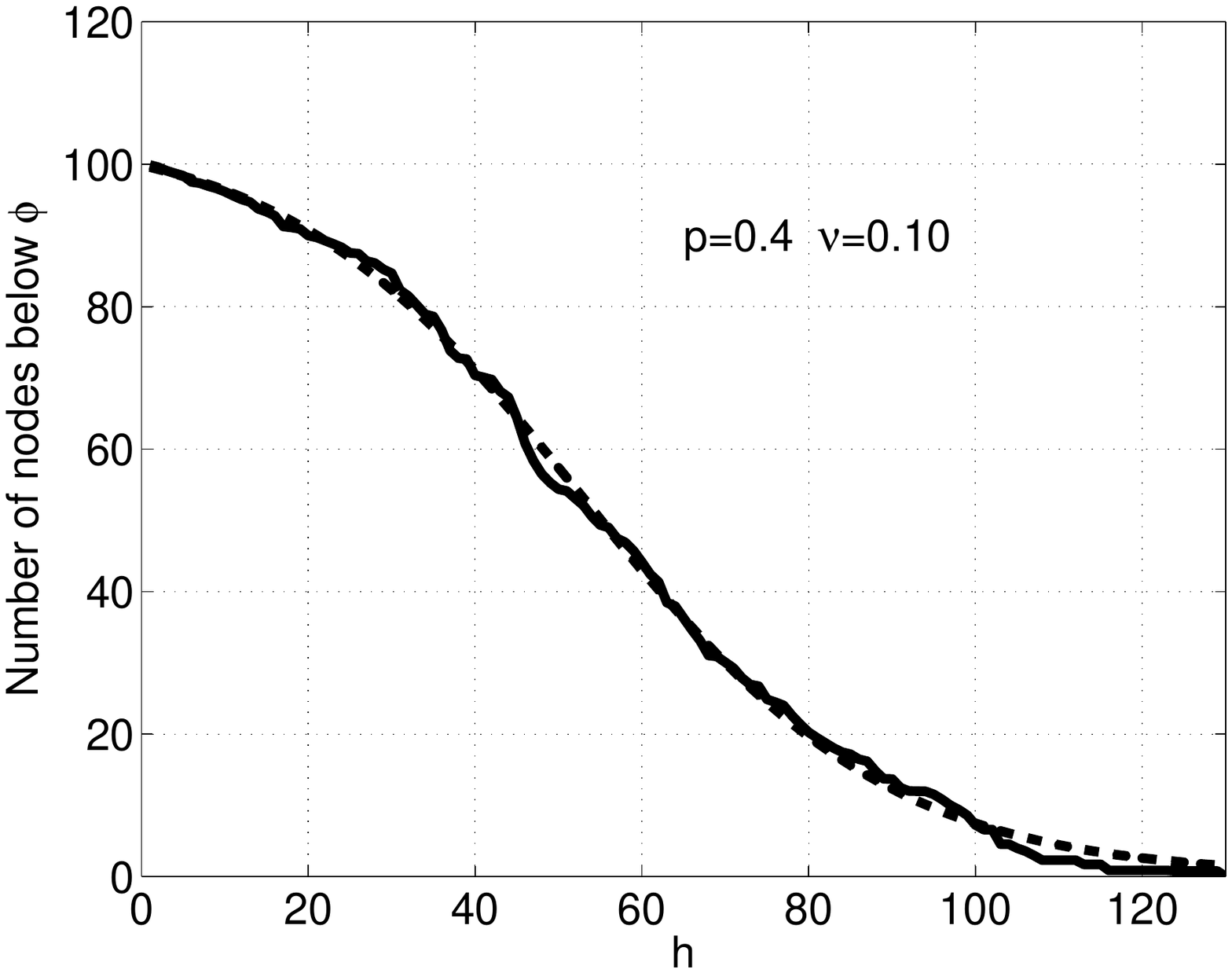}
\includegraphics[height=5cm,width=4cm]{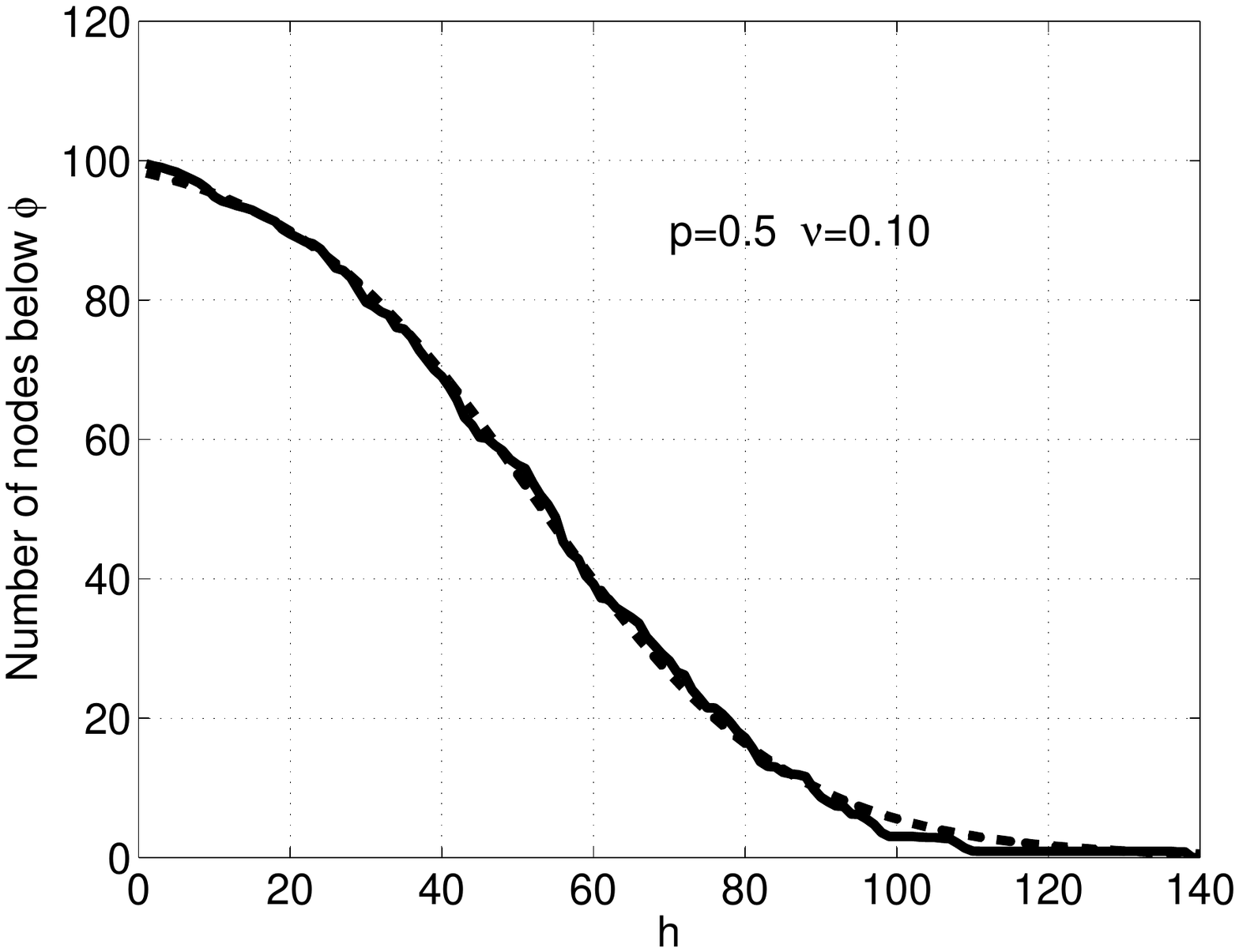}
\includegraphics[height=5cm,width=4cm]{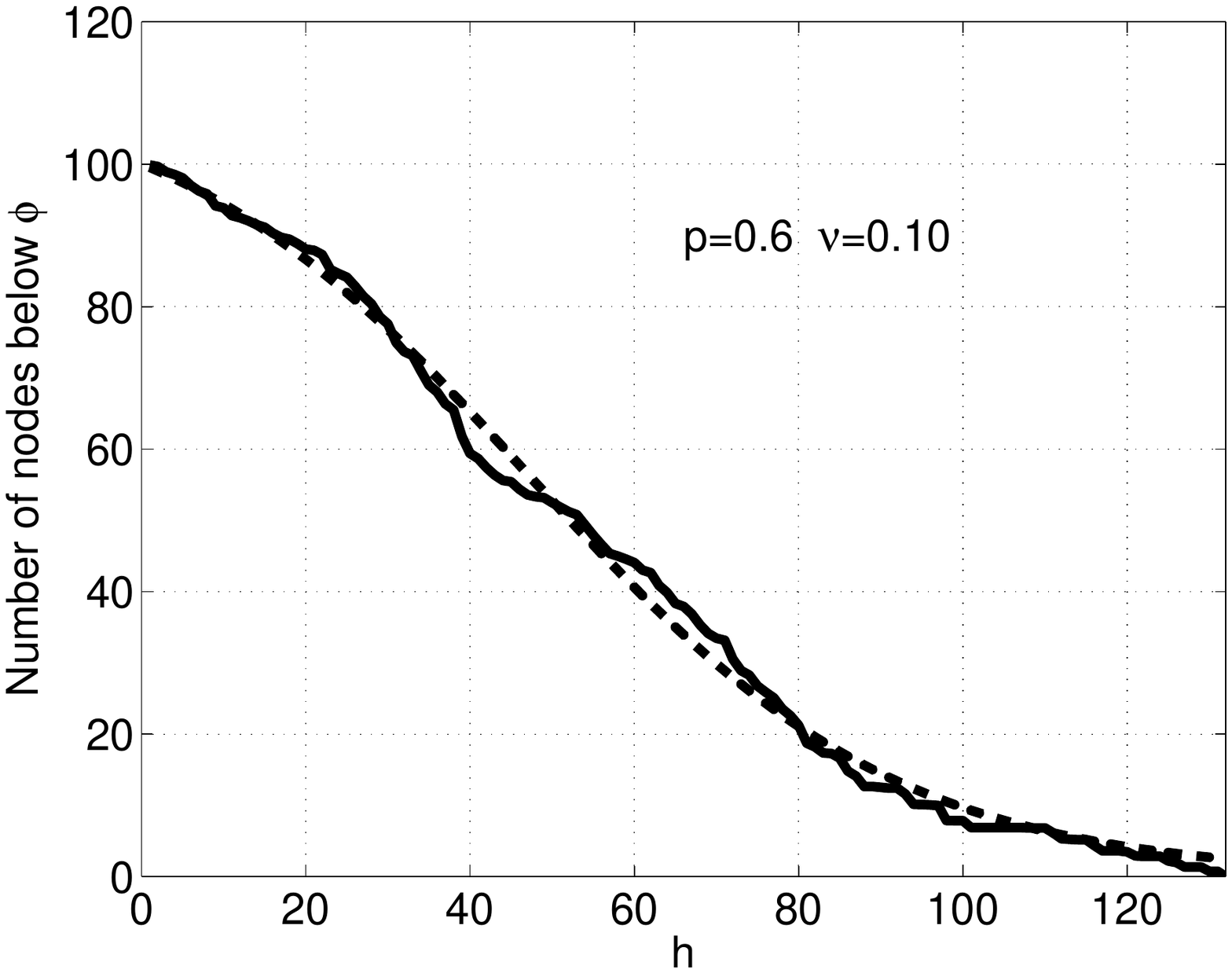}
\includegraphics[height=5cm,width=4cm]{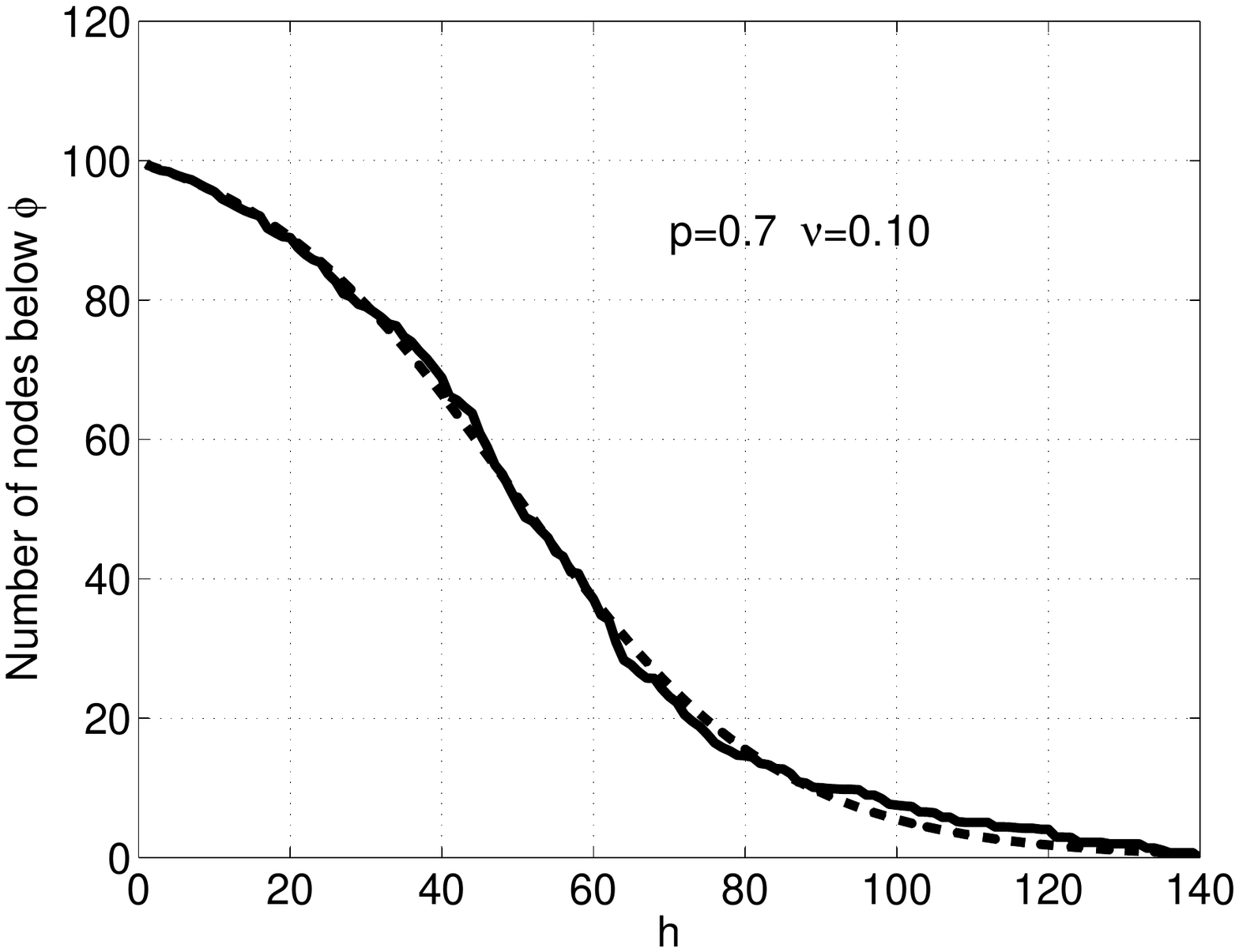}
\includegraphics[height=5cm,width=4cm]{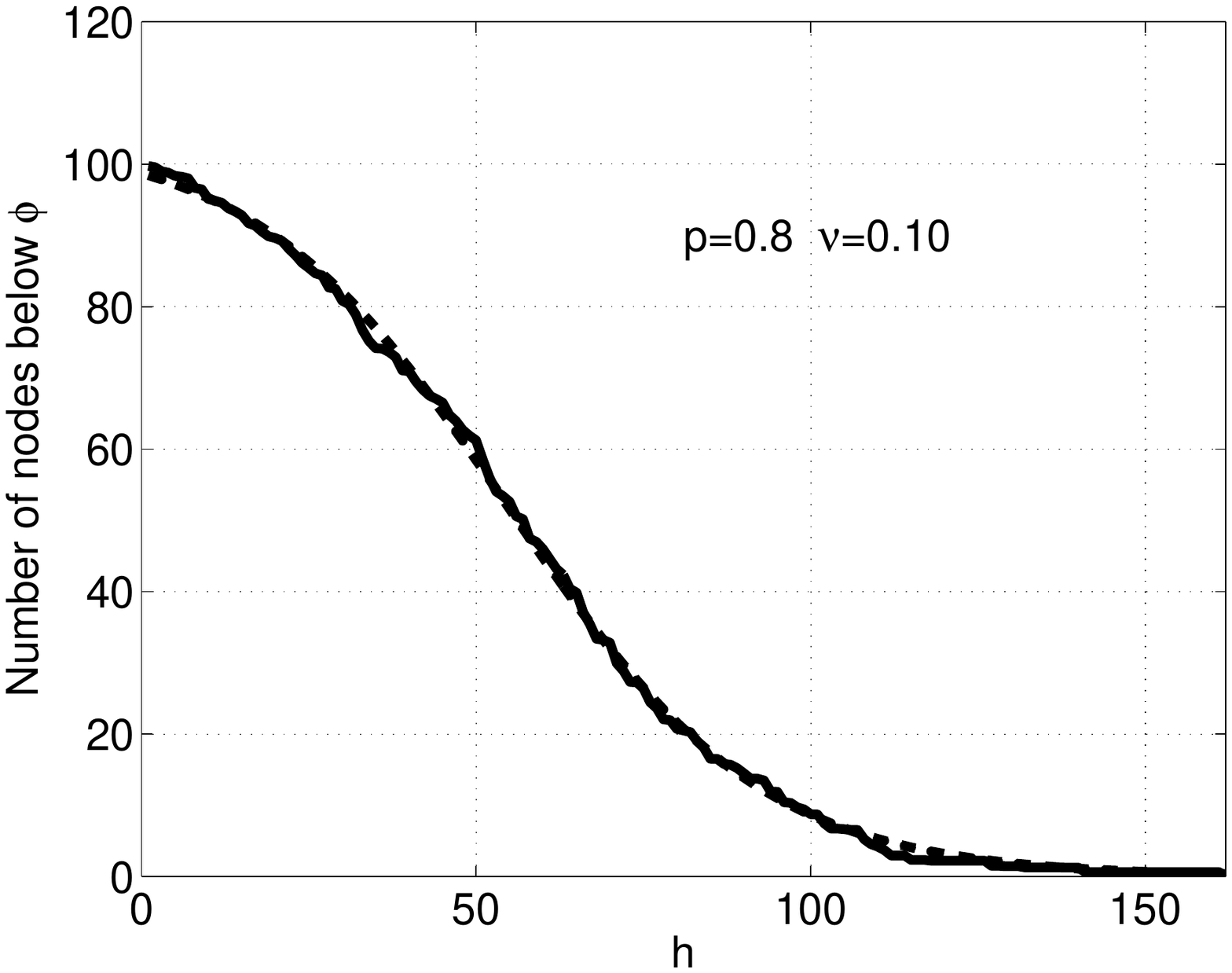}
\includegraphics[height=5cm,width=4cm]{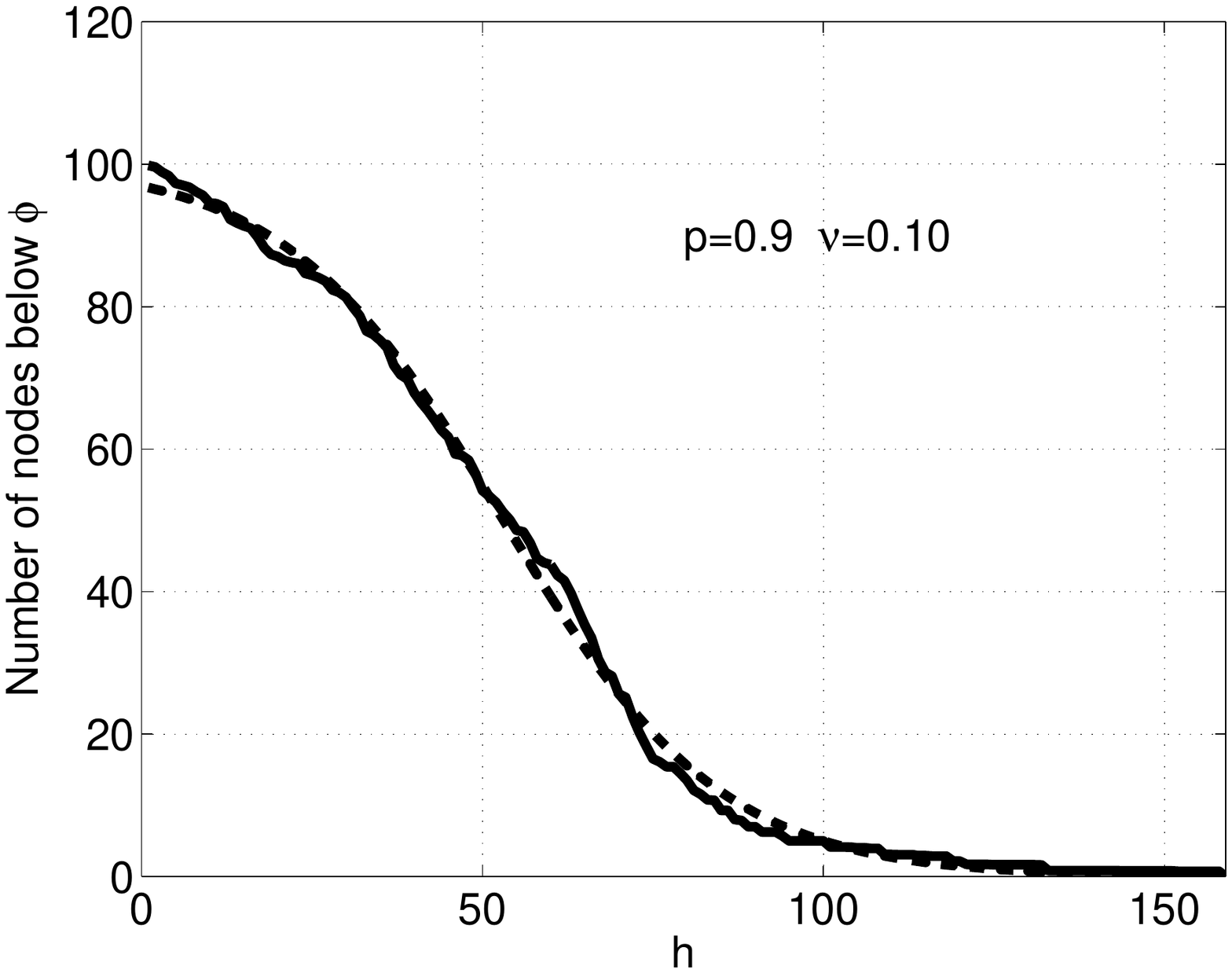}
\includegraphics[height=5cm,width=4cm]{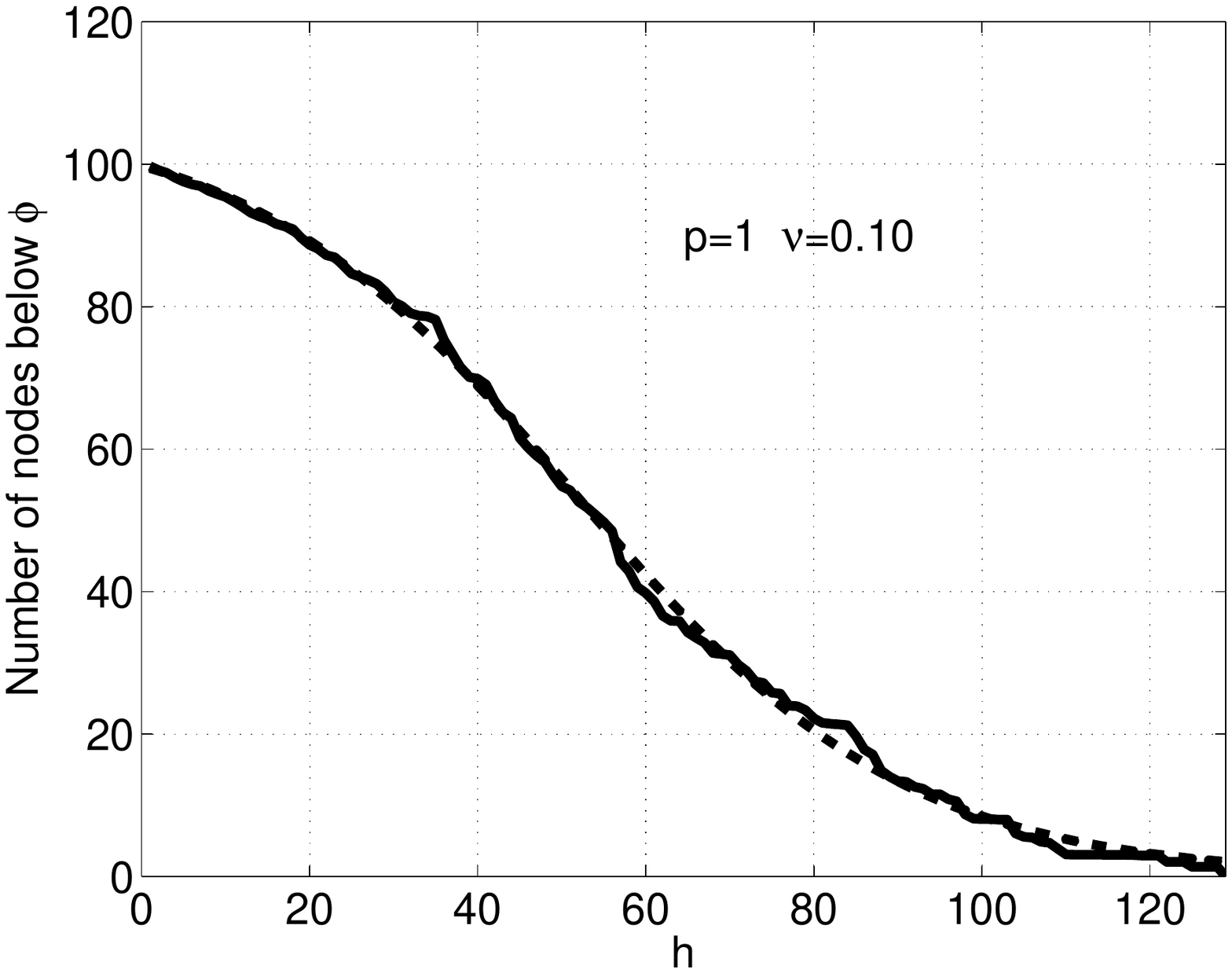}
\caption{Number of nodes with $a$ below $\phi$ as function of $h$. In all plots $\nu=0.10$ and for each plot the value of the connection probability $p$ is indicated. The curve is fitted with a logistic function as explained in the text.}\label{fignodenumber010}
\end{figure}

\section{Discussion}

 According to \cite{PetersPNAS}, catastrophic events share characteristic nonlinear behaviors that are often generated by cross-scale interactions and feedbacks among system elements. These events result in surprises that cannot easily be predicted based on information obtained at a single scale \cite{PetersPNAS}. Progress on catastrophic events has focused on one of the following two areas: nonlinear dynamics through time without an explicit consideration of spatial connectivity \cite{holling}   or spatial connectivity and the spread of contagious processes without a consideration of cross-scale interactions and feedbacks \cite{zeng}. The process dynamics should be correlated with the  structure of the network, i.e. to some law, sometimes through the exponents characterizing the latter. 

 In this paper we have concentrated on a network of (scientific or more generally opinion) communication and looked into the phenomena of  related avalanches.  It is intuitively known that (opinions like scientific) ones occur as a	 avalanches spreading out of  (fashion, paradigm) topics through (scientific) communities, in use of tools and instruments in scientific communities or in the sudden increase in cooperation structures (co-authorship, conferences) \cite{lambi2,lambi3}. Here we have presented an original model  implying  some spreading of information after triggering and excitation of some node over an awareness threshold.
  
This should be put in line with considerations on driven threshold systems which are now used to model sand piles, earthquakes, magnetic depinning transitions and driven foams \cite{rundle}, etc. usually on lattices. Indeed the model is reminiscent of the original BTW sand pile model \cite{BTW1,BTW2} as applied in recent papers  on a lattice \cite{BTWkoz} or on a network \cite{froncszak2} though the present pile toppling conditions differs from the usual ones, - which are related to the number of neighbors,  usually a fixed integer. Here the node ''number of grains'' is a continuous function,  there is no toppling but  some effect occurs when the height of the pile reaches a threshold $independent$ of the number of neighbors. 

On the other hand,  the connectivity of a landscape can influence the dynamics of disturbances, e.g. in internet (or neural or biological) epidemics and/or fire extension problems \cite{miller}.  Thus networks should be investigated. A question raised is often that whether the  spatial pattern rather than the disturbance dynamics (ordinarily) determines the total extent of a single disturbance event. We have shown that, for a Erd\"os-R\'enyi  network, a fixed {\it a priori} disturbance  perfectly scales the evolution process, in time $and$ in space.   We have found evolution laws similar, though with different exponents, to those found in other problems, e.g. even in earthquake dynamics \cite{rundle}. As an example recall, the frequency of spinodal fluctuations $n(A)$  of area $A$, which are realized here as clusters of  excited nodes above a threshold,   given by the Fisher-Stauffer relation \cite{49},
\begin{equation}
n(A,\delta t)= \frac{n_0}{A^\zeta} \; exp\left[\displaystyle -k[K_LV\delta t]^{1/\sigma_s}A\right].
\end{equation}
 where the interesting exponent $\sigma_s$ is the surface exponent. The  exponent $\zeta$ is either $\tau$ or $\tau-1$, in terms of the Fisher-Stauffer exponent  $\tau$   which  characterizes  the fluctuations about the spinodal or  describes the frequency of ``arrested'' nucleation events, respectively.
Here above we find  $\sigma_s=1$,  $\zeta=0$, - somewhat unspectacular results, because of the type of network being studied \cite{luczac}.  We emphasize that the sequence of avalanches much varies from one simulation case to another. However the $global$ dependence on $p$ disappears following case averaging, and only the $\nu$ dependence remains in a scaling way.

      Another aspect which might be also theoretically interesting is the relation between dynamic changes of the network and dynamics on the network. In science, and in many other interacting agent communities, diffusion of ideas is linked to the emergence of   coalitions \cite{drastic}. What predictions complex network theory can make for the coupling between epidemics, in a wide sense, on networks and the evolution of networks themselves is clearly of interest!
 The main challenge after this will be to make use of the various qualitative descriptions science and technology studies have produced describing such phenomena and to re-consider them in the light of complex network research, - through the parameters and variants that we have indicated.  At this time, the results seem qualitatively sound.

 
  
\vskip 1cm
{\bf Acknowledgements}
The authors would like to thank  European Commission Project 
CREEN FP6-2003-NEST-Path-012864  (CREEN : Critical Events on Evolving Networks) and
European Commission Project 
E2C2 FP6-2003-NEST-Path-012975   (E2C2: Extreme Events: Causes and Consequences)
for support. Comments by D. Stauffer are also much recognized.

\end{document}